\documentclass[11pt]{article}
\pdfoutput=1
\usepackage{amssymb}
\usepackage{amsmath}
\usepackage{ dsfont }
\usepackage{latexsym}
\usepackage[usenames]{color}
\usepackage{fancybox}
\usepackage{simplewick}
\usepackage{cite}
\usepackage{framed}
\definecolor{shadecolor}{rgb}{0.9,0.9,0.95}
\usepackage{setspace}
\usepackage{graphicx} 
\usepackage{epsfig}
\usepackage{physics}

\usepackage[utf8]{inputenc}
\usepackage[english]{babel}

\usepackage{simplewick}

\definecolor{darkred}{rgb}{0.7,0.1,.1}
\newcommand{\nullify}[1]{}

\usepackage[setpagesize=false,pagebackref=false, linktocpage, bookmarksopen=true, colorlinks=true, linkcolor=blue, filecolor=magenta,citecolor=magenta,urlcolor=darkred]{hyperref}
\numberwithin{equation}{section}

\usepackage[utf8]{inputenc}
\usepackage{tikz}
\usetikzlibrary{shapes.misc}
\usetikzlibrary{decorations.markings}
\tikzset{cross/.style={cross out, draw=black, ultra thick, minimum size=2*(#1-\pgflinewidth), inner sep=0pt, outer sep=0pt},
cross/.default={5pt}}

\usetikzlibrary{arrows.meta, positioning, calc, fit, decorations.pathreplacing}
\usepackage{amsmath}
\usepackage{tcolorbox}
\usepackage{xcolor}

\usetikzlibrary{decorations.pathmorphing}

\tikzset{snake it/.style={decorate, decoration=snake}}

\begin{document}
\thispagestyle{empty}

\renewcommand{\thefootnote}{\fnsymbol{footnote}}
\setcounter{page}{1}
\setcounter{footnote}{0}
\setcounter{figure}{0}
\definecolor{darkgreen}{cmyk}{0.9,0,0.9,0}
\newcommand{\SK}[1]{{\color{darkgreen} \bf \footnotesize{[{SK:} #1]}}}
\newcommand{\PM}[1]{{\color{red} \bf \footnotesize{[{PM:} #1]}}}
\begin{flushright}
\end{flushright}
\vspace{1cm}
\begin{center}
{\Large \bf
Chiral Composite Linear Dilaton as String Dual to Two-Dimensional Yang-Mills
\par}

\vspace{1cm}

\textrm{Shota Komatsu$^{a}$ and Pronobesh Maity$^{b}$}
\\ \vspace{1cm}
\footnotesize{$^{a}$\textit{
Department of Theoretical Physics, CERN, 1211 Meyrin, Switzerland
}} \\
\footnotesize{$^{b}$\textit{Laboratory for Theoretical Fundamental Physics, EPFL, Rte de la Sorge, Lausanne, Switzerland}}

\vspace{0.5cm}

{\tt  shota.komatsu$\otimes$cern.ch, pronobesh.maity$\otimes$epfl.ch}

\par\vspace{2cm}

\textbf{Abstract}\vspace{2mm}
\end{center}
\noindent
Two-dimensional Yang-Mills   theory (2d YM) is arguably the simplest confining gauge theory and its large $N_c$ expansion has a structure of the genus expansion in string theory.
Nevertheless various aspects of its worldsheet description have not been fully understood. 
In this paper, we elaborate on a bosonic string dual to large $N_c$ chiral 2d YM at finite 't Hooft coupling, proposed in our earlier work. The worldsheet theory consists of $\beta$-$\gamma$ system deformed by linear dilaton action built from a composite of $\gamma$. It can be seen as a noncritical version of nonrelativistic string theory introduced by Gomis and Ooguri. We provide a detailed analysis of the worldsheet operator product expansion and the computation of three- and four-point scattering amplitudes.

\setcounter{page}{1}
\renewcommand{\thefootnote}{\arabic{footnote}}
\setcounter{footnote}{0}
\setcounter{tocdepth}{2}
\numberwithin{equation}{section}
\newpage

{\color{blue!80!black}
\tableofcontents
}
\parskip 5pt plus 1pt   \jot = 1.5ex


\section{Introduction\label{sec:intro}}
String theory has been one of the most powerful frameworks in theoretical physics over the past five decades. Beyond its ambitious claim of providing a “theory of everything”  where strings replace point particles as the fundamental building blocks of the universe, it has also revealed a web of remarkable dualities. Among the most striking of these is the gauge/string duality, which relates certain gauge theories to string theories. Unlike the traditional reductionist approach underlying the “theory of everything”  program, the gauge/string duality realizes strings as emergent geometrical objects arising naturally from large $N_c$ gauge dynamics.

The idea of reformulating gauge theories in terms of strings is a longstanding one. Following ’t Hooft’s seminal observation \cite{tHooft:1973alw} that the large $N_c$ expansion of gauge theory amplitudes can be organized as a sum over two-dimensional surfaces, this intuition was given a concrete realization by Maldacena in the celebrated AdS/CFT correspondence \cite{Maldacena:1997re}. The correspondence provides a precise duality between $\mathcal{N}=4$ Super Yang–Mills theory at large $N_c$ and type IIB string theory on $AdS_5\times S^5$. Despite its success, extending this duality to more realistic, non-supersymmetric gauge theories such as four-dimensional Quantum Chromodynamics (QCD), which governs hadronic physics, remains an outstanding challenge. A concrete string dual of QCD would illuminate the strongly coupled infrared regime of the theory, where field-theoretic methods offer limited analytical control.

As a first step toward this goal, it is natural to seek the string dual of two-dimensional pure Yang–Mills theory (2d YM), which is exactly solvable on an arbitrary Riemann surface even at finite $N_c$. Furthermore, when coupled to matter fields, 2d QCD exhibits color confinement and an infinite tower of resonances \cite{tHooft:1974pnl}, thus serving as an ideal toy model to explore the stringy mechanisms underlying confinement in four dimensions.

Indeed, in their seminal work \cite{Gross:1993hu}, Gross and Taylor proposed a string interpretation of 2d YM by expanding its partition function in powers of $1/N_c$ and identifying it as a sum over branched coverings of the target space by an auxiliary worldsheet. However, the explicit worldsheet action governing the dynamics of these strings was not fully elucidated. There have been important subsequent developments \cite{Cordes:1994sd, Horava:1995ic, Vafa:2004qa, Benizri:2025xmz, Aharony:2023tam, Aharony:2025owv} but each of them has its own limitation; they are either limited (so far) to the zero-coupling limit of the theory \cite{Cordes:1994sd,Aharony:2023tam,Aharony:2025owv}, or the worldsheet description is not yet fully developed \cite{Benizri:2025xmz}, or the resulting worldsheet descriptions were too intricate for practical computations \cite{Horava:1995ic}. One important exception may be the topological string dual\footnote{There is also a closely related proposal on topological string dual to $q$-deformed 2d YM \cite{Aganagic:2004js}. However, the limit $q\to 1$ is subtle as mentioned already in \cite{Aganagic:2004js}.} to 2d YM on a torus, proposed in \cite{Vafa:2004qa}, which claimed to reproduce the full (non-chiral) partition function at finite coupling and finite $N_c$. However, its status at finite $N_c$ is a subject of current scrutiny \cite{Okuyama:2019rqn,Chen:2025yhg}  and further studies seem necessary. In addition, being formulated as topological string, it is not clear how to compute observables beyond the partition function.

In a recent proposal \cite{Komatsu:2025sqo}, we introduced a bosonic string dual to large $N_c$ chiral\footnote{At large $N_c$, the partition function of 2d YM factorizes into chiral and anti-chiral sectors: $$ \mathcal{Z}_{\mathcal{M}}\to \mathcal{Z}^{\text{Chiral}}_{\mathcal{M}} \times \mathcal{Z}^{\text{Anti-chiral}}_{\mathcal{M}} \, ,$$ corresponding to the orientation-preserving and orientation-reversing branched coverings, respectively, in the Gross-Taylor picture \cite{Gross:1993hu}. } 2d YM at finite ’t Hooft coupling. The proposed worldsheet theory is formulated as a $\beta$-$\gamma$ system deformed by a chiral analog of the Polchinski–Strominger term \cite{Polchinski:1991ax} in the Polyakov formalism \cite{Hellerman:2014cba}. Alternatively, it may be viewed as a noncritical version of the nonrelativistic string theory introduced by Gomis and Ooguri \cite{Gomis:2000bd, Danielsson:2000gi}. A crucial ingredient in this noncritical generalization is the introduction of a chiral composite linear dilaton (CLD) field, constructed as a composite of the $\gamma$ fields, whose detailed structure will be discussed in the main text. An advantage of our formulation is that it allows for explicit computation of observables beyond the partition function, including the scattering amplitudes. 

Our construction is distinctive in several ways. It introduces no additional scalar mode on the worldsheet \cite{Teper:2009uf,Athenodorou:2010cs,Athenodorou:2011rx,Dubovsky:2013gi,Dubovsky:2014fma}, unlike typical holographic models, and requires no worldsheet supersymmetry. However, the CLD action is highly non-linear, and certain operator product expansions (OPEs) are non-standard, involving inverse fields. A key result of this work is the demonstration that the string scattering amplitudes are explicitly computable owing to a remarkable localization property in the string path integral. Moreover, we show that the stress-tensor OPEs are fully consistent with conformal invariance, establishing our proposal as a consistent framework for a noncritical string theory.

Finally, we evaluate the moduli integral of the four-point string amplitude using a KLT-like factorization introduced in \cite{Komatsu:2025onf}. Although the amplitude diverges, by adapting the analysis of \cite{Komatsu:2025onf} to our setting, we extract the leading behavior in the $s$-, $t$-, and $u$-channels and find precise agreement with the corresponding behavior in the dual 2d YM theory. This provides a non-trivial consistency check and further evidence supporting our duality proposal \cite{Komatsu:2025sqo}.

The structure of this paper is as follows. In Sec.~\ref{sec:2dYM}, we elaborate on the worldsheet dual to chiral 2d YM proposed in our earlier work~\cite{Komatsu:2025sqo}, and present explicit expressions for the vertex operators associated with the winding strings. We also discuss the operator product expansion (OPE), postponing the derivation of the $\beta$-$\beta$ OPE to a later section. In Sec.~\ref{sec:amplitudes}, we compute the corresponding scattering amplitudes from both the Yang–Mills and the dual worldsheet descriptions, and match the results. The worldsheet analysis involves a detailed evaluation of the string path integral and a KLT-like factorization of the four-point amplitudes. A key ingredient in this computation is the evaluation of the localized CLD action originally due to Mandelstam \cite{Mandelstam:1985ww}. In Sec.~\ref{sec:Mandelstam}, we present the details of the evaluation of the CLD action including a subtle issue of regularizing and renormalizing the action using a cutoff. We also explain the derivation of $\beta$-$\beta$ OPE from the CLD action.   Sec.~\ref{sec:conclusion} contains our concluding remarks. Appendix~\ref{two-point-section} reviews the computation of the two-point string amplitudes, clarifying the Fadeev–Popov procedure, while Appendix~\ref{Alternative_Mandelstam} presents an alternative derivation of Mandelstam’s formula.

\section{Worldsheet dual to chiral 2d Yang-Mills\label{sec:2dYM}}
Here we explain details of the worldsheet action proposed in \cite{Komatsu:2025sqo}, which we conjecture to be dual to large $N_c$ chiral 2d Yang-Mills theory. We present the worldsheet action and its stress tensor, the operator product expansion (OPE) of fundamental fields, and vertex operators corresponding to winding strings in the chiral 2d YM.

\subsection{Worldsheet action and OPEs}

\paragraph{Action.}Our proposed worldsheet action is 
\begin{equation}\label{action1}
\begin{split}
S=&\int \frac{d^2z}{2\pi} \left[\beta\, \bar{\partial}X^{+}+\bar{\beta}\,\partial X^{-}+\frac{q}{2}\mathcal{L}_{CLD}+\frac{\lambda \pi}{2}\left(\partial X^{+} \bar{\partial} X^{-}-\bar{\partial}X^{+}\partial X^{-}\right)\right]+ \text{bc-ghost} \, ,
\end{split}
\end{equation}
where $\lambda=g_{\text{YM}}^2 N_c$ is the 't Hooft coupling, $X^{\pm}=(X^1\pm X^0)$ are light-cone co-ordinates of the target space, and $\mathcal{L}_{CLD}$ is the chiral composite linear dilaton (CLD) action  
\begin{align}\label{eq:CCLDaction}
    \mathcal{L}_{CLD}=2\,\partial \varphi\,\bar{\partial}\varphi+\hat{R}\,\varphi\,, \quad \varphi =\log (\frac{\partial X^{+}\bar{\partial}X^{-}}{R^2}) \,,
\end{align}
with $\hat{R}$ is worldsheet Ricci scalar, and $R$ is the length-scale of the target space. Here $q$ is not yet determined but we will fix it to one ($q=1$) below by imposing the conformal anomaly cancellation. 

\paragraph{Stress tensor and operator product expansion.}To compute the stress tensor, it is convenient to first express the action in a covariant form. We start with the $\beta$-$\gamma$ system, which reads
\begin{equation}
      \int \frac{d^2z}{2\pi}\beta_z\,\partial_{\bar{z}}X^{+}\to \int d^2\sigma \sqrt{-g}\,\frac{1}{\pi} \left( \frac{1}{4}g^{ab}\beta_a\,\partial_b X^{+}-\frac{1}{4i}\epsilon^{ab}\beta_a\,\partial_b X^{+}\right) \, ,
\end{equation}
and similarly for the $\bar{\beta}\partial X^{-}$ term. Taking a derivative with respect to $g^{zz}$, we obtain 
\begin{equation}
    T_{\beta\gamma}(z)=-\frac{4\pi}{\sqrt{-g}}\frac{\delta S_{\beta\gamma}}{\delta g^{zz}}=-\beta\, \partial X^{+} \, ,
\end{equation}
where we have used $\delta \epsilon^{ab}/ \delta g^{zz}=\frac{1}{2}\epsilon^{ab} g_{zz}$, with $g_{zz}=0$ in the conformal gauge. Next we consider the term proportional to $\lambda$, to be called the ``tensionfull term" in what follows:
\begin{equation}
	\left(\partial X^{+} \bar{\partial} X^{-}-\bar{\partial}X^{+}\partial X^{-}\right)\to\frac{i}{2}\,\epsilon^{cd} \partial_c X^{+}\partial_d X^{-}  \, ,
\end{equation}
Being given purely in terms of $\epsilon^{ab}$, this term does not contribute to the stress tensors. Finally, the covariant form of the CLD action is giiven by the following expression 
\begin{equation}\label{Gamma}
	\left(\Gamma[X^{\pm}]\right)_{\text{covariant}}=\frac{q}{2}\int \frac{d^2\sigma}{2\pi}\sqrt{-g}\left[g^{ab}\,\partial_a\Phi \, \partial_b \varphi+2\hat{R}\,\varphi\right] \, ,
\end{equation}
where 
\begin{equation}\label{Covariant-Phi}
\varphi = \log\left[ \frac{1}{4} \, g^{ab}\partial_a X^{+}\partial_b X^{-}+\frac{i}{4}\,\epsilon^{ab} \partial_a X^{+}\partial_b X^{-} \right] \, .
\end{equation}
One can now compute its contribution as follows
\begin{equation}
    \begin{split}
        T_{CLD}(z)=-\frac{4\pi}{\sqrt{-g}}\frac{\delta \Gamma}{\delta g^{zz}}=-q \,(\partial \varphi)^2+2q\,\partial^2\varphi=2q\,\{ X^{+},z\} \, ,
    \end{split}
\end{equation}
where $\{ X^{+},z\}:=\partial^3 X^{+}/\partial X^{+}-3(\partial^2 X^{+}/\partial X^{+})^2/2 $, is the Schwarzian derivative of $X^{+}$. Adding these contributions, the stress tensor of the matter sector takes the form
\begin{align}
\begin{aligned}\label{eq:stress}
    T(z)&=-\beta\,\partial X^+(z)+2q\,\{ X^{+},z\}  \, \\
    &=-\beta  \partial X^{+}+2q  \partial^2 \log \partial X^{+}-q(\partial \log \partial X^{+})^2\,.
    \end{aligned}
\end{align}


From the worldsheet action, one can compute the OPEs:
\begin{equation}\label{OPE_first}
\begin{split}
&\beta(z) X^{+}(w)\sim -\frac{1}{z-w},\quad \bar{\beta}(\bar{z})X^{-}(\bar{w})\sim-\frac{1}{\bar{z}-\bar{w}} \, ,\\
&\beta(z)\beta(w)\sim 2q \, \partial_z\partial_y \left[\frac{1}{(z-w)^2}\,\frac{1}{\partial_z X^{+}(z)\,\partial_w X^{+}(w)}\right],\quad \text{its cc} \, .
\end{split}
\end{equation}
The $\beta$-$\gamma$ OPEs are straightforward to derive, whereas the $\beta$-$\beta$ OPE requires a more elaborate computation, which we present in section \ref{beta-beta_OPE}. 

\paragraph{Computation of $TT$ OPE.}Using these OPEs, the OPE between stress tensors \eqref{eq:stress} can be computed. First consider the following OPE, 
\begin{equation}
	\begin{split}
	&   :\beta(z)\partial_z X^{+}(z):\,:\beta(w)\partial_w X^{+}(w):\\=&: \contraction{}{\beta(z)}{\partial_z X^{+}(z):\,:}{\beta(w)}\beta(z)\partial_z X^{+}(z):\,:\beta(w)\partial_w X^{+}(w):+:\contraction{}{\beta(z)}{\partial_z X^{+}(z):\,: \partial_w}{X^{+}(w)} \beta(z)\partial_z X^{+}(z):\,:\beta(w)\partial_w X^{+}(w):\\&+:\contraction{\beta(z)}{ \partial_z X^{+}(z)}{:\,:}{\beta(w)} \beta(z)\partial_z X^{+}(z):\,:\beta(w)\partial_w X^{+}(w):+: \contraction{}{\beta(z)}{ \partial_z X^{+}(z):\,:\beta(w)\partial_w}{X^{+}(w)} \contraction[2ex]{\beta(z)\partial_z }{X^{+}(z)}{:\,:}{\beta(w)}\beta(z)\partial_z X^{+}(z):\,:\beta(w)\partial_w X^{+}(w):\\=&2q \,\partial_z\partial_w \left[\frac{1}{(z-w)^2}\,\frac{1}{\partial_z X^{+}(z)\partial_w X^{+}(w)}\right]\,\partial_z X^{+}(z)\partial_w X^{+}(w)
	\\&-\frac{\partial_z X^{+}(z)\beta(w)+\beta(z)\partial_w X^{+}(w)}{(z-w)^2}+\frac{1}{(z-w)^4} \, .
	\end{split}
	\end{equation}
 
 Next for the $\partial^2$ part of the stress tensor, the relevant OPEs are
\begin{equation}\label{partial_2_part}
	-2q \, \beta(z) \partial_z X^{+}(z)\, \partial_w^2 \log \partial_w X^{+}(w) -2q\, \partial_z^2 \, [\log \partial_z X^{+}(z)]\, \beta(w)\partial_w X^{+}(w)  \, .
	\end{equation}
	For the first term we compute
	\begin{equation}
	\begin{split}
	-2q\, \beta(z) \partial_z X^{+}(z)\, \partial_w^2 \log \partial_w X^{+}(w)&=	-2q \, \partial_z X^{+}(z) \,\partial_w^2 \left[ \contraction{}{\beta(z)}{\,\log \partial}{  X^{+}(w)}\beta(z)\,\log \partial_w X^{+}(w)\right]\\
	&=	2q \, \partial_z X^{+}(z) \,\partial_w^2 \left[\frac{1}{\partial_w X^{+}(w)}\frac{1}{(z-w)^2}\right] \, ,
	\end{split}
	\end{equation}
    where we used the following OPE,\begin{equation}\label{eq:betaloggamma}
		\beta(z)\, \log\partial_w X^{+}(w)\sim -\frac{1}{(z-w)^2}\frac{1}{\partial_w X^+(w)} \, .
	\end{equation}
    A quick way\footnote{A similar analysis of OPE is discussed in Appendix A of \cite{Callebaut:2019omt}.} to derive this OPE \eqref{eq:betaloggamma} is to use the ``replica trick". Namely we consider the OPE between $\beta(z)$ and $\left(\partial_w X^{+}(w)\right)^{n}$ for $n\in \mathbb{Z}_{\geq 0}$, analytically continue in $n$, and take the limit  $\lim_{n\to 0}\left[\left(\partial_w X^{+}(w)\right)^{n}-1\right] /n$. This is of course not a fully rigorous procedure but the result can be confirmed by an independent path integral analysis.
    Similarly, the second term in \eqref{partial_2_part} yields
	\begin{equation}
	-2q\, \partial_z^2 \, [\log \partial_z X^{+}(z)]\, \beta(w)\partial_w X^{+}(w)=2q \, \partial_w X^{+}(w) \,\partial_z^2 \left[\frac{1}{\partial_z X^{+}(z)}\frac{1}{(z-w)^2}\right] \, .
	\end{equation}

 Finally, we consider $(\partial \log \partial X^{+})^2$ piece of the stress tensor, where relevant OPEs are
	\begin{equation}
	q \, \beta(z)\,\partial_z X^{+}(z) \, (\partial_w \log \partial_w X^{+}(w))^2	+q\, \beta(w)\partial_w X^{+}(w) \, (\partial_z \log \partial_z X^{+}(z))^2 \, .
	\end{equation}
	As before, for the first term we compute
	\begin{align}
	\begin{aligned}
	&q \, \beta(z)\,\partial_z X^{+}(z) \, (\partial_w \log \partial_w X^{+}(w))^2 = 2q \, \partial_z X^{+}(z) \,\partial_w \left[ \contraction{}{\beta(z)}{\,\log }{ \partial_w X^{+}(w)}  \beta(z)\,\log \partial_w X^{+}(w)\right] \,\partial_w \log \partial_w X^{+}(w) \\
	&=	-2q \, \partial_z X^{+}(z) \,\partial_w \left[\frac{1}{\partial_w X^{+}(w)}\frac{1}{(z-w)^2}\right]\,\partial_w \log \partial_w X^{+}(w) \, .
	\end{aligned}
	\end{align}
	Similarly we have
	\begin{equation}
	q \, \beta(w)\,\partial_w X^{+}(w) \, (\partial_z \log \partial_z X^{+}(z))^2= -2q \, \partial_w X^{+}(w) \,\partial_z \left[\frac{1}{\partial_z X^{+}(z)}\frac{1}{(z-w)^2}\right]\,\partial_z \log \partial_z X^{+}(z) \, .
	\end{equation}
Expanding in a Laurent series around $z=w$ and combining all contributions gives
\begin{equation}
	T(z)T(w)=\frac{1+12q}{(z-w)^4}+\frac{2T(w)}{(z-w)^2}+\frac{\partial_w T(w)}{z-w} \, .
\end{equation}
For conformal anomaly cancellation, we require $q=1$.

\paragraph{Comments.} As mentioned in \cite{Komatsu:2025sqo}, the $\beta$-$\gamma$ system along with the term proportional to $\lambda$ arises in non-relativistic string theory \cite{Gomis:2000bd,Danielsson:2000gi,Danielsson:2000mu}. The main difference from the standard non-relativistic string action is the presence of the CLD term, which allows its noncritical generalization. The latter may be seen as the chiral analogue of the Polchinski-Strominger term  \cite{Polchinski:1991ax} in the Polyakov formalism \cite{Hellerman:2014cba}. 
Integrate $\beta$'s out, $X^{+}\,(X^{-})$ are set to be (anti-)holomorphic --- which underlies the reason why we can perform the string path integral explicitly despite the highly non-linear terms introduced by $\mathcal{L}_{CLD}$.

Note that we always have the constant dilaton field $\Phi_0$, and $\varphi$ can be viewed as the ``dynamical correction" to the total dilaton field $\Phi=\frac{1}{2}\Phi_0+\varphi$, so that the constant field contributes to the usual factor of $g_s^{-\chi}$ in the string path integral with $g_s$ being closed string coupling: 
   \begin{equation}
      e^{-S}\to e^{-\frac{q}{2}\int \frac{d^2z}{2\pi}\hat{R}\frac{1}{2}\Phi_0}=\left( g_s\right)^{-q\,\chi},\quad \text{with}\quad g_s=e^{\Phi_0} \, .
   \end{equation}

\subsection{Vertex operators}
With no local dynamics, the physical degrees of freedom of chiral 2d YM on a cylinder are its holonomies which correspond to winding string states\cite{Minahan:1993np}, as we will discuss in sec \ref{sec:Amplitudesofchiral2dYM}. The vertex operator for such a winding string reads
\begin{align}\label{vertex_operator}
\mathcal{V}_k (z_k,\bar{z}_k)=e^{iw_kR\int^{z_k} (\beta dz- \bar{\beta}d\bar{z})}   e^{-i E_kX^{0}},\quad \text{with}\quad E_k=\lambda\pi Rw_k\,.
\end{align}
This creates a string state winding $w_k$ times around the $X^1$ direction in the target space, as is evident from the following OPEs:
\begin{equation}
\begin{split}
   & X^{+}(z,\bar{z}) \mathcal{V}_{k}(z_k,\bar{z}_k) \sim -iw_kR\, \log(z-z_k) \mathcal{V}_{k}(z_k,\bar{z}_k),\\
   &X^-(z,\bar{z}) \mathcal{V}_k(z_k,\bar{z}_k)\sim +iw_k R\, \log(\bar{z}-\bar{z}_k) \mathcal{V}_k(z_k,\bar{z}_k)\, ,
\end{split}    
\end{equation}
which imply:
\begin{equation}
X^1(z,\bar{z})\mathcal{V}_k(z_k,\bar{z}_k)\sim -\frac{i}{2}w_kR\,\log\left[ \frac{z-z_k}{\bar{z}-\bar{z}_k}\right]    \mathcal{V}_k(z_k,\bar{z}_k)\, ,
\end{equation}
thus, as we encircle around $(z_k,\bar{z}_k)$, i.e under $z\to z_k+(z-z_k)\,e^{2\pi i},\; \bar{z}\to \bar{z}_k + (\bar{z}-\bar{z}_k)\, e^{-2\pi i}$, the field $X^1(z,\bar{z})$ acquires a monodromy of $2\pi Rw_k$.

In Secs.~\ref{Two_and_Three-point_amplitudes} and \ref{four-point_amplitude}, we will verify that the vertex operators \eqref{vertex_operator} are conformal primaries of dimension $(1,1)$ by analyzing their worldsheet correlators.

\section{Scattering amplitudes \label{sec:amplitudes}}
In the previous section, we proposed a worldsheet dual to chiral 2d YM and introduced the vertex operators for the winding strings. In this section we compute the corresponding scattering amplitudes in both Yang–Mills theory and its dual string description, and match them to find the parameter dictionary of our duality. In particular, we evaluate the four-point string amplitude using a KLT-like factorization and show how it matches the amplitude obtained from the dual Yang–Mills theory.  We start below with the YM amplitudes.

\subsection{Amplitudes of chiral 2d Yang-Mills}\label{sec:Amplitudesofchiral2dYM}
The wave function of chiral 2d YM on $\mathbb{S}^1$ is defined by the path integral with holonomy $U$ as the boundary condition. As a class function of $U$, it can be expanded in irreducible representations of $U(N_c)$. A convenient basis for this at large $N_c$ is the multi-trace basis $\{|\sigma\rangle\}$,
\begin{align}
		\Upsilon_{\sigma}(U)= \langle U | \sigma\rangle =\prod_{\ell} \left(\text{tr} (U^\ell)\right)^{k_{\ell}}\, ,
	\end{align}
where $k_{\ell}$ denotes the number of cycles of length $\ell$ in $\sigma \in S_{r}$ with $r=\sum_{\ell} \ell k_{\ell}$. As shown by Minahan and Polychronakos in \cite{Minahan:1993np}, $|\sigma\rangle$ can be interpreted as a multi-string state with $k_{\ell}$ strings winding $\ell$ times around $\mathbb{S}^1$. In the canonical picture, one can express it with string creation operators acting on the vacuum of the free theory $|0 \rangle$: 
\begin{align}
|\sigma\rangle =\prod_{\ell}(a_{\ell}^{\dagger})^{k_{\ell}}\, | 0\rangle\,,
\end{align}
where $[a_{k},a_{l}^{\dagger} ]=k\,\delta_{k,l}$. The Hamiltonian on $S^1$ of radius $R$ is then given by $H=H_{\rm free}+H_{\rm int}$ with \cite{Minahan:1993np}
\begin{align}
\begin{aligned}
H_{\rm free}=\pi \lambda R \sum_{m}a_{m}^{\dagger}a_m\, ,\quad \text{and}\quad
H_{\rm int}=\frac{\pi R \lambda}{N_c}\sum_{n}\sum_{m} \left( a_{m+n}^{\dagger}a_m a_n +a_m^{\dagger}a_n^{\dagger}a_{m+n}\right)  \,.
\end{aligned}
\end{align}
The free hamiltonian $H_{\rm free}$ counts the winding number as expected for the Nambu-Goto contribution, while the interaction term describes single branch points in the Gross–Taylor covering map picture. Concretely, the operator
\begin{equation}
    a_{m+n}^{\dagger}a_m a_n 
\end{equation}
splits two strings of windings $m,n$ and glues them into a single string of winding $(m+n)$, while its Hermitian conjugate implements the inverse process.

The two-point amplitude of these strings with winding numbers $w_1$ and $w_2$ is 
\begin{equation}\label{YM_two-point_amplitude}
    \mathcal{S}_2= \langle 0| a_{w_1} a_{w_2}^{\dagger}| 0 \rangle = w_1\, \delta_{w_1,w_2}\, .
\end{equation}
Between an in-state $|\Psi_{\text{in}}\rangle=a^{\dagger}_{w_1}a^{\dagger}_{w_2}| 0 \rangle$, and out-state $ |\Psi_{\text{out}}\rangle = a^{\dagger}_{w_3}|0\rangle$, the three-point amplitude for $2\to 1$ scattering process reads
\begin{equation}
    \mathcal{S}_3= \langle \Psi_{\text{out}}| \exp\left[-i\int_{-\infty}^{\infty}H_I(t')\, dt'\right] |\Psi_{\text{in}}\rangle \, ,
\end{equation}
with $H_I(t')=e^{iH_{\rm free}t'} \,H_{\rm int}\, e^{-iH_{\rm free}t'}$. Evaluating the relevant matrix element,
\begin{equation}
    \langle 0| a_{w_3}   \sum_{n}\sum_{m} \left( a_{m+n}^{\dagger}a_m a_n +a_m^{\dagger}a_n^{\dagger}a_{m+n}\right)  \, a_{w_1}^\dagger  a_{w_2}^\dagger|0 \rangle = 2w_1w_2w_3\, \delta_{w_1+w_2,w_3} \, ,
\end{equation}
the S-matrix becomes
\begin{equation}\label{YM_three-point_amplitude}
   \mathcal{S}_3=-i\frac{4\lambda\pi^2 R}{N_c}w_1w_2w_3\,\delta(E_1+E_2-E_3)\,\delta_{w_1+w_2,w_3}\, .
\end{equation}
In what follows, we will reproduce it from the three-point string scattering amplitudes. In Sec.~\ref{four-point_amplitude}, we will show that the four-point amplitude factorizes into these three-point amplitudes in the $s$-, $t$-, and $u$-channel exchanges.

\subsection{Worldsheet correlator}
In this subsection, we evaluate the worldsheet correlator of vertex operators \eqref{vertex_operator} via the string path integral, and show that localization of the worldsheet matter fields allows us to obtain an explicit expression for the correlator with $n$ insertions. 

We start with the path-integral expression of $n$-point worldsheet correlator of vertex operators \eqref{vertex_operator} which reads
\begin{equation}\label{n-point_matter}
	G^{\text{matter}}_n=g_s^{n-2}\,\langle \prod_{k=1}^{n} \, V_k(z_k,\bar{z}_k)\rangle = g_s^{n-2}\int [\mathfrak{D}X^{+}\mathfrak{D}X^{-}\mathfrak{D}\beta \,\mathfrak{D}\bar{\beta}] \, e^{-\boldsymbol{S}} \, .
\end{equation}
The net worldsheet action $\boldsymbol{S}$, including the contributions from the vertex operators, can be decomposed into three parts:
\begin{equation}
    \begin{split}
      \boldsymbol{S}= \left( S_{\beta \text{-}\bar{\beta}  }+S_\lambda+\Gamma \right)[X^\pm] \, ,
    \end{split}
\end{equation}
where $S_{\beta \text{-}\bar{\beta}  }$ is $\beta$ and $\bar{\beta}$ dependent part, while $S_\lambda$ depends on the string tension $\lambda$:
\begin{equation}
    \begin{split}
         & S_{\beta \text{-}\bar{\beta}  }[X^\pm]:= \int\frac{d^2z}{2\pi} \left(\beta\,\bar{\partial} X^{+}+\bar{\beta}\,\partial X^{-} \right) -i\sum_{k=1}^{n}w_kR\int^{z_k}dz'\beta(z')+i\sum_{k=1}^{n}w_kR\int^{\bar{z}_k}d\bar{z}'\bar{\beta}(\bar{z}') \, ,\\
         & S_\lambda [X^\pm] := \frac{\lambda}{4}\int \frac{d^2z}{2\pi} \left( \partial X^{+} \bar{\partial}X^{-}- \bar{\partial}X^{+}\partial X^{-} \right) +i\sum_{k=1}^{n}E_k X^0(z_k,\bar{z}_k) \, ,
    \end{split}
\end{equation}
and $\Gamma[X^\pm]$ is the CLD action already defined in \eqref{Gamma}. We can rephrase $S_{\beta \text{-}\bar{\beta}  }[X^\pm]$ as
\begin{equation}
    S_{\beta \text{-}\bar{\beta}  }[X^\pm]= \int\frac{d^2z}{2\pi} \beta(z)\bar{\partial} [X^{+}-R\rho] +\int\frac{d^2z}{2\pi} \beta(z)\partial [X^{-}-R\bar{\rho}] \, ,
\end{equation}
where $\rho$ and $\bar{\rho}$ are Mandelstam maps\cite{Mandelstam:1973jk}: 
\begin{equation}\label{Mandelstam_maps}
    \rho(z):= -i\sum_{k=1}^{n}w_k \log(z-z_k),\quad \bar{\rho}(\bar{z}):= i\sum_{k=1}^{n}w_k \log(\bar{z}-\bar{z}_k) \, .
\end{equation}
The string coupling is denoted by $g_s$. Integrating out $\beta$ and $\bar{\beta}$ in the string path integral \eqref{n-point_matter}, $X^\pm$ gets localized onto $(\rho,\bar{\rho})$: 
\begin{equation}
    \int [\mathfrak{D}\beta \, \mathfrak{D}\bar{\beta}] \,e^{-  S_{\beta \text{-}\bar{\beta}  }[X^\pm]}=\mathcal{N}\left[\text{det}'\left(\frac{\partial\bar{\partial}}{4\pi^2}\right)\right]^{-1} \delta(X^+-R\rho)\,\delta(X^{-}-R\bar{\rho}) \, ,
\end{equation}
where $\mathcal{N}$ denotes the normalization factor arising from the measure, and and the prime on the determinant indicates that the zero modes of $\partial\bar{\partial}$ are excluded. 

We can perform the integrals over zero modes of $X^0$ and $X^1$:
\begin{equation}
    \int_{0}^{2\pi R}dx^1\int_{-\infty}^{\infty}dx^0\, e^{-i\left(\sum_{k=1}^{n}E_k\right) x^0}=4\pi^2 R\,\delta\left(\sum_k E_k\right) \, .
\end{equation}
Thus, $G^{\text{matter}}_n$ can be expressed as 
\begin{equation}
    G^{\text{matter}}_n=ig_s^{n-2}\mathcal{N}_{S^2}\cdot4\pi^2 R\,\delta\left(\sum_k E_k\right) \delta_{\sum_k w_k}\left[\text{det}'\left(\frac{\partial\bar{\partial}}{4\pi^2}\right)\right]^{-1} e^{-S_\lambda[\rho]-\Gamma[\rho]} \, ,
\end{equation}
where we have absorbed sphere partition function and other constants from the measure in $\mathcal{N}\to \mathcal{N}_{S^2}$. The factor $\delta_{\sum_k w_k}$ reflects the requirement that, after localizing onto $(\rho,\bar{\rho})$, the fields $X^\pm$ should remain analytic at $\infty$ in the absence of any vertex operator insertion there. The correlator of ghost insertions at $z_a,z_b$ and $z_c$ has the familiar expression given by
\begin{equation}
     G^{\text{ghost}}_n= \left[\text{det}'\left(\frac{\partial\bar{\partial}}{4\pi^2}\right)\right]|z_{ab}z_{bc}z_{ca}|^2 \, .
\end{equation}
Combining the contributions from both the matter and ghost sectors, the $n(>2)$-point worldsheet correlator with fixed points at $z_a,z_b$ and $z_c$, takes the form
\begin{equation}\label{G_n-1}
    G_n=ig_s^{n-2}\mathcal{N}_{S^2}\cdot4\pi^2 R\,\delta\left(\sum_k E_k\right) \delta_{\sum_k w_k} |z_{ab}z_{bc}z_{ca}|^2\, e^{-S_\lambda[\rho]-\Gamma[\rho]} \, .
\end{equation}
In the following we compute $S_\lambda[\rho]$. 
\paragraph{Computation of $S_\lambda[\rho]$:}
We first compute the tensionful term of our worldsheet action \eqref{action1} with $(X^{+}, X^{-})$ localized at the Mandelstam maps $(\rho,\bar{\rho})$: 
\begin{equation}\label{Mandelstam_general_maps}
	X^{+}(z)\to\rho(z)=\sum_{k=1}^{n}\alpha_k \log(z-z_k),\quad X^{-}(\bar{z})\to \bar{\rho}(\bar{z})=\sum_{k=1}^{n}\bar{\alpha}_k\log(\bar{z}-\bar{z}_k)\, ,
\end{equation}
where we introduced $\alpha_k =-iw_k$ and $\bar{\alpha}_k=iw_k$ to simplify the expressions below.
For convenience we strip off the factor of $i\lambda/4$ from the tensionful term and consider the worldsheet integral
\begin{equation}
	S_{t}[\rho]=-iR^2\int_{\mathcal{M}} d^2z \left( \partial\rho \, \bar{\partial} \bar{\rho}-\bar{\partial}\rho \, \partial \bar{\rho}\right)=R^2\int_{\mathcal{M}}d\rho\wedge d\bar{\rho} \, .
\end{equation}
Note that we have identified $dz\wedge d\bar{z}=-id^2z$. We can make the above integrand into a total derivative and evaluate it as a boundary integral: 
\begin{equation}\label{boundary_integral}
	\frac{S_{t}[\rho]}{R^2}=\frac{1}{2}\int_{\mathcal{M}} d\left(\rho\, d\bar{\rho}-\bar{\rho}\,d\rho\right)=\frac{1}{2}\int_{\partial\mathcal{M}} \left(\rho\, d\bar{\rho}-\bar{\rho}\,d\rho\right) \, .
\end{equation}

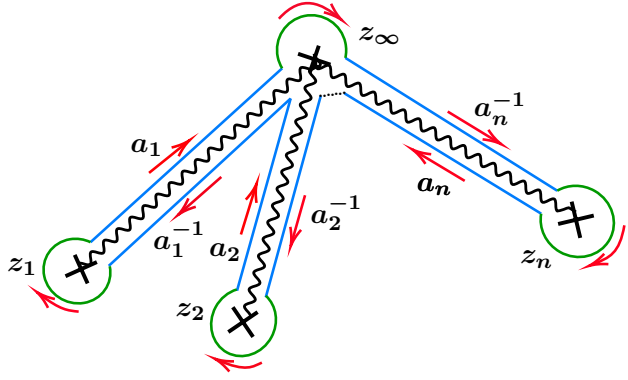
\begin{figure}[!htb]

\hspace{2cm}
\resizebox{0.6\linewidth}{!}{%
\begin{tikzpicture}[x=0.75pt,y=0.75pt,yscale=-1,xscale=1]

\useasboundingbox (0cm, 0cm) rectangle (15cm,9cm); 


\draw  [line width=2.25]  (295.66,92.32-4) -- (320.4,100.62-4) (311.5,86.14-4) -- (304.56,106.8-4) ;
\draw  [line width=2.25]  (117.93,258.41) -- (141.92,247.59)(125,242.09) -- (134.85,263.92) ;
\draw  [line width=2.25]  (242.36,301.1) -- (264.74,285.85)(245.81,282.13) -- (261.29,304.82) ;
\draw  [line width=2.25]  (494.38,218.48) -- (521.62,212.02)(504.71,201.39) -- (511.29,229.11) ;

\draw [color={rgb, 255:red, 0; green, 124; blue, 252 }  ,draw opacity=1 ][fill={rgb, 255:red, 30; green, 123; blue, 234 }  ,fill opacity=1 ][line width=1.5]    (283.51,99.42) -- (137.07,232.4) ;
\draw [color={rgb, 255:red, 0; green, 124; blue, 252 }  ,draw opacity=1 ][line width=1.5]    (290,122.5) -- (151,250.5) ;
\draw [color={rgb, 255:red, 0; green, 124; blue, 252 }   ,draw opacity=1 ][line width=1.5]    (290,122.5) -- (249,269.5) ;
\draw [color={rgb, 255:red, 0; green, 124; blue, 252 }   ,draw opacity=1 ][line width=1.5]    (313,120.5) -- (271,274.5) ;
\draw [color={rgb, 255:red, 0; green, 124; blue, 252 }  ,draw opacity=1 ][line width=1.5]    (331.79,91.45) -- (493.7,193.04) ;
\draw [color={rgb, 255:red, 0; green, 124; blue, 252 }   ,draw opacity=1 ][line width=1.5]    (331,118.5) -- (481.43,210.59) ;

\draw  [draw opacity=0][line width=1.5]  (151.75,250.83) .. controls (154.42,264.11) and (145.68,276.59) .. (132.05,278.83) .. controls (118.27,281.09) and (104.7,272.02) .. (101.75,258.58) .. controls (98.8,245.14) and (107.57,232.41) .. (121.36,230.15) .. controls (126.85,229.25) and (132.31,230.14) .. (137.07,232.4) -- (126.7,254.49) -- cycle ;

\draw  [green!60!black ,draw opacity=1 ][line width=1.5]  (151.75,250.83) .. controls (154.42,264.11) and (145.68,276.59) .. (132.05,278.83) .. controls (118.27,281.09) and (104.7,272.02) .. (101.75,258.58) .. controls (98.8,245.14) and (107.57,232.41) .. (121.36,230.15) .. controls (126.85,229.25) and (132.31,230.14) .. (137.07,232.4) ;

\draw  [draw opacity=0][line width=1.5]  (283.51,99.42) .. controls (281.97,96.89) and (280.82,94.05) .. (280.17,90.98) .. controls (277.11,76.57) and (286.19,62.43) .. (300.46,59.4) .. controls (314.73,56.37) and (328.77,65.6) .. (331.83,80.02) .. controls (332.66,83.92) and (332.6,87.8) .. (331.79,91.45) -- (306,85.5) -- cycle ;

\draw  [green!60!black ,draw opacity=1 ][line width=1.5]  (283.51,99.42) .. controls (281.97,96.89) and (280.82,94.05) .. (280.17,90.98) .. controls (277.11,76.57) and (286.19,62.43) .. (300.46,59.4) .. controls (314.73,56.37) and (328.77,65.6) .. (331.83,80.02) .. controls (332.66,83.92) and (332.6,87.8) .. (331.79,91.45) ;

\draw  [draw opacity=0][line width=1.5]  (271.07,273.41) .. controls (279.21,280.89) and (281.26,293.91) .. (275.33,304.96) .. controls (268.41,317.83) and (253.25,323.15) .. (241.47,316.82) .. controls (229.7,310.49) and (225.75,294.92) .. (232.67,282.04) .. controls (236.32,275.25) and (242.27,270.56) .. (248.81,268.57) -- (254,293.5) -- cycle ; 

\draw  [green!60!black ,draw opacity=1 ][line width=1.5]  (271.07,273.41) .. controls (279.21,280.89) and (281.26,293.91) .. (275.33,304.96) .. controls (268.41,317.83) and (253.25,323.15) .. (241.47,316.82) .. controls (229.7,310.49) and (225.75,294.92) .. (232.67,282.04) .. controls (236.32,275.25) and (242.27,270.56) .. (248.81,268.57) ;

\draw  [draw opacity=0][line width=1.5]  (493.7,193.04) .. controls (502.02,188.15) and (512.96,187.86) .. (522.31,193.2) .. controls (535.9,200.96) and (540.96,217.69) .. (533.61,230.55) .. controls (526.26,243.42) and (509.29,247.56) .. (495.69,239.8) .. controls (484.84,233.6) and (479.43,221.69) .. (481.43,210.59) -- (509,216.5) -- cycle ; 

\draw [green!60!black ,draw opacity=1 ][line width=1.5]  (493.7,193.04) .. controls (502.02,188.15) and (512.96,187.86) .. (522.31,193.2) .. controls (535.9,200.96) and (540.96,217.69) .. (533.61,230.55) .. controls (526.26,243.42) and (509.29,247.56) .. (495.69,239.8) .. controls (484.84,233.6) and (479.43,221.69) .. (481.43,210.59) ;

\draw [color={rgb, 255:red, 254; green, 0; blue, 5 }  ,draw opacity=1 ][line width=1.5]    (181.79,179.41) -- (210.46,156.29) ;

\draw [shift={(212.79,154.41)}, rotate = 141.12] [color={rgb, 255:red, 254; green, 0; blue, 5 }  ,draw opacity=1 ][line width=1.5]    (14.21,-4.28) .. controls (9.04,-1.82) and (4.3,-0.39) .. (0,0) .. controls (4.3,0.39) and (9.04,1.82) .. (14.21,4.28)   ;
\draw [color={rgb, 255:red, 242; green, 1; blue, 26 }  ,draw opacity=1 ][line width=1.5]    (237,182.5) -- (206.25,209.52) ;
\draw [shift={(204,211.5)}, rotate = 318.69] [color={rgb, 255:red, 242; green, 1; blue, 26 }  ,draw opacity=1 ][line width=1.5]    (14.21,-4.28) .. controls (9.04,-1.82) and (4.3,-0.39) .. (0,0) .. controls (4.3,0.39) and (9.04,1.82) .. (14.21,4.28)   ;
\draw [color={rgb, 255:red, 255; green, 0; blue, 33 }  ,draw opacity=1 ][line width=1.5]    (128,285.5) .. controls (119.41,284.55) and (114.46,284.5) .. (98.35,272.31) ;
\draw [shift={(96,270.5)}, rotate = 37.87] [color={rgb, 255:red, 255; green, 0; blue, 33 }  ,draw opacity=1 ][line width=1.5]    (14.21,-4.28) .. controls (9.04,-1.82) and (4.3,-0.39) .. (0,0) .. controls (4.3,0.39) and (9.04,1.82) .. (14.21,4.28)   ;
\draw [color={rgb, 255:red, 255; green, 0; blue, 4 }  ,draw opacity=1 ][line width=1.5]    (252,226.5) -- (263.12,190.37) ;
\draw [shift={(264,187.5)}, rotate = 107.1] [color={rgb, 255:red, 255; green, 0; blue, 4 }  ,draw opacity=1 ][line width=1.5]    (14.21,-4.28) .. controls (9.04,-1.82) and (4.3,-0.39) .. (0,0) .. controls (4.3,0.39) and (9.04,1.82) .. (14.21,4.28)   ;
\draw [color={rgb, 255:red, 255; green, 0; blue, 30 }  ,draw opacity=1 ][line width=1.5]    (301,196.5) -- (290.81,232.61) ;
\draw [shift={(290,235.5)}, rotate = 285.75] [color={rgb, 255:red, 255; green, 0; blue, 30 }  ,draw opacity=1 ][line width=1.5]    (14.21,-4.28) .. controls (9.04,-1.82) and (4.3,-0.39) .. (0,0) .. controls (4.3,0.39) and (9.04,1.82) .. (14.21,4.28)   ;
\draw [color={rgb, 255:red, 254; green, 0; blue, 29 }  ,draw opacity=1 ][line width=1.5]    (423,185.5) -- (385.57,163.04) ;
\draw [shift={(383,161.5)}, rotate = 30.96] [color={rgb, 255:red, 254; green, 0; blue, 29 }  ,draw opacity=1 ][line width=1.5]    (14.21,-4.28) .. controls (9.04,-1.82) and (4.3,-0.39) .. (0,0) .. controls (4.3,0.39) and (9.04,1.82) .. (14.21,4.28)   ;
\draw [color={rgb, 255:red, 255; green, 0; blue, 34 }  ,draw opacity=1 ][line width=1.5]    (410,132.5) -- (448.44,155.94) ;
\draw [shift={(451,157.5)}, rotate = 211.37] [color={rgb, 255:red, 255; green, 0; blue, 34 }  ,draw opacity=1 ][line width=1.5]    (14.21,-4.28) .. controls (9.04,-1.82) and (4.3,-0.39) .. (0,0) .. controls (4.3,0.39) and (9.04,1.82) .. (14.21,4.28)   ;
\draw [color={rgb, 255:red, 247; green, 5; blue, 26 }  ,draw opacity=1 ][line width=1.5]    (268,324.5) .. controls (261.25,327.4) and (256.35,332.15) .. (234.45,322.6) ;
\draw [shift={(232,321.5)}, rotate = 24.62] [color={rgb, 255:red, 247; green, 5; blue, 26 }  ,draw opacity=1 ][line width=1.5]    (14.21,-4.28) .. controls (9.04,-1.82) and (4.3,-0.39) .. (0,0) .. controls (4.3,0.39) and (9.04,1.82) .. (14.21,4.28)   ;
\draw [color={rgb, 255:red, 244; green, 7; blue, 41 }  ,draw opacity=1 ][line width=1.5]    (281,61.5) .. controls (298.96,41.65) and (314.23,50.39) .. (326.83,60.69) ;
\draw [shift={(329,62.5)}, rotate = 220.24] [color={rgb, 255:red, 244; green, 7; blue, 41 }  ,draw opacity=1 ][line width=1.5]    (14.21,-4.28) .. controls (9.04,-1.82) and (4.3,-0.39) .. (0,0) .. controls (4.3,0.39) and (9.04,1.82) .. (14.21,4.28)   ;
\draw [color={rgb, 255:red, 255; green, 0; blue, 29 }  ,draw opacity=1 ][line width=1.5]    (545,219.5) .. controls (541.24,241.12) and (533.95,244.18) .. (520.63,249.46) ;
\draw [shift={(518,250.5)}, rotate = 338.2] [color={rgb, 255:red, 255; green, 0; blue, 29 }  ,draw opacity=1 ][line width=1.5]    (14.21,-4.28) .. controls (9.04,-1.82) and (4.3,-0.39) .. (0,0) .. controls (4.3,0.39) and (9.04,1.82) .. (14.21,4.28)   ;
\draw  [color={rgb, 255:red, 7; green, 111; blue, 239 }  ,draw opacity=1 ][line width=1.5] [line join = round][line cap = round] (708,567.5) .. controls (708,567.5) and (708,567.5) .. (708,567.5) ;
\draw  [color={rgb, 255:red, 7; green, 111; blue, 239 }  ,draw opacity=1 ][line width=1.5] [line join = round][line cap = round] (316,122.5) .. controls (316,122.5) and (316,122.5) .. (316,122.5) ;
\draw  [color={rgb, 255:red, 7; green, 111; blue, 239 }  ,draw opacity=1 ][line width=1.5] [line join = round][line cap = round] (713,555.5) .. controls (713,555.5) and (713,555.5) .. (713,555.5) ;
\draw  [color={rgb, 255:red, 7; green, 111; blue, 239 }  ,draw opacity=1 ][line width=1.5] [line join = round][line cap = round] (329,120.5) .. controls (329,120.5) and (329,120.5) .. (329,120.5) ;
\draw  [color={rgb, 255:red, 7; green, 111; blue, 239 }  ,draw opacity=1 ][line width=1.5] [line join = round][line cap = round] (320,121.5) .. controls (320,121.5) and (320,121.5) .. (320,121.5) ;

\draw  [color={rgb, 255:red, 7; green, 111; blue, 239 }  ,draw opacity=1 ][line width=1.5] [line join = round][line cap = round] (324,121.5) .. controls (324,121.5) and (324,121.5) .. (324,121.5) ;


\draw[ultra thick, snake it ] (308.03,96.47-2) -- (129.925,253-2)   (308.03,96.47-8) -- (253.55, 293.475-8) (308.03,96.47-6) --  (508, 215.25-6) ;


\node at (180,160)  {\scalebox{1.5}{$\boldsymbol{a_1}$}}; 

\node at (205,230)  {\scalebox{1.5}{$\boldsymbol{a_1^{-1}}$}}; 

\node at (240,240)  {\scalebox{1.5}{$\boldsymbol{a_2}$}}; 

\node at (325,210)  {\scalebox{1.5}{$\boldsymbol{a_2^{-1}}$}}; 

\node at (400,190)  {\scalebox{1.5}{$\boldsymbol{a_n}$}}; 

\node at (400,190)  {\scalebox{1.5}{$\boldsymbol{a_n}$}}; 

\node at (450,130)  {\scalebox{1.5}{$\boldsymbol{a_n^{-1}}$}};

\node at (129.925-45,253-2)  {\scalebox{1.5}{$\boldsymbol{z_1}$}}; 

\node at (253.55-40, 293.475-8)  {\scalebox{1.5}{$\boldsymbol{z_2}$}};

\node at (508-30, 215.25+30) {\scalebox{1.5}{$\boldsymbol{z_n}$}};

\node at (308.03+50,96.47-22) {\scalebox{1.5}{$\boldsymbol{z_\infty}$}};






\node at  (313,120.5)   {\scalebox{1.2}{.}};

\node at  (313+3,120.167)   {\scalebox{1.2}{.}};

\node at  (313+6,119.833)   {\scalebox{1.2}{.}};

\node at  (313+9,119.5)   {\scalebox{1.2}{.}};

\node at  (313+12,119.167)   {\scalebox{1.2}{.}};

\node at  (313+15,118.833)   {\scalebox{1.2}{.}};

\node at  (331,118.5)  {\scalebox{1.2}{.}};


\end{tikzpicture}}

 \caption{Octopus diagram computing the tensionful term in the worldsheet action, expressed as a boundary integral along the cuts and encircling the branch points.}
    \label{fig:Octopus_tensionful_term}
\end{figure}

The maps $(\rho,\bar{\rho})$ have $(n+1)$ branch points: at the insertion points of the vertex operators $(z_k,\bar{z}_k),\; k=1,\cdots,n$, and at the point at infinity $(z_\infty,\bar{z}_\infty)$. We introduce $n$ branch cuts $\{C_j\}_{j=1}^{n}$ joining $(z_\infty,\bar{z}_\infty)$ to each of other branch points $(z_k,\bar{z}_k),\; k=1,\cdots,n$, and make the integrand of the area integral single-valued on $\mathcal{M}\backslash C$ with $C=\cup_j C_j$. Then $\partial \mathcal{M}$ consists of both sides of the cuts, and boundaries of small circles around $(z_k,\bar{z}_k)$ with $k=1,\cdots,n$ and $(z_\infty,\bar{z}_\infty)$ --- all traversed with positive orientation, i.e the interior domain is always on the left as we walk along the boundaries (see Fig.~\ref{fig:Octopus_tensionful_term}). Below we compute the boundary integral from each component of $\partial \mathcal{M}$. For the computation, we introduce regulator for $(\rho, \bar{\rho} )(z_j,\bar{z}_j)$ given by the radius  $\epsilon_{z_j}$ of small circle around $(z_j,\bar{z}_j)$ with $j=1,\cdots,n$. The net $\lambda$-dependent contribtion $S_\lambda[\rho]$ is independent of these regulators, since, as we will see, their contributions cancel between the tension term of the worldsheet action and the vertex operator insertions.

\paragraph{Along $C_j$:} Along $C_j\equiv a_j+a_j^{-1}$ in Fig.~\ref{fig:Octopus_tensionful_term}, first term of the boundary integral \eqref{boundary_integral} reads
\begin{equation}
	\begin{split}
		&\int_{a_j+a_j^{-1}}\rho \, d\bar{\rho}=\int_{a_j} \left(\oint_{C_{z_j}}d\rho\right) d\bar{\rho} =\oint_{C_{z_j}}d\rho \cdot \int_{(z_j,\bar{z}_j)}^{(z_{\infty},\bar{z}_{\infty})}d\bar{\rho}
		=-2\pi i \alpha_j \sum_{k=1}^{n}\bar{\alpha}_k \int_{(z_j,\bar{z}_j)}^{(z_{\infty},\bar{z}_{\infty})} \frac{d\bar{z}}{\bar{z}-\bar{z}_k}\\&=-2\pi i \alpha_j\bar{\alpha}_j\log\left[\frac{\bar{z}_\infty-\bar{z}_j}{\epsilon_{z_j} e^{-i\theta_j}}\right]-2\pi i \alpha_j\sum_{k(\neq j)}^{n}\bar{\alpha}_k \log\left[\frac{\bar{z}_\infty-\bar{z}_k}{\bar{z}_j-\bar{z}_k}\right] \, ,
	\end{split}
\end{equation}
where we have used $\oint_{C_{z_j}}d\rho=\alpha_j\oint_{C_{z_j}}\frac{dz}{z-z_j}=-2\pi i \alpha_j $ (since the contour around $(z_j,\bar{z}_j)$ is clockwise).  The first equality above follows from 
\begin{equation}
	\begin{split}
		&\int_{a_j} \rho(a_j)\,d\bar{\rho}+\int_{a_j^{-1}} \rho(a_j^{-1})\, d\bar{\rho} = \int_{a_j} \rho(a_j)\, d\bar{\rho}-\int_{a_j}\rho(a_j^{-1})\, d\bar{\rho} = \int_{a_j} \left[\rho(a_j)-\rho(a_j^{-1})\right]d\bar{\rho}\\
		&\text{and}\quad \delta \rho = \left[\rho(a_j)-\rho(a_j^{-1}) \right]= \oint_{C_{z_j}}d\rho \, .
	\end{split}
\end{equation}
Summing it over all cuts, we thus obtain
\begin{equation}
	\sum_{j=1}^{n}\int_{a_j+a_j^{-1}}\rho \, d\bar{\rho} =-2\pi i\sum_{j=1}^{n} \alpha_j\bar{\alpha}_j\log\left[\frac{\bar{z}_\infty-\bar{z}_j}{\epsilon_{z_j} e^{-i\theta_j}}\right]-2\pi i \sum_{j\neq k}^{n}\alpha_j\bar{\alpha}_k \log\left[\frac{\bar{z}_\infty-\bar{z}_k}{\bar{z}_j-\bar{z}_k}\right] \, .
\end{equation}
Similarly for the other term of \eqref{boundary_integral}, we have
\begin{equation}
		\sum_{j=1}^{n}\int_{a_j+a_j^{-1}}\bar{\rho} \, d\rho =2\pi i\sum_{j=1}^{n} \bar{\alpha}_j\alpha_j\log\left[\frac{z_\infty-z_j}{\epsilon_{z_j} e^{i\theta_j}}\right]+2\pi i \sum_{j\neq k}^{n}\bar{\alpha}_j\alpha_k \log\left[\frac{z_\infty-z_k}{z_j-z_k}\right] \, ,
\end{equation}
where we have used $\oint_{C_{z_j}}d\bar{\rho}=+2\pi i \bar{\alpha}_j$. Adding both terms of the boundary integral, we get
\begin{equation}
		\sum_{j=1}^{n}\int_{a_j+a_j^{-1}}\left( \rho \, d\bar{\rho} -\bar{\rho} \, d\rho \right)=-2\pi i \sum_{j=1}^{n}|\alpha_j|^2\log\left[\frac{|z_\infty-z_j|^2}{\epsilon_{z_j}^2}\right]-2\pi i \sum_{j\neq k}^{n} \alpha_j\bar{\alpha}_k \log \left[\frac{(\bar{z}_\infty-\bar{z}_k)(z_\infty-z_j)}{(\bar{z}_j-\bar{z}_k)(z_k-z_j)}\right] \, .
\end{equation} 
Using the fact that $\sum_{j=1}^{n}\alpha_j=0$, the dependence of the above integral on $z_\infty$ and $\bar{z}_\infty$  cancel out, and we are left with the expression
\begin{equation}
		\sum_{j=1}^{n}\int_{a_j+a_j^{-1}}\left( \rho \, d\bar{\rho} -\bar{\rho} \, d\rho \right)=2\pi i\sum_{j=1}^{n}|\alpha_j|^2 \log [\epsilon_{z_j}^2]+2\pi i  \sum_{j\neq k} \alpha_j\bar{\alpha}_k \log[-|z_j-z_k|^2] \, .
\end{equation}

\paragraph{Around $(z_k,\bar{z}_k)$: } Next we compute the integral around $(z_k,\bar{z}_k)$ for $k=1,\cdots,n$: 
\begin{equation}
	\sum_{l=1}^{n}\oint_{C_{z_l}} \rho \, d\bar{\rho}=\sum_{l}\sum_{k}\alpha_k\sum_j \bar{\alpha}_j \oint_{C_{z_l}} d\bar{z} \,\frac{\log(z-z_k)}{\bar{z}-\bar{z}_j} \, .
\end{equation}
We address this on a case-by-case basis:

\begin{enumerate}
    \item For $j\neq l$ and $k\neq l$, the above expression clearly vanishes. 

    \item For $j=l$ and $k\neq l$, we obtain 
\begin{equation}
	\sum_{k\neq l} \alpha_k \bar{\alpha}_l \oint_{C_{z_l}} d\bar{z}\, \frac{\log(z-z_k)}{\bar{z}-\bar{z}_l}=2\pi i \sum_{k\neq l} \alpha_k \bar{\alpha}_l\, \log(z_l-z_k) \, .
\end{equation}

\item For $j\neq l$ and $k=l$, we have
\begin{equation}
	\sum_{j\neq l} \alpha_l \bar{\alpha}_j \oint_{C_{z_l}}\frac{\log(z-z_l)}{\bar{z}-\bar{z}_j}=	\sum_{j\neq l} \frac{ \alpha_l \bar{\alpha}_j}{\bar{z}_l-\bar{z}_j} \oint_{C_{z_l}}d\bar{z}\, \log(z-z_l) =0 \, .
\end{equation}

\item For $j=k=l$, we have
\begin{equation}
	\begin{split}
	\sum_{l=1}^{n}\alpha_l\bar{\alpha}_l \oint_{C_{z_l}} d\bar{z}\, \frac{\log(z-z_l)}{\bar{z}-\bar{z}_l} =2\pi i \sum_{l=1}^{n}|\alpha_l|^2 \log(\epsilon_{z_l}) -2\pi^2\sum_{l}|\alpha_l|^2 \, .
	\end{split}	
\end{equation}
\end{enumerate}
We thus obtain:
\begin{equation}
		\sum_{l=1}^{n}\oint_{C_{z_l}} \rho \, d\bar{\rho}=2\pi i \sum_{k\neq l}^{n} \alpha_k \bar{\alpha}_l\, \log(z_l-z_k)+2\pi i \sum_{l=1}^{n}|\alpha_l|^2 \log(\epsilon_{z_l}) -2\pi^2\sum_{l=1}^{n}|\alpha_l|^2 \, .
\end{equation}
Similarly for the other term of the boundary integral, we get
\begin{equation}
	\sum_{l=1}^{n}\oint_{C_{z_l}} \bar{\rho} \, d\rho=-2\pi i \sum_{k\neq l}^{n} \bar{\alpha}_k \alpha_l\, \log(\bar{z}_l-\bar{z}_k)-2\pi i \sum_{l=1}^{n}|\alpha_l|^2 \log(\epsilon_{z_l}) -2\pi^2\sum_{l}|\alpha_l|^2 \, .
\end{equation}
Adding both terms, the contribution from circles around insertion points is
\begin{equation}
	\begin{split}
			\sum_{l=1}^{n}\oint_{C_{z_l}} \left( \rho \, d\bar{\rho} - \bar{\rho} \, d\rho \right)= 2\pi i \sum_{k\neq l}^{n} \alpha_k \bar{\alpha}_l \log\left[ -|z_k-z_l|^2\right] +2\pi i \sum_{l=1}^{n} | \alpha_l|^2\, \log(\epsilon_{z_l}^2) \, .
	\end{split}
\end{equation}

\paragraph{Around $(z_\infty,\bar{z}_\infty)$:} To compute $\oint_{C_{z_\infty}}\left(\rho\,d\bar{\rho}-\bar{\rho}\, d\rho\right)$, we first note that  near $(z_\infty,\bar{z}_\infty)$: 
\begin{equation}
	\begin{split}
	&\rho(z)= \sum_{k=1}^{n}\alpha_k \left[\log z+ \log\left(1-\frac{z_k}{z}\right)\right]\sim -\frac{\left(\sum_{k=1}^{n}\alpha_k z_k \right) }{z},\quad \text{and} \quad d\bar{\rho}=\frac{ \left(\sum_{k=1}^{n}\bar{\alpha}_k \bar{z}_k\right) }{\bar{z}^2} \, d\bar{z} \, .
	\end{split}
\end{equation}
Using this, the contribution from around $(z_\infty,\bar{z}_\infty)$ to the first term reads
\begin{equation}
	\begin{split}
		\oint_{C_{z_\infty}} \rho \, d\bar{\rho}=-\left|\sum_{k=1}^{n}\alpha_k z_k\right|^2 \oint_{C_{z_\infty}} \frac{d\bar{z}}{|z|^2\, \bar{z}}=2\pi i\, \frac{\left|\sum_{k=1}^{n}\alpha_k z_k\right|^2}{|z_\infty|^2} \, ,
	\end{split}
\end{equation}
where we have used $\oint_{C_{z_\infty}} \frac{d\bar{z}}{\bar{z}}=-2\pi i$. Similarly,
\begin{equation}
		\oint_{C_{z_\infty}} \bar{\rho} \, d\rho = -2\pi i \frac{\left|\sum_{k=1}^{n}\alpha_k z_k\right|^2}{|z_\infty|^2}  \, .
\end{equation}
Since $|z_{\infty}|\to \infty$, these contributions vanish.

Hence the net contribution to the boundary integral $S_t$ is
\begin{align}
	\begin{aligned}
	S_t[\rho]&=2\pi i\sum_{j=1}^{n}|\alpha_j|^2 \log[\epsilon_{z_j}^2]+2\pi i  \sum_{j\neq k} \alpha_j\bar{\alpha}_k \log[-|z_j-z_k|^2]\\
    &=2\pi i\sum_{k=1}^{n}\alpha_k \, \bar{\rho}(\bar{z}_k)+2\pi i \sum_{k=1}^{n}\bar{\alpha}_k\, \rho(z_k) \, .
\end{aligned}
\end{align}
We can now uplift to the computation of $S_\lambda[\rho]$ where $\alpha_k=-iw_k$ and $\bar{\alpha}_k=iw_k$ for $k=1,\cdots,n$:
\begin{equation}
\begin{split}
    S_\lambda[\rho] &= i\frac{\lambda R^2}{4}S_t[\rho] +\frac{i}{2}\sum_{k=1}^{n}E_k R(\rho-\bar{\rho})(z_k,\bar{z}_k)\\
    &=i\frac{\lambda\pi R^2}{2} \sum_{k=1}^{n} w_k \, \bar{\rho}(\bar{z}_k) -i\frac{\lambda\pi R^2}{2} \sum_{k=1}^{n} w_k \, \rho(z_k) +\frac{i}{2}\sum_{k=1}^{n}E_k R(\rho-\bar{\rho})(z_k,\bar{z}_k)\\
    &=0 \quad \, .
\end{split}    
\end{equation}
Hence  $S_\lambda[\rho]$  vanishes. 
\paragraph{Final expression of $G_n$:} Since $S_\lambda[\rho]$ vanishes, the only contribution to the worldsheet correlator \eqref{G_n-1} comes from $\Gamma[\rho]$. Since its evaluation is technically complicated, here we simply quote the result postponing the derivation to section \ref{sec:Mandelstam}. As explained there, after regularization and applying the Mandelstam formula \eqref{Mandelstam_final} together with the explicit evaluation of $\partial^2\rho(Z_I)$ in \cite{Komatsu:2025onf}, $\Gamma[\rho]$ takes the form 
\begin{equation}\label{localization_result_1}
   e^{-\Gamma[\rho]}\to e^{-\Gamma^{\rm ren}[\rho]/} = \prod_{k=1}^{n}(w_k)^{q} \prod_{1\leq i< j\leq n} |z_i-z_j|^{-2q}\cdot  |\Delta (P_{n})|^{q} \, ,
\end{equation}
where $\Delta (P_{n})$ is discriminant of the $(n-2)$-degree polynomial 
\begin{equation}\label{polynomial}
    P_{n}(z):= \sum_{k=1}^{n}w_k \prod_{i(\neq k)}^{n} (z-z_i) \, .
\end{equation}
Note that the coefficient of leading term $z^{n-1}$ in $P_{n}(z)$ above vanishes since $\sum_k w_k=0$. Thus the expression of worldsheet correlator \eqref{G_n-1} with fixed points at $z_a,z_b$ and $z_c$, takes the form
\begin{equation}\label{G_n-final}
     G_n=ig_s^{n-2}\mathcal{N}_{S^2}\cdot4\pi^2 R\,\delta\left(\sum_k E_k\right) \delta_{\sum_k w_k} |z_{ab}z_{bc}z_{ca}|^2\, \prod_{k=1}^{n}(w_k)^{q} \prod_{1\leq i< j\leq n} |z_{ij}|^{-2q}\cdot  |\Delta (P_{n})|^{q} \, .
\end{equation}
The $n$-point string amplitude is obtained by integrating the above correlator over the unfixed coordinates on the complex plane $\mathbb{C}$:
\begin{equation}
\begin{split}
\mathcal{A}_n= ig_s^{n-2}\mathcal{N}_{S^2}\cdot4\pi^2 R\,&\delta\left(\sum_k E_k\right) \delta_{\sum_k w_k} |z_{ab}z_{bc}z_{ca}|^2\, \prod_{k=1}^{n}(w_k)^{q} \\ &\times \left[\prod_{l\neq a,b,c}^{n}\int_{\mathbb{C}} d^2 z_{l}\right]\prod_{1\leq i< j\leq n} |z_{ij}|^{-2q}\cdot  |\Delta (P_{n})|^{q} \, .
\end{split}   
\end{equation}
In the following two sections, we apply this formula to evaluate the three- and four-point string amplitudes.

\subsection{Two- and three-point string amplitudes}\label{Two_and_Three-point_amplitudes}
Using the general $n$-point amplitude derived in the previous subsection, the three-point amplitude can be computed directly. The two-point amplitude, however, requires a more careful analysis, for which we will occasionally draw on results from Appendix \ref{two-point-section}, where an analogous computation is carried out for the critical closed string. 

\subsubsection{Two-point amplitude}
The two-point string amplitude in our set up takes the form
\begin{equation}
    \mathcal{A}_2=\frac{1}{V_{\text{CKG}}}\int [\mathfrak{D}X^+\mathfrak{D}X^{-}\mathfrak{D}\beta\,\mathfrak{D}\bar{\beta}\,]\, e^{-S_{\hat{g}}}\, \mathcal{F}[\hat{g}]\, \int d^2z_1\, d^2 z_2\,  V_1(z_1,\bar{z}_1)\, V_1(z_2,\bar{z}_2) \, ,
\end{equation}
with $\mathcal{F}[\hat{g}]$ is the Fadeev Popov determinant to fix the worldsheet metric, and $V_{CKG}$ is the volume of the conformal killing group. Note that, two-point amplitude carries no explicit factor of $g_s$.

As discussed in Appendix \ref{two-point-section}, we fix the conformal Killing group symmetry by inserting the identity \eqref{Delta_closed_string} in the path integral. After relabelling the insertion points $z_1\to \alpha\circ z_1$ and  $z_2\to \alpha\circ z_2$, the amplitude reduces to
\begin{align}\label{A_2-2}
\begin{aligned}
\mathcal{A}^{\text{closed}}_2=&i\mathcal{F}[\hat{g}]\,\Big\langle \Delta_{X^0} V_1(z_1^0,\bar{z}_1^0)\, V_2(z_2^0,\bar{z}_2^0)\Big\rangle' (z_2^0)^2 (\bar{z}_2^0)^2\\
&\times \int_{-\infty}^{\infty} dx^0  \, e^{-i(E_1+E_2) x^0} \delta(x^0-\#) \int_{0}^{2\pi R} dx^1    \, ,
\end{aligned}
\end{align}
where the prime on the correlator indicates that the zero-mode integrals $(\int dx^0 dx^1)$ are taken out. 

Evaluating $\Delta_{X^0}$ and inserting it into the worldsheet correlator above, we recover the structure in \eqref{Correlator_after_decomposition}.  Thus, we are led to compute
\begin{equation}\label{correlator_2d-YM_2_pt}
    \begin{split}
        &\Big\langle \partial X^0(|A|^2w,|A|^2\bar{w})\, V_1(z_1^0,\bar{z}_1^0)\, V_2(z_2^0,\bar{z}_2^0)\Big\rangle' \\&= \int [\mathfrak{D}X^+\mathfrak{D}X^{-}]'\,[\mathfrak{D}\beta\,\mathfrak{D}\bar{\beta}\,]\, e^{-S_{\hat{g}}}\, \partial X^0(|A|^2w,|A|^2\bar{w})\, V_1(z_1^0,\bar{z}_1^0)\, V_2(z_2^0,\bar{z}_2^0)\\
        &= \mathcal{N}_{S^2}\,\text{det}'\left(\frac{\partial\bar{\partial}}{4\pi^2} \right)^{-1}\,  e^{-\Gamma[\rho]}   \Big[\partial X^0(|A|^2w,|A|^2\bar{w}) \Big]_{X^+=\rho,\,X^-=\bar{\rho}}\, \delta_{w_1+w_2,0} \, ,
    \end{split}
\end{equation}
together with its anti-holomorphic counterpart with $\bar{\partial}X^0$ replaced by $\partial X^0$. As discussed in Appendix \ref{two-point-section}, $A$ is a solution to the delta function constranit, which will drop out from the final answer as we see below.  $(\rho,\bar{\rho})$ are the Mandelstam maps \eqref{Mandelstam_maps} with $n=2$ and $\mathcal{N}_{S^2}$ is the overall normalization coming from the sphere partition function.

Using
\begin{equation}
    \begin{split}
        & e^{-\Gamma[\rho]}=\frac{1}{|z_{12}^0|^4}\to \frac{1}{(z_2^0)^4}, \quad  [\partial X^0]_{\rho,\, \bar{\rho}} =-\frac{i}{2} w_1R\, \frac{z_1^0-z_2^0}{(z-z_1^0)(z-z_2^0)}\to -\frac{i}{2} w_1R\,\frac{1}{z}  \, , 
    \end{split}
\end{equation}
and similarly $ [\partial X^0]_{\rho,\, \bar{\rho}} = -\frac{i}{2} w_1R\,\frac{1}{\bar{z}}$ (with $z_1^0=0$, $z_2^0=\infty$), the correlator \eqref{correlator_2d-YM_2_pt} becomes
\begin{equation}
    \begin{split}
        &\Big\langle \partial X^0(|A|^2w,|A|^2\bar{w})\, V_1(z_1^0,\bar{z}_1^0)\, V_2(z_2^0,\bar{z}_2^0)\Big\rangle' = \mathcal{N}_{S^2}\,\text{det}'\left(\frac{\partial\bar{\partial}}{4\pi^2} \right)^{-1}\, \left( -\frac{i}{2}w_1R\frac{1}{|A|^2w}\frac{1}{|z_{12}^0|^4}\right)\, \delta_{w_1+w_2,0} \, ,\nonumber
    \end{split}
\end{equation}
Combining the holomorphic and anti-holomorphic pieces in the analogue of \eqref{Correlator_after_decomposition}, we obtain
\begin{align}
\begin{aligned}
    \Big\langle \Delta_{X^0} V_1(z_1^0,\bar{z}_1^0)\, V_2(z_2^0,\bar{z}_2^0)\Big\rangle'=& \mathcal{N}_{S^2}\,\frac{|A|^2}{4\pi}\left( -\frac{i}{2}w_1R\frac{1}{|A|^2}\frac{1}{|z_{12}^0|^4}\times 2\right)\\
    &\times \frac{\int d^2w\sqrt{|g_{S^2}(w)|}}{4\pi R_{S^2}^2}\,\text{det}'\left(\frac{\partial\bar{\partial}}{4\pi^2} \right)^{-1}\delta_{w_1+w_2,0}\\
    =& -\frac{i}{2\pi}w_1R\,\text{det}'\left(\frac{\partial\bar{\partial}}{4\pi^2} \right)^{-1}\frac{\mathcal{N}_{S^2}\,}{|z_{12}^0|^4}\,\delta_{w_1+w_2,0} \, .
\end{aligned}    
\end{align}
Substituting into \eqref{A_2-2}, the two-point amplitude simplifies to
\begin{equation}\label{string_two-point_amplitude}
    \mathcal{A}_2=\frac{1}{2} w_1R^2  \mathcal{N}_{S^2}\,\delta_{w_1+w_2,0} \, ,
\end{equation}
where we have used the relation,
\begin{equation}
    \mathcal{F}[\hat{g}]\, \text{det}'\left(\frac{\partial\bar{\partial}}{4\pi^2} \right)^{-1}=1 \, .
\end{equation}
In the following, we match the two-point string amplitude \eqref{string_two-point_amplitude} to its dual Yang–Mills counterpart and determine the normalization factor $\mathcal{N}_{S^2}$.

\subsubsection{Three-point amplitude:}\label{three-point_amplitude}

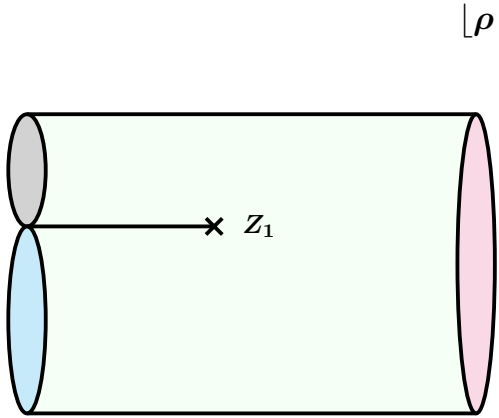
\begin{figure}[!htb]
    \centering
\resizebox{0.4\linewidth}{!}{%
\begin{tikzpicture} [scale = 0.55]


\draw[white, fill=green!4!]  (0,1.5) -- (12, 1.5) -- (12,-6.5) -- (0, -6.5) --(0,1.5);



\draw[ultra thick, fill=gray!35!] (0,0) ellipse (0.5 and 1.5);


\draw[ultra thick, fill=cyan!20!] (0,-4) ellipse (0.5 and 2.5);


\draw[ultra thick, fill=magenta!18!] (12,-2.5) ellipse (0.5 and 4);


\draw[ultra thick] (0,1.5) -- (12,1.5);
\draw[ultra thick] (0,-6.5) -- (12,-6.5);
\draw[ultra thick] (0,-1.5) -- (5,-1.5);

\draw (5,-1.5) node[cross] {};

\node at (5+1.2,-1.5) {$\boldsymbol{Z_1}$};

\node at (12,4) {\scalebox{1.3}{$\lfloor \boldsymbol{\rho}$}};

\end{tikzpicture}}

\caption{The string diagram for three point amplitude with $Z_1$ being the interaction point where $\partial \rho$ vanishes.}
\end{figure}

For three point string amplitudes, the polynomial in $\eqref{polynomial}$ is linear, and its discrminant is $1$. Using this, and putting $q=1$ in \eqref{G_n-final}, we obtain the following expression for three-point string amplitude
\begin{equation}\label{string_three-point_amplitude}
   \mathcal{A}_3= -i(2\pi)^2Rg_s\mathcal{N}_{S^2}\,w_1w_2w_3\,\delta\left(\sum_k E_k\right) \delta_{\sum_k w_k} \, .
\end{equation}
The position dependence of the worldsheet matter correlator, $|z_{12}z_{23}z_{31}|^{-2}$ confirms that our vertex operators in \eqref{vertex_operator} are conformal primaries with dimension $(1,1)$.

\subsubsection{Duality dictionary}
Matching the two-point amplitudes in chiral 2d YM \eqref{YM_two-point_amplitude} and in dual string theory \eqref{string_two-point_amplitude}, one finds 
\begin{equation}
    \frac{1}{2} R^2\mathcal{N}_{S^2} =1\, , 
\end{equation}
while matching the three-point amplitudes \eqref{YM_three-point_amplitude}  and \eqref{string_three-point_amplitude}, we obtain
\begin{equation}
    g_s \mathcal{N}_{S^2} = \frac{\lambda}{N_c} \, . 
\end{equation}
Allowing the scaling ambiguity in operator normalizations, parameterized by a factor $r$, the dictionary between the two theories reads
\begin{equation}
   \mathcal{N}_{S^2}  = \frac{2r^2}{R^2} \, , \quad \text{and} \quad g_s=\frac{r\lambda R^2}{2N_c} \, .
\end{equation}
This dictionary was presented in~\cite{Komatsu:2025sqo}. Note that, although the dictionary involves the ambiguity $r$, the kinematical dependence for the three-point function $w_1w_2w_3$ is free from this ambiguity and its matching provides a nontrivial test of our duality.

In what follows, we compute the four-point amplitudes on both sides of our proposed duality, which provide an additional consistency check of the proposal.

\subsection{Four-point string amplitude}\label{four-point_amplitude}
For four point string amplitudes, the polynomial in \eqref{polynomial} takes the form
\begin{equation}
    P_4(z)=Az^2+Bz+C \, ,
\end{equation}
with the coefficients given by
\begin{equation}
A=\sum_{i=1}^{4}w_i\,z_i,\quad B=-\sum_{i<j}^{4}(w_i+w_j)\,z_iz_j,\quad C=-\sum_{i\neq j,\,j<k<l}^{4} w_iz_jz_kz_l \, .
\end{equation}
We can compute its discriminant and obtain
\begin{equation}
    \begin{split}
        \Delta(P_4) &= B^2-4AC= (w_1+w_3)^2 z_{13}^2 z_{24}^2 \,(z_c-a)(z_c-b) \, ,
    \end{split}
\end{equation}
with $z_c$ being the conformal cross ratio, $ z_c=z_{12}z_{34}/(z_{13} z_{24})$, and
\begin{equation}\label{ab_1}
   a,b=-\frac{w_1w_4+w_2w_3 \pm 2\sqrt{w_1w_2w_3w_4}}{(w_1+w_3)^2} \, .
\end{equation}
The 4-point worldsheet correlator then reads
\begin{equation}
    \begin{split}
        G_4= i(2\pi)^2 Rg_s^2\mathcal{N}_{S^2} \,&\delta\left(\sum_k E_k\right) \delta_{\sum_k w_k} |z_{ab}z_{bc}z_{ca}|^2 (w_1w_2w_3w_4)^q \\
        &\times \prod_{1\leq i<j\leq 4} |z_{ij}|^{-2q} |w_1+w_3|^{2q}|z_{13}z_{24}|^{2q} |z_c-a|^q|z_c-b|^q \, .
    \end{split}
\end{equation}
Fixing three insertion points $z_1\to 0, z_3\to 1$ and $z_4\to\infty$, and defining the vertex operator and ghost insertion at $\infty$ according to the standard conformal prescription, the above expression becomes
\begin{equation}
    G_4= i(2\pi)^2 Rg_s^2\mathcal{N}_{S^2} \,\delta\left(\sum_k E_k\right) \delta_{\sum_k w_k}(w_1w_2w_3w_4)^q |w_1+w_3|^{2q}|z_2|^{-2q}|1-z_2|^{-2q} |z_2-a|^q |z_2-b|^q  \, .
\end{equation}
This expression for the worldsheet correlator further confirms that the vertex operators \eqref{vertex_operator} are conformal primaries of dimension $(q,q) = (1,1)$. 

The string amplitude is obtained by integrating this correlator over the remaining unfixed coordinate $z_2$ on the complex plane $\mathbb{C}$
\begin{equation}
    \mathcal{A}_4= \int_{\mathbb{C}} d^2 z_2\, G_4 \, ,
\end{equation}
which requires evaluating the integral
\begin{equation}
    \mathcal{I}_4= \int_{\mathbb{C}} d^2 z_2 \,|z_2|^{-2q}|1-z_2|^{-2q} |z_2-a|^q |z_2-b|^q  \, .
\end{equation}
In the $q=1$ limit that is relevant for 2d YM, the integral is divergence. This is because the conservation of the winding $w$'s forces the intermediate state to be always on-shell. Note that this property is also true on the 2d YM side. 

To perform a more quantitative comparison (beyond $\infty =\infty$),  we will now regulate the integral and evaluate its leading behavior near the $s$-, $t$-, and $u$-channel poles and compare the result with the corresponding contributions on the dual Yang–Mills side.

\paragraph{Evaluating the moduli integral:} We rewrite the moduli integral,
\begin{equation}
    \mathcal{I}_4= \int_{\mathbb{C}} d^2 z \,|z|^{-2s-2}|1-z|^{-2t-2} |z_2-a|\, |z_2-b|  \, ,
\end{equation}
by introducing two regularization parameters, $s$ and $t$ in the integrand and setting $q=1$. This integral has been evaluated in \cite{Komatsu:2025onf} using a KLT-like factorization of the complex area integral, yielding 
\begin{equation}
     \mathcal{I}_4=i\, \mathcal{I}_\zeta \mathcal{I}_\xi \, ,
\end{equation}
where $\zeta$ and $\xi$ are real light-cone-like coordinates on the worldsheet. Depending on the range of $\zeta$, the integral $\mathcal{I}_\xi$ receives contributions from different contours and thus takes different forms. In what follows, we list these cases and analyze the behavior of the integral in the limits $s\to 0_{-}$, $t\to 0_{-}$, and $u=-(s+t)\to0_{-}$, which will be our main focus. In \cite{Komatsu:2025onf}, the computation was carried out under the assumption that $a,b$ are real with $0<a<b<1$. However, the final expressions for the integrals admit an analytic continuation to arbitrary complex values of $a$ and $b$.

\paragraph{ 1. $\boldsymbol{0<\zeta <a}$ :} In this range of $\zeta$, the integrals take the form
\begin{equation}
\begin{split}
    \mathcal{I}_\xi &= 2i\sin(\pi s)\int_{-\infty}^{0}d\xi' \, (-\xi')^{-s-1} (1-\xi')^{-t-1} (a-\xi')^{1/2} (b-\xi')^{1/2} \, ,\\
     \mathcal{I}_\zeta &= \int_{0}^a d\zeta \, \zeta^{-\alpha's/4-1} (1-\zeta)^{-\alpha't/4-1} (a-\zeta)^{1/2} (b-\zeta )^{1/2} \, .
\end{split}    
\end{equation}
As $s\to 0_{-}$, the simple pole in $s$ from the integral in $\mathcal{I}_\xi$ is canceled by the zero of the prefactor $\sin(\pi s)$, so $\mathcal{I}_\xi$ remains finite. In contrast, $\mathcal{I}_\zeta$ develops a simple pole in $s$. We now analyze this limits in detail.

As $s\to 0_{-}$, the integral in $\mathcal{I}_\xi$ gets contribution near $\xi'\sim 0$, where it behaves as (with $\epsilon>0$)
\begin{equation}
    \begin{split}
        \mathcal{I}_\xi &\sim  2i\sin[\pi s]\int_{-\epsilon }^{0}d\xi' \, (-\xi')^{-s-1} (1-\xi')^{-t-1} (a-\xi')^{1/2} (b-\xi')^{1/2}\\
        &\sim 2i\pi s \cdot a^{1/2} \cdot b^{1/2} \int_{-\epsilon }^{0}d\xi' \, (-\xi')^{-s-1} \sim -2\pi i \cdot  (ab)^{1/2} \, .
    \end{split}
\end{equation}
Similarly integral in $\mathcal{I}_\zeta$ receives a contribution from $\zeta\sim 0$:
\begin{equation}
    \begin{split}
        \mathcal{I}_\zeta & \sim \int_{0}^{\epsilon} d\zeta \, \zeta^{-s-1} (1-\zeta)^{-t-1} (a-\zeta)^{1/2} (b-\zeta )^{1/2}\\
        &\sim a^{1/2} \cdot b^{1/2} \int_0^{\epsilon} d\zeta \, \zeta^{-s-1} \sim -\frac{(ab)^{1/2}}{s} \, .
    \end{split}
\end{equation}
where we have used, 
\begin{equation}
    \int_{-\epsilon }^{0}d\xi' \, (-\xi')^{-s-1} = \int_0^\epsilon d\xi' \,(\xi')^{-s-1}= \frac{ (\xi')^{-s}}{-s}\Big|_{0}^{\epsilon}=\frac{\epsilon^{-s}-0}{-s}\sim -\frac{1}{s}  \, .
\end{equation}
Thus, near $s\to 0_{-}$, the integral behaves as
\begin{equation}
    i\,\mathcal{I}_\xi \,\mathcal{I}_\zeta \sim -2\pi \frac{ab}{s} \quad \text{as}\quad s\to 0_{-} \, .
\end{equation}

On the other hand, as $t\to 0_{-}$, both $\mathcal{I}_\xi$ and $\mathcal{I}_\zeta$ remain finite. However $\mathcal{I}_\xi$ develops a simple pole in $u$ as $u\to 0_{-}$. o make this explicit, we change variables $\xi'=-x/(1-x)$, so that 
\begin{equation}
    d\xi'=-dx(1-x)^{-2},\quad 1-\xi'=(1-x)^{-1},\quad a-\xi'=[a+(1-a)x](1-x)^{-1},\quad b-\xi'=[b+(1-b)x](1-x)^{-1}\, .
\end{equation}
 The integral $\mathcal{I}_\xi$ then takes the form
\begin{equation}
    \begin{split}
        \mathcal{I}_\xi
        & = 2i\sin(\pi s) \int_0^1 dx \, x^{-s-1} (1-x)^{s+t-1} [a+(1-a)x]^{1/2}  [b+(1-b)x]^{1/2} \, .
    \end{split}
\end{equation}
As $u=-(s+t)\to 0_{-}$, the dominant contribution comes from $x\sim 1$, where the integral goes as
\begin{equation}
    \begin{split}
         & \int_{1-\epsilon}^{1} dx\, x^{-s-1} (1-x)^{-u-1} [a+(1-a)x]^{1/2}  [b+(1-b)x]^{1/2} \\
         \sim  & \int_{1-\epsilon}^{1} dx\,  (1-x)^{-u-1}  = \int_{\epsilon}^0 (-dx')\, (x')^{-u-1}= \int_0^\epsilon dx' \, (x')^{-u-1}\sim -\frac{1}{u} \, .
    \end{split}
\end{equation}
Hence,
\begin{equation}
      \mathcal{I}_\xi \sim -\frac{2i\sin(\pi s)}{u}, \quad \text{as}\quad u\to 0_{-} \, .
\end{equation}

\paragraph{2. $\boldsymbol{a<\zeta <b}$ :}
In this regime of $\zeta$, the integrals are
\begin{equation}
    \begin{split}
        \mathcal{I}_\xi &=
         2i \int_{0}^{a}d\xi\, \xi^{-s-1} (1-\xi)^{-t-1} (a-\xi)^{1/2} (b-\xi)^{1/2}-2i\cos(\pi s) \int_{-\infty}^{0}d\xi\, (-\xi)^{-s-1} (1-\xi)^{-t-1} (a-\xi)^{1/2} (b-\xi)^{1/2} \, ,\\
          I_\zeta &=\int_a^b d\zeta\, \zeta^{-s-1} (1-\zeta)^{-t-1} (\zeta-a)^{1/2} (b-\zeta)^{1/2} \, .
    \end{split}
\end{equation}
It is clear that as $t\to 0_{-}$, neither $\mathcal{I}_\xi$ nor $\mathcal{I}_\zeta$ produces a $t^{-1}$ pole.  Similarly, $\mathcal{I}_\zeta$ doesn't generate a $s^{-1}$ pole. We examine the behavior of $\mathcal{I}_\xi$ as $s\to 0_{-}$ where dominant contribution comes from $\xi\sim 0$: 
\begin{equation}
    \begin{split}
         \mathcal{I}_\xi\sim & 2i\int_0^\epsilon d\xi\, \xi^{-s-1} -2i\cos(\pi s)\int_{-\epsilon}^{0}d\xi \, (-\xi)^{-s-1}\\
         =& 2i \cdot \frac{\epsilon^{-s}}{(-s)}-2i\cos(\pi s) \cdot \frac{\epsilon^{-s}}{(-s)} \sim 0 \, ,
    \end{split}
\end{equation}
showing that $\mathcal{I}-\xi$ doesn't develop a $s^{-1}$ pole either.

Poles in $u=-(s+t)$ from $\mathcal{I}_\xi$ arise when $\xi$ gets close to $\infty$. The first integral in $\mathcal{I}_\xi$ doesn't contribute, while the second integral produces a $u^{-1}$ pole (as computed previously). Thus,
\begin{equation}
    \mathcal{I}_\xi \sim +\frac{2i\cos(\pi s)}{u},\quad \text{as} \quad u\to 0_{-} \, .
\end{equation}

\paragraph{ 3. $\boldsymbol{b<\zeta <1}$ :}
In this regime, the integrals take the form
\begin{equation}
\begin{split}
    \mathcal{I}_\xi &= 2i\sin(\pi t) \int_0^\infty d\xi' \, (1+\xi')^{-s-1} (\xi')^{-t-1} (\xi'+1-a)^{1/2} (\xi'+1-b)^{1/2} \, , \\
     \mathcal{I}_\zeta &= \int_b^1 d\zeta \, \zeta^{-s-1} (1-\zeta)^{-t-1} (\zeta-a)^{1/2} (\zeta-b)^{1/2}  \, .
\end{split}    
\end{equation}
As $t\to 0_{-}$, $\mathcal{I}_\zeta$ develops a simple pole in $t$, while $\mathcal{I}_\xi$ remains due to the zero of the prefactor $\sin(\pi t)$. In what follows, we analyze these limits explicitly. 

As $t\to 0_{-}$, the dominant contribution from the integral in $\mathcal{I}_\xi$ comes from $\xi'\sim 0$:
\begin{equation}
    \begin{split}
        \mathcal{I}_\xi &\sim   2i\sin(\pi t) \int_0^{\epsilon} d\xi' \, (1+\xi')^{-s-1} (\xi')^{-t-1} (\xi'+1-a)^{1/2} (\xi'+1-b)^{1/2}\\
        &\sim 2i\cdot \pi t \cdot (1-a)^{1/2}(1-b)^{1/2} \int_0^{\epsilon} d\xi' \,  (\xi')^{-t-1} \sim -2\pi i \, (1-a)^{1/2}(1-b)^{1/2} \, ,
    \end{split}
\end{equation}
where we have used $\int_0^{\epsilon} d\xi' \,  (\xi')^{-t-1} \sim -1/t$. 

The $\mathcal{I}_\zeta$ integral gets contribution from $\zeta \sim 1$, and we get 
\begin{equation}
    \begin{split}
        \mathcal{I}_\zeta & \sim \int_{1-\epsilon}^1 d\zeta \, \zeta^{-s-1} (1-\zeta)^{-t-1} (\zeta-a)^{1/2} (\zeta-b)^{1/2} \\
        &= \int_{\epsilon}^{0} (-d\zeta') \, (1-\zeta')^{-s-1} (\zeta')^{-t-1} (1-a-\zeta)^{1/2} (1-b-\zeta)^{1/2}\\
        &\sim (1-a)^{1/2}(1-b)^{1/2} \int_{\epsilon}^{0} (-d\zeta') (\zeta')^{-t-1}\sim - \frac{(1-a)^{1/2}(1-b)^{1/2} }{t} \, ,
    \end{split}
\end{equation}
where we have used 
\begin{equation}
    \int_{\epsilon}^{0} (-d\zeta') (\zeta')^{-t-1} = \int_0^\epsilon d\zeta' (\zeta')^{-t-1}= \frac{ (\zeta')^{-t}}{-t} \Big|_{0}^{\epsilon}= \frac{\epsilon^{-t}-0}{-t}=-\frac{1}{t} \, .
\end{equation}
Combining both contributions, we obtain
\begin{equation}
   \mathcal{I}_\xi \,    \mathcal{I}_\zeta \sim  2\pi i \frac{(1-a)(1-b)}{t} \quad \text{as}\quad t\to 0_{-}  \, .
\end{equation}
Next we examine the $u\to 0_{-}$ behavior of $\mathcal{I}_\xi$. Making a change of variable $\xi'=x/(1-x)$ in the integral of $\mathcal{I}_\xi$, we have
\begin{equation}
    d\xi'=dx(1-x)^{-2},\quad 1+\xi'=(1-x)^{-1},\quad 1+\xi'-a=[1-a(1-x)](1-x)^{-1},\quad 1+\xi'-b=[1-b(1-x)](1-x)^{-1}  \, .
\end{equation}
Using these, the integral becomes
\begin{equation}
    \begin{split}
        \mathcal{I}_\xi 
        =  2i\sin(\pi t) \int_0^1 dx \, x^{-t-1} (1-x)^{s+t-1} [1-a(1-x)]^{1/2} [1-b(1-x)]^{1/2}  \, .
    \end{split}
\end{equation}
As $u=-(s+t)\to 0_{-}$, the dominant contribution arises from $x\sim 1$: 
\begin{equation}
   \begin{split}
         &  \int_{1-\epsilon}^1 dx \, x^{-t-1} (1-x)^{-u-1} [1-a(1-x)]^{1/2} [1-b(1-x)]^{1/2} \\
        \sim & \int_{1-\epsilon}^1 dx \, (1-x)^{-u-1}\sim\int_{\epsilon}^0 (-dx') (x')^{-u-1}= \int_0^\epsilon dx'\, (x')^{-u-1}=-\frac{1}{u}
   \end{split}
\end{equation}
Hence,
\begin{equation}
      \mathcal{I}_\xi \sim \frac{2i\sin(\pi s)}{u}, \quad \text{as}\quad u\to 0_{-} \, ,
\end{equation}
where we used $\sin (\pi t )\sim -\sin (\pi s)$ as $u\to 0_{-}$.

\paragraph{Matching with 2d YM amplitudes:} We now collect the poles in $s,t$ and $u$ from the integrals for different ranges of $\zeta$, and compute the full string amplitude,
\begin{equation}
    \mathcal{A}_4 =  i(2\pi)^2 Rg_s^2\mathcal{N}_{S^2} \,\delta_{\sum_k w_k}\,\delta\left(\sum_k E_k\right) w_1w_2w_3w_4 (w_1+w_3)^{2} \times \mathcal{I}_4 \, .
\end{equation}
We then compare these singular contributions with the corresponding four-point amplitudes in the dual chiral 2d YM theory. Although four-point amplitudes diverge on both sides, as $s,t,u \to 0_{-}$ the leading behavior factorizes into a product of two three-point amplitudes, consistent with the expected structure of exchange of intermediate winding states. 

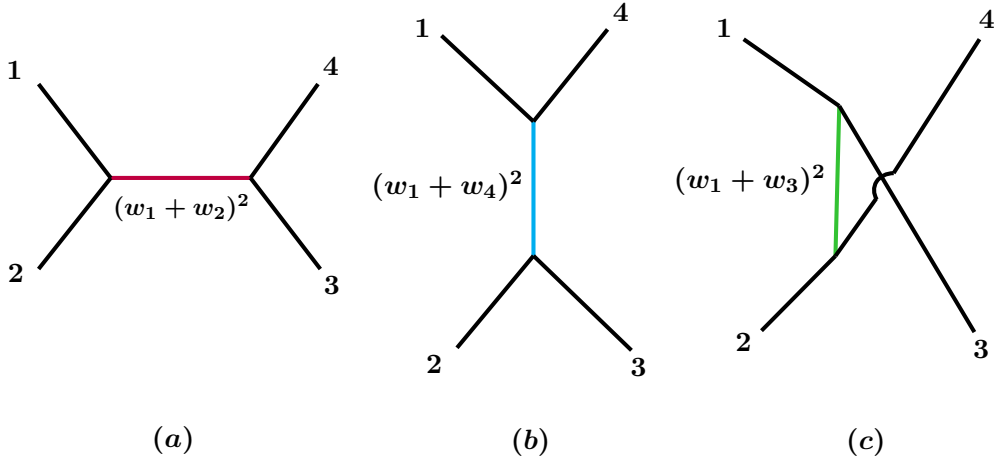
\begin{figure}
    \centering
\resizebox{0.85\linewidth}{!}{%

\begin{tikzpicture}[x=0.50pt,y=0.55pt, yscale=-1,xscale=1]

\draw [purple , ultra thick]  (71.62,235.5) -- (181.21,235.5) ;
\draw [ultra thick]    (15,168.5) -- (71.62,235.5) ;
\draw [ultra thick]    (15,300.5) -- (71.62,235.5) ;
\draw [ultra thick]    (234.17,168.5) -- (181.21,235.5) ;
\draw [ultra thick]    (236,300.5) -- (181.21,235.5) ;
\draw [cyan , ultra thick ] (403.16,195.18) -- (403.16,291.04) ;
\draw [ultra thick]    (331,133.94) -- (403.16,195.18) ;
\draw [ultra thick]    (461.26,129.5) -- (403.16,195.18) ;
\draw [ultra thick]    (403.16,291.04) -- (480,358.5) ;
\draw [ultra thick]    (403.16,291.04) -- (343.18,356.72) ;
\draw [green!70!black!80!, ultra thick ]    (643,184.18) -- (640,291.19) ;
\draw [ultra thick]    (568,136.5) -- (643,184.18) ;
\draw [ultra thick]    (640,291.19) -- (582,344.51) ;
\draw [ultra thick]    (643,184.18) -- (749,345.5) ;
\draw [ultra thick]    (672,250.23) -- (640,291.19) ;
\draw  [draw opacity=0][line width=1.5]  (672.38,250.86) .. controls (669.77,246.79) and (669.49,242.07) .. (672.19,238.2) .. controls (674.92,234.27) and (680.07,232.13) .. (685.7,232.08) -- (688.25,246.01) -- cycle ; \draw  [ultra thick]  (672.38,250.86) .. controls (669.77,246.79) and (669.49,242.07) .. (672.19,238.2) .. controls (674.92,234.27) and (680.07,232.13) .. (685.7,232.08) ;  
\draw [ultra thick]    (753,136.5) -- (685.63,232.08) ;

\draw (124.5-20-10,256.75) node   [align=left] {\begin{minipage}[lt]{21.08pt}\setlength\topsep{0pt}
{\scalebox{0.9}{$\boldsymbol{(w_1+w_2)^2}$}}
\end{minipage}};
\draw (378-50-30,241.25) node   [align=left] {\begin{minipage}[lt]{21.76pt}\setlength\topsep{0pt}
{\scalebox{1}{$\boldsymbol{(w_1+w_4)^2}$}}
\end{minipage}};
\draw (618.5-50-30,236.75) node   [align=left] {\begin{minipage}[lt]{25.16pt}\setlength\topsep{0pt}
{\scalebox{1}{$\boldsymbol{(w_1+w_3)^2}$}}
\end{minipage}};
\draw (-5.5+15,158.75) node   [align=left] {\begin{minipage}[lt]{19.72pt}\setlength\topsep{0pt}
{\scalebox{1}{$\boldsymbol{1}$}}
\end{minipage}};
\draw (-5+15,303.25) node   [align=left] {\begin{minipage}[lt]{19.04pt}\setlength\topsep{0pt}
{\scalebox{1}{$\boldsymbol{2}$}}
\end{minipage}};
\draw (256+5,160.75-5) node   [align=left] {\begin{minipage}[lt]{23.12pt}\setlength\topsep{0pt}
{\scalebox{1}{$\boldsymbol{4}$}}
\end{minipage}};
\draw (257.5+5,307.25+5) node   [align=left] {\begin{minipage}[lt]{23.8pt}\setlength\topsep{0pt}
{\scalebox{1}{$\boldsymbol{3}$}}
\end{minipage}};
\draw (318+10+5,118.75+10) node   [align=left] {\begin{minipage}[lt]{23.12pt}\setlength\topsep{0pt}
{\scalebox{1}{$\boldsymbol{1}$}}
\end{minipage}};
\draw (477.76+10,117.25+5-5) node   [align=left] {\begin{minipage}[lt]{22.44pt}\setlength\topsep{0pt}
{\scalebox{1}{$\boldsymbol{4}$}}
\end{minipage}};
\draw (328.68+10,367.97) node   [align=left] {\begin{minipage}[lt]{19.72pt}\setlength\topsep{0pt}
{\scalebox{1}{$\boldsymbol{2}$}}
\end{minipage}};
\draw (498.5+6,370.25) node   [align=left] {\begin{minipage}[lt]{25.16pt}\setlength\topsep{0pt}
{\scalebox{1}{$\boldsymbol{3}$}}
\end{minipage}};
\draw (556.5+10+5,123.75+5) node   [align=left] {\begin{minipage}[lt]{25.16pt}\setlength\topsep{0pt}
{\scalebox{1}{$\boldsymbol{1}$}}
\end{minipage}};
\draw (777.5,121.25+5-4) node   [align=left] {\begin{minipage}[lt]{25.16pt}\setlength\topsep{0pt}
{\scalebox{1}{$\boldsymbol{4}$}}
\end{minipage}};
\draw (570.5+10+5,352.75) node   [align=left] {\begin{minipage}[lt]{23.8pt}\setlength\topsep{0pt}
{\scalebox{1}{$\boldsymbol{2}$}}
\end{minipage}};
\draw (765.5+5,357.25) node   [align=left] {\begin{minipage}[lt]{22.44pt}\setlength\topsep{0pt}
{\scalebox{1}{$\boldsymbol{3}$}}
\end{minipage}};
\draw (127,422.25) node   [align=left] {\begin{minipage}[lt]{23.12pt}\setlength\topsep{0pt}
{\scalebox{1}{$\boldsymbol{(a)}$}}
\end{minipage}};
\draw (406.5,423.75) node   [align=left] {\begin{minipage}[lt]{21.08pt}\setlength\topsep{0pt}
{\scalebox{1}{$\boldsymbol{(b)}$}}
\end{minipage}};
\draw (673,424.25) node   [align=left] {\begin{minipage}[lt]{24.48pt}\setlength\topsep{0pt}
{\scalebox{1}{$\boldsymbol{(c)}$}}
\end{minipage}};

\end{tikzpicture}}

\caption{Four-point amplitudes in chiral 2d YM with $s-, t-$ and $u$-channel exchanges }
\label{fig:YM_amplitudes}
\end{figure}

\begin{itemize}

\item The $s$-channel pole comes only from the region $0<\zeta<a$ and reads
\begin{equation}
 \mathcal{I}_4\sim-2\pi\frac{ab}{s}=-\frac{2\pi}{s}\frac{(w_1+w_2)^2}{(w_1+w_3)^2},\quad \text{as}\quad s\to 0_{-} \, ,
\end{equation}
The string amplitude as $ s\to 0_{-}$ thus reads
\begin{equation}
     \mathcal{A}_4 \sim -i\frac{(2\pi)^3}{s} Rg_s^2\mathcal{N}_{S^2} \,\delta_{\sum_k w_k}\,\delta\left(\sum_k E_k\right) w_1w_2 (w_1+w_2)^2 w_3 w_4 \, .
\end{equation}
This matches with the dual YM amplitude with an $s$-channel exchange of a state of winding $(w_1+w_2)$  (see Fig. \ref{fig:YM_amplitudes}).
\item The $t$-channel pole comes from the region $b<\zeta<1$ and is given by 
\begin{equation}
 \mathcal{I}_4\sim - 2\pi \frac{(1-a)(1-b)}{t} =-\frac{2\pi}{t}\frac{(w_1+w_4)^2}{(w_1+w_3)^2}, \quad \text{as}\quad t\to 0_{-}\, .
\end{equation}
The string amplitude as $t\to 0_{-}$ reads
\begin{equation}
     \mathcal{A}_4 \sim -i\frac{(2\pi)^3}{t} Rg_s^2\mathcal{N}_{S^2} \,\delta_{\sum_k w_k}\,\delta\left(\sum_k E_k\right) w_1w_4 (w_1+w_4)^2 w_2 w_3
\end{equation}
which gives the dual YM amplitude with a $t$-channel exchange of a string state of winding $(w_1+w_4)$ (see Fig. \ref{fig:YM_amplitudes}).
\item The $u$-channel pole comes from all three ranges of $\zeta$. Collecting them gives
\begin{equation}
  \begin{split}
         \mathcal{I}_4 \sim  \frac{2}{u} \,F(a,b,s), \quad \text{as}\quad u\to 0_{-}
    \end{split}
\end{equation}
where $ F(a,b,s)$ is defined by
\begin{equation}
    \begin{split}
        F(a,b,s):=\Big[ & \sin(\pi s)  \int_{0}^a d\zeta \, \zeta^{-s-1} (1-\zeta)^{s-1} (a-\zeta)^{1/2} (b-\zeta)^{1/2} \\
         &-\cos(\pi s) \int_a^b d\zeta\, \zeta^{-s-1} (1-\zeta)^{s-1} (\zeta-a)^{1/2} (b-\zeta)^{1/2}  \\
         &- \sin(\pi s) \int_b^1 d\zeta \, \zeta^{-s-1} (1-\zeta)^{s-1} (\zeta-a)^{1/2} (\zeta-b)^{1/2}  \Big]
    \end{split}
\end{equation}
It is easy to recognize that $F(a,b,s)$ is the integral of the discontinuity of 
\begin{equation}
    G(\zeta)=(-\zeta)^{-s-1}(1-\zeta)^{s-1}(a-\zeta)^{1/2}(b-\zeta)^{1/2} \, ,
\end{equation}
across the cut between $0$ and $1$. Concretely
\begin{equation}
F(a,b,s) =\frac{1}{2i} \left( \int_0^a+\int_a^b +\int_b^1 \right) d\zeta\, \text{Disc}[ G(\zeta)]=-\frac{1}{2i} \oint_{C_{[0,1]}} d\zeta \, G(\zeta) =\frac{1}{2i} \oint_{C_\infty} d\zeta \, G(\zeta) \, ,
\end{equation}
where in the second equality we have expressed the discontinuity as a contour integral around the cut, and in the last equality we have deformed the contour into the one around $\infty$ where $G(\zeta)$ has a pole. The latter can be evaluated by making a change of variable $\zeta=-1/\zeta'$. Near $\zeta'\sim 0$, one finds $G(\zeta)\sim \zeta'$ as $\zeta'\to 0$, so that
\begin{equation}
    \oint_{C_\infty}d\zeta \, G(\zeta)= -\oint_{0}\frac{d\zeta'}{\zeta'}=-2\pi i
\end{equation}
This implies $F(a,b,s)=-\pi$, and thus the $u$-channel contribution simplifies to
\begin{equation}
    \mathcal{I}_4\sim -\frac{2\pi}{u},\quad \text{as}\quad u\to 0_{-}\, .
\end{equation}
Hence the string amplitude as $u\to 0_{-}$ reads
\begin{equation}
     \mathcal{A}_4 \sim -i\frac{(2\pi)^3}{u} Rg_s^2\mathcal{N}_{S^2} \,\delta_{\sum_k w_k}\,\delta\left(\sum_k E_k\right) w_1w_3 (w_1+w_3)^2 w_2 w_4
\end{equation}
This yields the dual YM amplitude in the $u$-channel with an exchange of winding string with winding number $(w_1+w_3)$ (see Fig. \ref{fig:YM_amplitudes}).

\end{itemize}

\section{Derivation of the Mandelstam formula and $\beta$-$\beta$ OPE \label{sec:Mandelstam}}
Thus far we have computed the scattering amplitudes of our proposed string dual and matched them with those of Yang-Mills. In this section we now present the details of the Mandelstam formula that we used in \eqref{localization_result_1} to compute the string amplitudes, along with the derivation of $\beta$-$\beta$ OPE following the discussions in the literature \cite{Mandelstam:1985ww, Baba:2009ns} and adjusting them to our setup. We start below with the Mandelstam formula. 

\subsection{Mandelstam formula and its derivation}\label{Mandelstam_formula_and_its_derivation}
We want to evaluate the chiral composite linear dilaton action \eqref{eq:CCLDaction} with light-cone fields $(X^{+}, X^{-})$ localized at the Mandelstam maps $(\rho,\bar{\rho})$ in \eqref{Mandelstam_general_maps}. $\Gamma[\rho]$ can be viewed as the Liouville action associated with the conformal map from worldsheet to target-space coordinates $(\rho,\bar{\rho})$, and its evaluation has been discussed in \cite{Mandelstam:1985ww, Baba:2009ns}. Here we review the derivation with a careful treatment of regularization, which will be needed later for computing the $\beta$–$\beta$ OPE.

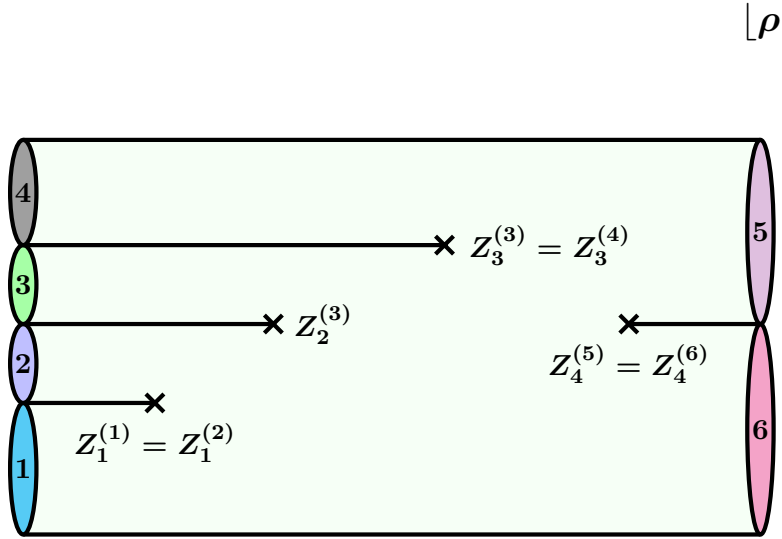
\begin{figure}[!htb]
    \centering
\begin{tikzpicture} [scale = 0.35]


\draw[white, fill=green!3.5!] (0,8.5) -- (28, 8.5) -- (28,-6.5) -- (0, -6.5) -- (0,8.5);


\draw[ultra thick, fill=cyan!55!] (0,-4) ellipse (0.5 and 2.5);

\draw[ultra thick, fill=blue!25!] (0,0) ellipse (0.5 and 1.5);

\draw[ultra thick, fill=green!35!] (0,0+3) ellipse (0.5 and 1.5);

\draw[ultra thick, fill= gray!75!] (0,0+3+3.5) ellipse (0.5 and 2);


\draw[ultra thick, fill=magenta!45!] (28,-2.5) ellipse (0.5 and 4);

\draw[ultra thick, fill= violet!25!] (28,5) ellipse (0.5 and 3.5);


\draw[ultra thick] (0,-6.5) -- (28,-6.5);
\draw[ultra thick] (0,-1.5) -- (5,-1.5);
\draw[ultra thick] (0,1.5) -- (9.5,1.5);
\draw[ultra thick] (0,1.5+3) -- (16,1.5+3);
\draw[ultra thick] (0,3+3.5+2) -- (28,3+3.5+2);
\draw[ultra thick] (28,1.5) -- (28-5,1.5);

\draw (5,-1.5) node[cross] {};
\draw (9.5,1.5) node[cross] {};
\draw (16,1.5+3) node[cross] {};
\draw (28-5,1.5) node[cross] {};

\node at (5,-1.5-1.5) {$\boldsymbol{Z_1^{(1)}=Z_1^{(2)}}$};
\node at (9.5+1.9,1.5) {$\boldsymbol{Z_2^{(3)}}$};
\node at (16+4,1.5+3) {$\boldsymbol{Z_3^{(3)}=Z_3^{(4)}}$};
\node at (23,1.5-1.5) {$\boldsymbol{Z_4^{(5)}=Z_4^{(6)}}$};

\node at (0,-4) { \textbf{1} };
\node at (0,0) { \textbf{2} };
\node at (0,0+3) { \textbf{3} };
\node at (0,6.5) { \textbf{4} };

\node at (28, -2.5) { \textbf{6} };
\node at (28, 5) { \textbf{5} };

\node at (28,13) {\scalebox{1.3}{$\lfloor \boldsymbol{\rho}$}};
\end{tikzpicture}
\caption{A six-point string tree diagram where the interaction point at which the $k$-th string interacts is denoted by $Z_I^{(k)}$ .}
\label{Light_cone_string_diagram}
\end{figure}

\begin{figure}
    \centering

\begin{tikzpicture} [scale = 0.6]


\draw[white, fill=green!7!]  (0,0) -- (23,0) -- (23,10) -- (0,10) -- (0,0);


\draw[red, ultra thick, fill=white] (3,5) circle (1);

\draw[orange, ultra thick, fill=white] (7,2) circle (1.5);

\draw[violet, ultra thick, fill=white] (10.7,8) circle (0.7);

\draw[blue, ultra thick, fill=white] (14.1,4) circle (1.1);

\draw[olive, ultra thick, fill=white] (17.2,7) circle (0.8);

\draw[teal, ultra thick, fill=white] (20.3,2.5) circle (1.3);

\draw (3,5) node[cross] {};

\draw (7,2) node[cross] {};

\draw (10.7,8) node[cross] {};

\draw (14.1,4) node[cross] {};

\draw (17.2,7) node[cross] {};

\draw (20.3,2.5) node[cross] {};


\draw[thick, ->] (3,5) -- (2.54,0.89+5);

\draw[ thick, ->] (7,2) -- (7.68, 1.336+2);  

\draw[ thick, ->] (10.7,8) -- (10.3822, 0.624+8);  

\draw[ thick, ->] (14.1,4) -- (13.606, 0.98+4);  

\draw[ thick, ->] (17.2,7) -- (17.563, 0.713+7);  

\draw[ thick, ->] (20.3,2.5) -- (19.709, 1.158+2.5);



\node at (3,-0.6+5)  {\scalebox{1}{$\boldsymbol{z_1}$}}; 


\node at (7,-0.8+2)  {\scalebox{1}{$\boldsymbol{Z_1}$}}; 


\node at (10.7,-0.42+8)  {\scalebox{0.7}{$\boldsymbol{x_2}$}};


\node at (14.1,-0.62+4)  {\scalebox{1}{$\boldsymbol{z_3}$}}; 


\node at (17.2,-0.5+7)  {\scalebox{0.7}{$\boldsymbol{Z_2}$}};


\node at (20.3,-0.7+2.5)  {\scalebox{1}{$\boldsymbol{z_4}$}};

\node at (3,1.6+5)  {\scalebox{1.3}{$\boldsymbol{\epsilon_{z_1}}$}}; 

\node at (7,2.1+2)  {\scalebox{1.3}{$\boldsymbol{\epsilon_{Z_1}}$}}; 

\node at (10.7,1.3+8)  {\scalebox{1.3}{$\boldsymbol{\epsilon_{z_2}}$}}; 

\node at (14.1,1.7+4)  {\scalebox{1.3}{$\boldsymbol{\epsilon_{z_3}}$}}; 

\node at (17.2,1.4+7)  {\scalebox{1.3}{$\boldsymbol{\epsilon_{Z_2}}$}}; 

\node at (20.3,1.8+2.5)  {\scalebox{1.3}{$\boldsymbol{\epsilon_{z_4}}$}};

\node at (22.2,9.3) {\scalebox{1.3}{$\lfloor \boldsymbol{z}$}};

\end{tikzpicture}
 
\caption{To regulate the divergences in $\Gamma[\rho]$ to compute the string amplitudes, we excise small discs on the worldsheet surface around the insertion points of the vertex operators $\{z_k\}_{k=1}^{n}$, the insertion points $\{Z_I\}_{I=1}^{n-2}$, and the point at infinity $(z_{\infty},\bar{z}_{\infty})$ . }
\label{excising_discs}
\end{figure}
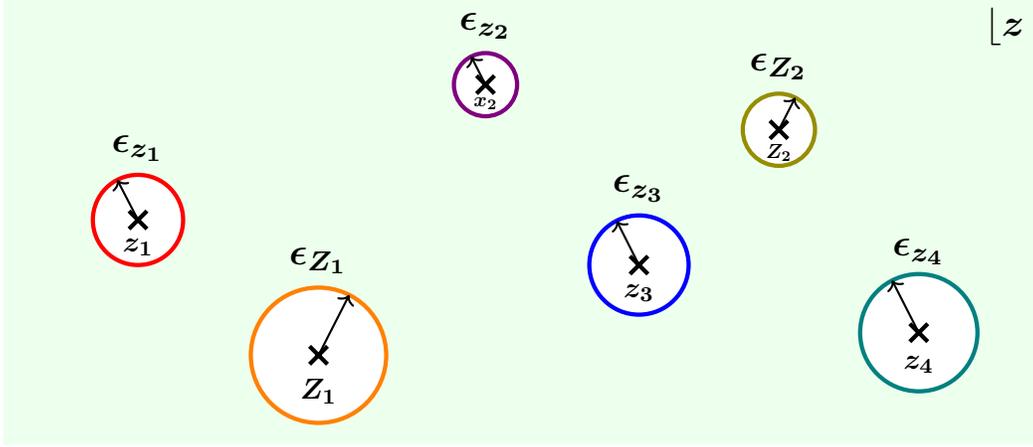

From its expression in \eqref{eq:CCLDaction}, $\Gamma[\rho]$ exhibits divergences around the vertex operator insertion points $\{z_k\}_{k=1}^{n}$, the zeros of $\partial \rho$ vanishes, called ``insertion points" $\{Z_I\}_{I=1}^{n-2}$, and the point at infinity $(z_{\infty},\bar{z}_{\infty})$.  We regulate these divergences by excising small discs around each of them (see Fig.~\ref{excising_discs}) and compute the contributions from the kinetic and curvature parts of $\Gamma[\rho]$ separately. 

Before proceeding, let us rephrase $\partial \rho (z)$ in terms of its zeros $\{Z_I\}_{I=1}^{n-2}$ and simple poles $\{ z_k\}_{k=1}^{n}$:
\begin{equation}\label{partial-rho-form}
	\partial\rho(z)= \left(\sum_{k=1}^{n}\alpha_k z_k\right)\frac{\prod_{I=1}^{n-2}(z-Z_I)}{\prod_{k=1}^{n}(z-z_k)} \, .
\end{equation}	
Similar expression holds for $\bar{\partial}\bar{\rho} (\bar{z})$.

\paragraph{Kinetic part of CLD action:}
We begin by considering the kinetic term of the CLD action. Performing integration by parts, we obtain
\begin{equation}
\begin{split}
\left(\Gamma[\rho]\right)_{\text{kinetic}}&=\frac{q}{2} \int \frac{d^2\sigma}{2\pi} \, \partial_{\alpha}\varphi\,\partial^{\alpha}\varphi=-\frac{q}{4\pi} \int d^2\sigma\, \varphi\, \partial_{\alpha}\partial^{\alpha}\varphi+ \frac{q}{4\pi} \int_{\partial} ds\,\varphi\, \partial_n \varphi \, ,
\end{split}
\end{equation} 
where the coordinates $(\sigma^1,\sigma^2)$ are related to  $(z,\bar{z})$ by, $z=\sigma^1+i\sigma^2$ and $\bar{z}=\sigma^1-i\sigma^2$. The equation of motion for the field $\varphi$ reads, $\partial_\alpha\partial^\alpha\varphi=0$. Consequently, only the boundary terms of the above integrals contribute and we thus obtain,
\begin{equation}\label{kinetic}
	\left(\Gamma[\rho]\right)_{\text{kinetic}}= \frac{q}{4\pi} \int_{\partial} ds\,\varphi\, \partial_n \varphi =\frac{q}{4\pi} \left(\int_{\cup_{I} \partial D_{Z_I}} +\int_{\cup_k \partial D_{z_k}}+\int_{\partial D_{z_\infty}}\right)ds\,\varphi\, \partial_n \varphi \, .
\end{equation}
For a disc boundary, the outward normal derivative takes the form
\begin{equation}
    \partial_n=-\frac{1}{|z|}(z\,\partial+\bar{z}\,\bar{\partial}) \, .
\end{equation}
Substituting this into the boundary integrals, the line integral can be rewritten in terms of holomorphic and anti-holomorphic contour integrals as 
\begin{equation}
\int_{\partial} ds\,\varphi\, \partial_n \varphi =i\oint dz \,\varphi\, \partial \varphi -i\oint d\bar{z} \, \varphi \,\bar{\partial} \varphi\, ,
\end{equation}
with contours encircling the centers of the excised discs\footnote{This follows directly by parameterizing the boundary as $z=re^{i\theta}$ with $ds=rd\theta$: $\int_{\partial}ds\, \left(-\frac{z}{|z|}\right)\varphi \partial \varphi=\int_0^{2\pi} rd\theta \,(-e^{i\theta}) \,\varphi \partial \varphi=i\int_{\theta=0}^{2\pi}d(re^{i\theta}) \, \varphi \partial \varphi =i\oint dz\, \varphi \partial \varphi \, ,$ and similarly $\int_{\partial}ds\, \left(-\frac{\bar{z}}{|z|}\right)\varphi \bar{\partial} \varphi= \int_0^{2\pi}r d\theta \, (- e^{-i\theta})\, \varphi \bar{\partial} \varphi=-i\int_{\theta=0}^{2\pi}d(re^{-i\theta})\, \varphi \bar{\partial} \varphi =-i\oint d\bar{z}\, \varphi \bar{\partial} \varphi \, . $}. In what follows we evaluate these integrals separately for $\cup_{I} \partial D_{Z_I}$, $\cup_{k} \partial D_{z_k}$, and $\partial D_{z_\infty}$.

\paragraph{Around $Z_I$:} We first consider the regions around the interaction points $z=Z_I$. In the $z$-plane, we excise discs of radius $\epsilon_{Z_I}$ around the interaction points $z=Z_I$. In the $\rho$-space it would corresponds to discs of radius $$r_I:= |\rho(Z_I+\epsilon_{Z_I}e^{i\theta_I})-\rho(Z_I)| \, ,$$ which are related by
\begin{equation}\label{regulator_map}
\begin{split}
r_I=\frac{1}{2}|\partial^2\rho(Z_I)|\,  \epsilon_{Z_I}^2 
&\Rightarrow \log \epsilon_{Z_I}=\frac{1}{2}\left(\log(2\,r_I)-\log|\partial^2 \rho(Z_I)|\right) \, .
\end{split}
\end{equation}
Note that, unlike the worldsheet regularization paramere $\epsilon_I$, the target space regulartion $r_I$ is independent of the choice of the coordinates on the worldsheet.
As we will see later, we use this target space regularization $r_I$ to remove the divergence and renormalize it by absorbing it to the string coupling $g_s$. 

Near the interaction point $z=Z_I$, the field $\partial \varphi$ exhibits a simple pole,
\begin{equation}
	\partial \varphi\approx\frac{\partial^2\rho(Z_I)}{(z-Z_I) \,\partial^2\rho(Z_I)+\cdots}=\frac{1}{z-Z_I}+\mathcal{O}(1) \, .
\end{equation}
Thus the boundary integral around $Z_I$ evaluates to
\begin{equation}
\begin{split}
&  \oint_{z=Z_I} dz \,\varphi \,\partial \varphi= \left(\log(2r_I)+\log|\partial^2\rho(Z_I)|\right)\oint_{z=Z_I}\frac{dz}{z-Z_I}=2\pi i \left(\log(2r_I)+\log|\partial^2\rho(Z_I)|\right)\, .
\end{split}
\end{equation}
Similarly, the anti-holomorphic counterpart gives
\begin{equation}
    \oint_{\bar{z}=\bar{Z}_I} d\bar{z} \,\varphi \,\bar{\partial} \varphi=-2\pi i \left(\log(2r_I)+\log|\partial^2\rho(Z_I)|\right) \, ,
\end{equation}
and combining both holomorphic and anti-holomorphic parts, we obtain:
\begin{equation}
\int_{\cup_{I} \partial D_{Z_I}} ds\,\varphi\, \partial_n \varphi =-4\pi\sum_{I=1}^{n-2}\left( \log|\partial^2\rho(Z_I)|+\log(2r_I)\right)\, .
\end{equation}

\paragraph{Around $z_k$:} Next, we consider the vicinities of the external insertion points $z=z_k$, where we introduce circular cutoffs of radius $\epsilon_{z_k}$. Near such a insertion point $z=z_k$, $\phi(z,\bar{z})$ goes as
\begin{equation}\label{near_z_k}
	\begin{split}
	 \varphi(z,\bar{z})=\log[\partial\rho \, \bar{\partial}\rho]\, \approx 2\log\left(\frac{|\alpha_k|}{|z-z_k|}\right) \, ,
		\end{split}
\end{equation}
and so the field $\partial \phi (z)$ behaves as
\begin{equation}
	\partial\varphi =\frac{\partial^2\rho(z)}{\partial \rho(z)}\approx -\frac{1}{z-z_k}\, .
\end{equation}
Thus the boundary integral takes the form  
\begin{equation}
	\oint_{z=z_k} dz\, \varphi \, \partial \varphi =-2\log\left(\frac{|\alpha_k|}{\epsilon_{z_k}}\right) \oint_{z=z_k}\frac{dz}{z-z_k}=-4\pi i \log\left(\frac{|\alpha_k|}{\epsilon_{z_k}}\right)  \, .
\end{equation}
Combining the similar contribution from anti-holomorphic part, we obtain:
\begin{equation}
	\int_{\cup_k \partial D_{z_k}} ds\, \varphi \,\partial_n \varphi =8\pi \log\left(\prod_{k=1}^{n}\frac{|\alpha_k|}{\epsilon_{z_k}}\right) \, .
\end{equation}

\paragraph{Around $z_\infty$:} Finally, we examine the region near infinity. We cut the $z$-space at $|z|=\frac{1}{\delta_\infty}$ with $\delta_\infty$ being small positive real number. Around infinity $(z_\infty,\bar{z}_{\infty})$, we have the following behaviours:
\begin{equation}\label{near_z_infinity}
\begin{split}
&\varphi(z,\bar{z}) \sim \log\left|\sum_{k=1}^{n}\alpha_k z_k\right|^2-\log|z|^4+\mathcal{O}\left(\frac{1}{|z|}\right)\, ,\\
&\partial\varphi \sim -\frac{2}{z}+\mathcal{O}\left(\frac{1}{z^2}\right)\, .
\end{split}
\end{equation}
Using these, we can evaluate the contour integral around $(z_\infty,\bar{z}_\infty)$:
\begin{equation}
\begin{split}
\oint_{C_{z_\infty}}dz\,\varphi\, \partial\varphi&=-2\log\left|\sum_{k=1}^{n}\alpha_k z_k\right|^2\underbrace{\oint_{C_{z_\infty}}\frac{dz}{z}}_{-2\pi i}+2\underbrace{\oint_{C_{z_\infty}} \frac{dz}{z}\, \log|z|^4}_{-2\pi i \,\log(1/\delta_{\infty}^4)}\\
&=4\pi i \log\left|\sum_{k=1}^{n}\alpha_k z_k\right|^2+16\pi i \log \delta_{\infty} \, .
\end{split} 
\end{equation}
Similarly, for the anti-holomorphic counterpart, we have
\begin{equation}
	\oint_{C_{z_\infty}}d\bar{z}\,\varphi\, \bar{\partial}\varphi=-4\pi i \log\left|\sum_{k=1}^{n}\alpha_k z_k\right|^2-16\pi i \log \delta_{\infty} \, .
\end{equation}
Adding both we thus obtain:
\begin{equation}
 \int_{D_{z_\infty}} ds\, \varphi \partial_n \varphi =-8\pi \log \left[\left|\sum_{k=1}^{n}\alpha_k z_k\right|^2\delta_\infty^{4}\right]\, .
\end{equation}

By summing these three boundary integral contributions in \eqref{kinetic}, we obtain the CLD kinetic term evaluated at the Mandelstam maps:
\begin{equation}\label{canonical_1}
  \frac{1}{q}  \left(\Gamma[\rho]\right)_{\text{kinetic}} = \log \left[\left|\sum_{k=1}^{n}\alpha_k z_k\right|^{-4} \delta_{\infty}^{-8}\,\prod_{I=1}^{n-2}\left|\partial^2\rho(Z_I)\right|^{-1}(2r_I)^{-1}\prod_{k=1}^{n}\left|\alpha_k\right|^{+2}\epsilon_{z_k}^{-2}\right] \, .
\end{equation}

\paragraph{Curvature part of CLD action:} We next evaluate the curvature contribution of $\Gamma[\rho]$ at the Mandelstam maps. It is given by
\begin{equation}
	\left(\Gamma[\rho]\right)_{\text{curvature}} =\frac{q}{2\pi}\int d^2z\, 	\left(\sqrt{-g}\,\hat{R}\right)_{(z,\bar{z})} \,\varphi (z,\bar{z}) \, .
\end{equation}
In computing the string amplitudes, the curvature of the worldsheet can be taken to be concentrated along the circular boundaries of the discs excised around the insertion points $\{z_k \}_{k=1}^{n}$ and the point at infinity $(z_\infty,\bar{z}_\infty)$. This is achieved by capping the holes with flat patches. Concretely, the metric near infinity reads
\begin{equation}
	ds^2=\begin{cases}
		dz \, d\bar{z},\quad |z|<\frac{1}{\delta_\infty}\\
		d\tilde{z}\, d\bar{\tilde{z}}, \quad |\tilde{z}|<\frac{1}{\delta_\infty} \, ,
	\end{cases}
\end{equation}
with $ \widetilde{z}=1/(\delta_\infty^2 z)$, while around each puncture $z_k$ one has
\begin{equation}
	ds^2=\begin{cases}
		d\xi_k \, d\bar{\xi}_k, \quad |\xi_k|<\epsilon_{z_k} \\
		d\tilde{\xi}_k\, d\bar{\tilde{\xi}}_k,  \quad |\tilde{\xi}_k|<\epsilon_{z_k} \, ,
	\end{cases}
\end{equation}
with $\xi_k=(z-z_k)$ and $\tilde{\xi}_k=\epsilon_{z_k}^2/\xi_k$ for $k=1,\cdots,n$. With this, the curvature is supported on the boundary circles and takes the distributional form 
\begin{equation}
    \left( \sqrt{-g}\, R \right)_{(z,\bar{z})} = 4\delta_\infty \cdot \delta\left( |z|-\frac{1}{\delta_\infty}\right) -2\sum_{k=1}^{n} \frac{1}{\epsilon_{z_k}} \delta\left(|z-z_k|-\epsilon_{z_k} \right) \, ,
\end{equation}
which clearly satisfies the Gauss-Bonnet theorem: $\int d^2z \, (\sqrt{|g|}\,\hat{R})_{(z,\bar{z})}=4\pi (2-n)$.

Using behaviours of $\varphi$ near $z_\infty$ \eqref{near_z_infinity}, and near the insertion points $z_k$  \eqref{near_z_k}, we find
\begin{equation}\label{canonical_2}
	\frac{1}{q}\left(\Gamma[\rho]\right)_{\text{curvature}}=8\log\left[\left|\sum_{k=1}^{n}\alpha_kz_k\right|\delta_\infty^2\right]-4\log\left[\prod_{k=1}^{n}\frac{|\alpha_k|}{\epsilon_{z_k}}\right] \, .
\end{equation}

\paragraph{Total contribution:} Adding the kinetic and curvature contributions to the CLD action, we obtain
\begin{equation}\label{Mandelstam_1}
\frac{1}{q}\Gamma[\rho]= \log\left[\prod_{k=1}^{n}|\alpha_k|^{-2}\left|\sum_{k=1}^{n}\alpha_kz_k\right|^4  \prod_{I=1}^{n-2}\left|\partial^2\rho\left(Z_I\right)\right|^{-1}\frac{ \delta_\infty^8\prod_{k=1}^{n}\epsilon_{z_k}^2}{\prod_{I=1}^{n-2}(2r_I)}\right] \,.
\end{equation}
Setting $r_I=r,\, \forall I$, the on-shell CLD action contributes to the worldsheet correlator as
\begin{equation}
    e^{-(\Gamma[\rho]-\Gamma_0)}= \prod_{k=1}^{n}|\alpha_k|^{2q} \left|\sum_{k=1}^{n}\alpha_k z_k\right|^{-4q} \prod_{I=1}^{n-2}\left|\partial^2\rho(Z_I)\right|^{q}   \cdot (2r)^{q(n-2)} \prod_{k=1}^{n} \epsilon_{z_k}^{-2q}\, ,
\end{equation}
where we subtracted $\Gamma_0$, the on-shell CLD action in the absence of vertex operator insertions. This cancels the divergent term $8q\log \delta_{\infty}$ in $\Gamma[\rho]$. The correlator is still singular because of the dependence on the regularization parameters $r$ and $\epsilon_k$. Below we explain how to remove them by appropriate renormalizations.

\paragraph{Renormalization:} The correlator comes with a prefactor $g_s^{,n-2}$ from the closed string coupling constant. Using this, one can remove the dependence on the regularization parameter $r$ by renormalizing $g_s$:
\begin{equation}
    \boldsymbol{g}_s=(2r)^{q} g_s\, ,
\end{equation}
and interpret $\boldsymbol{g}_s$ as the finite, physical closed string coupling. A similar renormalization has been discussed in the context of lightcone-gauge string (field) theory \cite{Mandelstam:1973jk}. On the other hand, the factors $\prod_{k=1}^{n}\epsilon_{z_k}^{-2q}$ can simply be absorbed into the renormalization of the vertex operators. After these redefinitions, the renormalized on-shell CLD action becomes
\begin{equation}\label{Mandelstam_final}
      e^{-\Gamma^{\rm ren}[\rho]}= \prod_{k=1}^{n}|\alpha_k|^{2q} \left|\sum_{k=1}^{n}\alpha_k z_k\right|^{-4q} \prod_{I=1}^{n-2}\left|\partial^2\rho(Z_I)\right|^{q}  \, .
\end{equation}

\paragraph{Different regularization to compute the $\beta$-$\beta$ OPE:} As we see shortly, to compute the $\beta$-$\beta$ OPE for the worldsheet CFT \eqref{action1}, we  insert winding vertex operators $e^{\alpha_k \int^{z_k}\beta}$ ($k=1,\dots,n$), differentiate with respect to $\alpha_k$'s and send $\alpha_k \to 0$. However, it turns out that the Mandelstam formula computed above is not suitable for this purpose since the result for the $n$ insertions of $e^{\alpha_k \int^{z_k}\beta}$ does not smoothly go over to that of the $n-1$ insertions  when one of $\alpha_k$'s is sent to zero. This is because $e^{\alpha_k \int^{z_k}\beta}$ behaves as a primary with a fixed non-zero dimension $q$ regardless of the value of $\alpha_k$ and does not trivialize in the limit $\alpha_k\to 0$. Technically, this peculiarity arises because the singularities around $Z_I$ are regulated and renormalized by the target space coordinate $\rho$, while the divergences around $z_k$ are regulated by the worldsheet cut-offs $\epsilon_k$'s. This makes the worldsheet theory ``non-local"; in other words, the CLD with the aforementioned renormalization is strictly speaking not a local worldsheet CFT. It should instead be viewed more as string {\it field} theory.

Thus, to properly compute the $\beta$-$\beta$ OPE, which is a local property of the worldsheet CFT rather than string {\it field} theory, one needs a different regularization in which the $n$-point function smoothly becomes $(n-1)$-point function. One such regularization was discussed in \cite{Baba:2009ns}. Instead of introducing a worldsheet cut-off $\epsilon_k$, we cut off the $k$-th external string propagator with its length in {\it target space} parameterized by a large positive number $l_k$, defined as
\begin{equation}\label{new_regularization}
	 l_k=\text{Re}\left[\Big(\rho(Z_I^{(k)})-\rho(z_k+\epsilon_{z_k} e^{i\theta_k})\Big)/\alpha_k \right] \, ,
\end{equation}
where $Z_I^{(k)}$ denotes the interaction point on the worldsheet at which the $k$-th external string interacts, see figure \ref{Light_cone_string_diagram}. (We will show that $l_k$ is independent of $\theta_k$.) Using this formula, we trade the worldsheet cut-off in the Mandelstam formula with the target space cut-off $l_k$. In this regularization, vertex operators are now reparameterization invariant as they are defined purely in terms of the target space, i.e.~they behave as primary operators with dimension 0. Hence they become trivial (identity operators) smoothly in the limit $\alpha_k\to 0$.
  We can indeed a posteriori justify\footnote{Of course, this argument does not explain why we employ this specific regularization rather than some other regularization that may have the same property. Ultimately, we justify our choice through the consistency of various computations such as the OPE of stress tensor being correctly reproduced. It would be desirable to develop deeper understanding on this point.} the regularization by computing the $n$-point function and verifying that it indeed reduces to the $(n-1)$-point function in the limit $\alpha_k\to 0$.

To relate $\epsilon_{z_k}$ and $l_k$, we first parameterize the neighborhood of the $k$-th interaction in the $\rho$-plane as
\begin{equation}
	\begin{split}
		 \rho \sim \alpha_k \log w_k+(\tau_0^{(k)}+i\beta_k)\quad \text{near}\quad z\sim z_k \, ,
	\end{split}
\end{equation}
with $w_k$ being the local coordinate near $z=z_k$, and $w_k=0$ corresponds to $z=z_k$, while $w_k=1$ corresponds to the interaction point $z=Z_I^{(k)}$: 
\begin{equation}
w_k(z_k)=0,\quad	w_k(Z_I^{(k)})=1 \Rightarrow\rho(Z_I^{(k)})=\tau_0^{(k)}+i\beta_k \, . 
\end{equation}
Thus expanding $\rho(z_k+\epsilon_{z_k}e^{i\theta_k})$ around $z=z_k$, we find
\begin{equation}
	\begin{split}
		\rho(z_k+\epsilon_ke^{i\theta_k})&=\alpha_k \log\Big[w_k(z_k)+\epsilon_{z_k}e^{i\theta_k}\cdot\frac{\partial w_k}{\partial z}\Big|_{z=z_k}+\mathcal{O}(\epsilon_{z_k}^2)\Big] +\rho(Z_I^{(k)})\\&\approx\alpha_k \log \epsilon_{z_k}+i\alpha_k\theta_k+\alpha_k \log\left[ \frac{\partial w_k}{\partial z}\Big|_{z=z_k}\right]+\rho(Z_I^{(k)}) \, ,
	\end{split}
\end{equation}
Substituting this into \eqref{new_regularization}, we obtain
\begin{equation}
\begin{split}
    l_k= \text{Re} \Big[- \log \epsilon_{z_k} - \log\left[ \frac{\partial w_k}{\partial z}\Big|_{z=z_k}\right] \Big]
   \Rightarrow \;  \epsilon_{z_k}= e^{- \text{Re}\left[ (\partial w_k/\partial z)_{z=z_k}\right]   -l_k} \, .
\end{split}    
\end{equation}
One can compute $\text{Re}\left[ (\partial w/\partial z_k)_{z=z_k}\right]$ explicitly, and it defines the ``Neumann coefficient" $\bar{N}_{00}^{kk}$ \cite{Mandelstam:1985ww}:
\begin{equation}\label{Neumann relation2}
\bar{N}_{00}^{kk}:=-\log\left[\frac{\partial w_k}{\partial z}\Big|_{z=z_k} \right]=-\frac{1}{\alpha_k}\sum_{j(\neq k)}^{n}\alpha_j \log(z_k-z_j)+\frac{\rho(Z_I^{(k)})}{\alpha_k} \, .
\end{equation} 
The second equality follows from writing
\begin{equation}
	\rho=\alpha_k \log w_k+(\tau_0^{(k)}+i\beta_k)=\sum_{j=1}^{n}\alpha_j\log(z-z_j)\quad \text{with}\quad z\sim z_k \, ,
\end{equation}
and taylor expanding the l.h.s around $z=z_k$ to collect the leading term. In terms of the Neumann coefficient, two regulators $\epsilon_{z_k}$ and $l_k$ are related by
\begin{equation}
    \epsilon_{z_k}=e^{\text{Re}[\bar{N}_{00}^{kk}]-l_k} \, .
\end{equation}
Substituting this in \eqref{Mandelstam_1} and omitting the curvature contrbutions from around $\{z_k\}_{k=1}^{n}$s, we obtain:
\begin{equation}\label{for_beta_beta}
		\frac{1}{q}\Gamma[\rho]= \log\left[\prod_{k=1}^{n}|\alpha_k|^{+2}\left|\sum_{k=1}^{n}\alpha_kz_k\right|^4\prod_{I=1}^{n-2}\left|\partial^2\rho\left(Z_I\right)\right|^{-1} e^{-2\sum_{k=1}^{n}\text{Re}\, \bar{N}^{kk}_{00}}\, e^{+2\sum_{k=1}^{n}l_k} \,\delta_\infty^8/(2r)^{n-2}\right]
\end{equation}
Thus, after subtracting the on-shell action $\Gamma_0$ without any insertions and dropping the target-space cutoff parameters $l_k$ and $r$, which are independent of the insertion points, the resulting expression reads
\begin{equation}\label{Gamma_for_beta_beta}
    -\frac{1}{q}\,(\Gamma[\rho]-\Gamma_0):= -2\sum_{r=1}^{n}\log |\alpha_r|-4\log\left|\sum_{r=1}^{n}\alpha_rz_r\right|  +2\sum_{r=1}^{n} \text{Re} \; \bar{N}^{rr}_{00}+\sum_{I=1}^{n-2}\log\left| \partial^2\rho(Z_I)\right| \, .
\end{equation}
As a consistency check of reparametrization invariance, consider the case with three insertions. In this case, the corresponding Neumann coefficients take the form
\begin{equation}
    \exp[-2\,\text{Re} \, \bar{N}^{33}_{00} ]= \frac{|z_{12}|^2}{|z_{13}z_{23}|^2} |\alpha_1|^{-2\alpha_1/\alpha_3}|\alpha_2|^{-2\alpha_2/\alpha_3}|\alpha_3|^{-2} \, ,
\end{equation}
with analogous expressions for $\text{Re} \, \bar{N}^{11}_{00}$ and $\text{Re} \, \bar{N}^{22}_{00}$. From \eqref{localization_result_1}, the contribution of the remaining factors in $\exp\left[\frac{1}{q} (\Gamma[\rho]-\Gamma_0) \right]$ gives the dependence $|z_{12}z_{23}z_{13}|^2$. Combining this with the Neumann coefficients, one finds that the on-shell CLD action becomes independent of the worldsheet insertion points $(z_i, \bar{z}_i)$, thereby confirming its invariance under reparametrizations.

\subsection{Computation of $\beta$-$\beta$ OPE}\label{beta-beta_OPE}
With the expression of the CLD action at Mandelstam localization maps, we now compute the $\beta$-$\beta$ OPE. We begin with the correlation function of vertex operators of the form $\exp\left[\alpha_r\int^{z_r}\beta\right]$, which acts as a generating functional for $\beta(z_r)$ insertions in the string path integral. Such correlation function takes the form
\begin{equation}
\begin{split}
	\mathcal{Z}\left(\{\alpha_r,\bar{\alpha}_r,z_r\}\right) &=\frac{1}{\mathcal{Z}_0}\int \left[\mathcal{D}\Phi\right]\, \exp[-S+\sum_{r=1}^{n}\alpha_r \int^{z_r}dz'\,\beta(z')+\sum_{r=1}^{n}\bar{\alpha}_r \int^{\bar{z}_r}d\bar{z}'\,\beta(\bar{z}')] \\
    &= \Big\langle \prod_{r=1}^{n} e^{\alpha_r \int^{z_r}\beta+\bar{\alpha}_r\int^{\bar{z}_r}\bar{\beta}}\Big\rangle \, ,
    \end{split}
\end{equation} 
where
$$\left[\mathcal{D}\Phi\right]=[\mathcal{D}\beta][ \mathcal{D}\bar{\beta}][\mathcal{D}X^{+}][ \mathcal{D} X^{-}]\, ,$$
and $\sum_{r=1}^{n}\alpha_r=\sum_{r=1}^{n}\bar{\alpha}_r=0$. Here $\mathcal{Z}_0$ is the vacuum partition function used to normalize the generating function to one without any insertions. Integrating out $(\beta,\bar{\beta})$, we find that $(X^+,X^-)$ localizes into the Mandelstam maps \eqref{Mandelstam_general_maps}. Then the generating functional reduces to the worldsheet CLD action localized at these maps $(\rho(z),\bar{\rho}(\bar{z}))$: 
\begin{equation}
    \mathcal{Z}\left(\{\alpha_r,\bar{\alpha}_r,z_r\}\right) = \exp\left[ -\Gamma [\rho]\right].
\end{equation}
Here $\Gamma[\rho]$ is the on-shell action $\frac{q}{2}\int\mathcal{L}_{CLD}$ with the regularization in \eqref{Gamma_for_beta_beta}. To compute the one-point function of $\beta(z)$, we first introduce two extra insertion points and consider the new generating functional
\begin{equation}
    \Big\langle \prod_{r=0}^{n+1} e^{\alpha_r \int^{z_r}\beta+\bar{\alpha}_r\int^{\bar{z}_r}\bar{\beta}}\Big\rangle=\exp\Big[ -\Gamma' [\rho']\Big] \, ,
\end{equation}
where the new Mandelstam map is given by
\begin{equation}
    \rho'(z)= \sum_{r=0}^{n+1}\alpha_r \log(z-z_r),\quad \text{with} \quad \alpha_{n+1}=-\alpha_0 \, .
\end{equation}
The one-point function $\langle \beta(z_0)\rangle$ is then obtained by bringing down $\int^{z_0}\beta$ in the path integral and taking another derivative $\partial_{z_0}$:
\begin{equation}
    \langle \beta(z_0)\rangle_\rho =  \partial_{z_0} \partial_{\alpha_0}  \Big( -\Gamma' [\rho']\Big)\Big|_{\alpha_0=0} \, ,
\end{equation}
where the subscript $\rho$ denotes the Mandelstam map used in computing the corresponding string path integral.  We can iterate this process and compute the two-point function
\begin{equation}
    \Big\langle \beta(z) \beta(z_0) \Big\rangle_\rho =  \partial_{z_0} \partial_{\alpha_0}   \langle \beta(z_0)\rangle_{\rho'}\Big|_{\alpha_0=0} \, .
\end{equation}
Thus to compute $\beta\beta$ OPE, we need to find derivative $\partial_{z_0}$ of various components of $\Gamma'[\rho']$ in analogue expression to \eqref{Gamma_for_beta_beta}, which we will be doing next.

\paragraph{Interaction points as series in $\alpha_0$:}
Two maps $\rho'(z)$ and $\rho(z)$ before and after the new insertions, are related by 
\begin{equation}\label{here12}
	\partial \rho'(z)=\partial \rho (z)+\frac{\alpha_0}{z-z_0}-\frac{\alpha_0}{z-z_{n+1}},\quad \text{with}\quad \partial\rho(z)=\sum_{r=1}^{n}\frac{\alpha_r}{z-z_r} \, .
\end{equation}
We denote the interaction points of $\rho'(z)$ by $\{Z'_I\}, Z'_{I^{(0)}}, Z'_{I^{(n+1)}}$, that is
\begin{equation}
	\partial \rho'(z)=0 \quad \text{for}\quad z=\{ Z'_I \}_{I=1}^{n-2}, Z'_{I^{(0)}}, Z'_{I^{(N+1)}} \,  .
\end{equation}
while those of $\rho(z)$ are $\{Z_I\}$:
\begin{equation}
	\partial\rho(z)=0 \quad \text{for}\quad z=\{ Z_I \}_{I=1}^{n-2} \, .
\end{equation}
We note that as $\alpha_0\to 0$, the following holds:
\begin{equation}
	Z'_I\to Z_I,\quad Z'_{I^{(0)}}\to z_0\quad \text{and} \quad Z'_{I^{(n+1)}} \to z_{n+1}\quad \text{as}\quad \alpha_0\to 0 \, . 
\end{equation}
This can be understood from the expression similar to \eqref{partial-rho-form} of $\partial \rho'(z)$ in terms of its zeros and poles. . In the limit $\alpha_0 \to 0$, $\partial \rho'$ must reduce to $\partial \rho$. This requires the two zeros $Z'_{I^{(0)}}$ and $Z'_{I^{(n+1)}}$ to approach the new poles $z_0$ and $z_{n+1}$ of $\partial \rho'(z)$, while the remaining zeros $\{Z'_I\}_{I=1}^{n-2}$ coincide with the original zeros $\{Z_I\}_{I=1}^{n-2}$ of $\partial \rho(z)$.

It is therefore convenient to expand the variations $(Z'_I-Z_I)$, $(Z'_{I^{(0)}}-z_0)$, and $(Z'_{I^{(n+1)}}-z_{n+1})$ as series in $\alpha_0$, which will then be used to compute the $\alpha_0$–derivative of $\Gamma'[\rho']$. We now turn to this computation.

Putting $z=Z'_I$ in \eqref{here12} and expanding $\partial\rho(Z'_I)=(Z'_I-Z_I)\, \partial^2\rho(Z_I)+\cdots$, we obtain:
\begin{equation}\label{DD.1-1}
 \quad	Z'_I-Z_I=-\frac{\alpha_0}{\partial^2\rho(Z_I)}\left(\frac{1}{Z_I-z_0}-\frac{1}{Z_I-z_{n+1}}\right)+\mathcal{O}\,(\alpha_0^2) \; .
\end{equation}
Next we put $z=Z'_{I^{0}}$ in \eqref{here12}, and expanding $\partial\rho(Z'_{I^{0}})=\partial \rho(z_0)+(Z'_{I^{(0)}}-z_0)\, \partial^2\rho(z_0)$, we get  
\begin{equation}\label{here14}
	\begin{split}
		-\frac{\alpha_0}{Z'_{I^{(0)}}-z_0} &= \partial\rho(z_0) \left[ 1+\frac{\partial^2\rho(z_0)}{\partial\rho(z_0)}(Z'_{I^{(0)}}-z_0)-\frac{\alpha_0}{\partial \rho(z_0)} \frac{1}{Z'_{I^{(0)}}-z_{n+1}} \right]\\
	\Rightarrow \quad  Z'_{I^{(0)}}-z_0 &=-\frac{\alpha_0}{\partial\rho(z_0)} \left[ 1-\frac{\partial^2\rho(z_0)}{\partial\rho(z_0)}(Z'_{I^{(0)}}-z_0)+\frac{\alpha_0}{\partial \rho(z_0)} \frac{1}{Z'_{I^{(0)}}-z_{n+1}} \right]\\
	&= -\frac{\alpha_0}{\partial\rho(z_0)} \left[ 1+\alpha_0\,\frac{\partial^2\rho(z_0)}{(\partial\rho(z_0))^2}+\frac{\alpha_0}{\partial \rho(z_0)} \frac{1}{z_0-z_{n+1}} +\mathcal{O}\,(\alpha_0^2)\right] \; ,
	\end{split}
\end{equation}
where in the last step, we used the step before it.  We can now replace the following in \eqref{here14}
\begin{equation}
	\alpha_0\to (-\alpha_0),\quad z_0\leftrightarrow z_{N+1}, \quad Z'_{I^{(0)}} \to Z'_{I^{(n+1)}}
\end{equation}
to obtain 
\begin{equation}\label{DD.1-3}
		Z'_{I^{(n+1)}}-z_{n+1}=\frac{\alpha_0}{\partial\rho(z_{n+1})} \left[ 1-\alpha_0\, \frac{\partial^2\rho(z_{n+1})}{(\partial\rho(z_{n+1}))^2}-\frac{\alpha_0}{\partial \rho(z_{n+1})} \frac{1}{z_{n+1}-z_{0}} +\mathcal{O}\, (\alpha_0^2)\right] \; .
\end{equation}
With these results in hand, we now evaluate the $\alpha_0$–variations of the Neumann coefficients and of $\partial^2\rho(Z_I)$, which will be needed for the variation of $\Gamma'[\rho']$. We begin with the Neumann coefficients.

\paragraph{Neumann coefficients as series in $\alpha_0$:}
From \eqref{Neumann relation2}, Neumann coefficients for $\rho(z)$ are 
\begin{equation}
	\bar{N}_{00}^{rr}=-\sum_{s(\neq r)}^{n} \frac{\alpha_s}{\alpha_r} \, \log(z_r-z_s)+\frac{\rho(Z_I^{(r)})}{\alpha_r},\quad r=1,\cdots,n.
\end{equation}
and similarly, for $\rho'(z)$ we have:
\begin{equation}
		\bar{N'}_{00}^{rr}= -\sum_{s(\neq r)}^{n} \frac{\alpha_s}{\alpha_r} \, \log(z_r-z_s)-\frac{\alpha_0}{\alpha_r} \log(z_r-z_0)+\frac{\alpha_0}{\alpha_r}\log(z_r-z_{n+1}) +\frac{\rho'({Z'}_I^{(r)})}{\alpha_r} \; ,
\end{equation}
for $r=1,\cdots,n$. Let's evaluate $\rho'({Z'}_I^{(r)})$ first:
\begin{equation}
	\begin{split}
		& 	\rho'(z)=\sum_{r=1}^{n} \alpha_r \log(z-z_r)+\alpha_0\log(z-z_0)-\alpha_0\log(z-z_{n+1})\\
		 \Rightarrow \; &\rho'({Z'}_I^{(r)}) = \rho(Z_I^{(r)})+\alpha_0\log\left[\frac{Z_I^{(r)}-z_0}{Z_I^{(r)}-z_{n+1}}\right]+\mathcal{O}(\alpha_0^2) \; .
	\end{split}
\end{equation}
Note that the term $({z'}_{I}^{(r)}-z_I^{(r)} ) \, \partial\rho(z_I^{(r)})$ from expansion of $\rho({z'}_I^{(r)})$ is absent here, since $\partial\rho(z_I^{(r)})=0$. Thus we obtain the series expansion of $\bar{N'}_{00}^{rr}$ in $\alpha_0$:
\begin{equation}\label{D.2-1}
		\bar{N'}_{00}^{rr}= 	\bar{N}_{00}^{rr}+ \frac{\alpha_0}{\alpha_r} \log\left[ \frac{(Z_I^{(r)}-z_0)(z_r-z_{n+1})}{(z_r-z_0)(Z_I^{(r)}-z_{n+1})}\right]+ \mathcal{O}\, (\alpha_0^2),\quad r=1,\cdots,n \; .
\end{equation}
Similarly, $\bar{N'}^{00}_{00}$ is explicitly given by
\begin{equation}
	\bar{N'}^{00}_{00}=-\sum_{s=1}^{n}\frac{\alpha_s}{\alpha_0}\log\left(z_0-z_s\right) +\log(z_0-z_{n+1})+\frac{\rho'({Z'}_I^{(0)})}{\alpha_0} \; .
\end{equation}
We need to expand $\rho({Z}_I^{(0)})$ upto $\mathcal{O}(\alpha_0^2)$ to get $	\bar{N'}^{00}_{00}$ upto $\mathcal{O}(\alpha_0)$. 
\begin{equation}
	\rho'({Z'}_{I}^{(0)})=\sum_{r=1}^{N}\alpha_r \log({Z'}^{(0)}_I-z_r)+\alpha_0 \log({Z'}^{(0)}_I-z_0)-\alpha_0\log\left( {Z'}_I^{(0)}-z_{n+1}\right)
\end{equation}
We can expand each term on the r.h.s into $\alpha_0$:
\begin{equation}
    \begin{split}
        i) \quad & \sum_{r=1}^{n}\alpha_r	\log({Z'}^{(0)}_I-z_r) =  \sum_{r=1}^{n}\alpha_r\log(z_0-z_r)+({Z'}_I^{(0)}-z_0)\, \partial \rho (z_0)+ \frac{1}{2}({Z'}_I^{(0)}-z_0)^2 \, \partial^2\rho(z_0) \\
& = \sum_{r=1}^{n}\alpha_r\log(z_0-z_r)-\alpha_0\left[ 1+\alpha_0\, \left( \frac{\partial^2\rho(z_0)}{(\partial\rho(z_0))^2}+\frac{1}{\partial\rho(z_0)}\frac{1}{z_0-z_{n+1}} \right)\right] +\frac{1}{2} \alpha_0^2 \,  \frac{\partial^2\rho(z_0)}{(\partial\rho(z_0))^2} \, ,\\
ii) \quad & \log({Z'}^{(0)}_I-z_0)=\log\left( -\frac{\alpha_0}{\partial\rho(z_0)}\right)+\alpha_0 \left(\frac{\partial^2\rho(z_0)}{(\partial\rho(z_0))^2}+\frac{1}{\partial\rho(z_0)}\frac{1}{z_0-z_{n+1}}\right)\, ,
\\ iii) \quad &\log({Z'}^{(0)}_I-z_{n+1}) =\log(z_0-z_{n+1})+\log\left(1+\frac{{Z'}_I^{(0)}-z_0}{z_0-z_{n+1}}\right)
	\\ & \quad \quad \quad \quad \quad \quad \quad \quad =\log\left(z_0-z_{n+1}\right) -\frac{\alpha_0}{\partial\rho(z_0)}\frac{1}{z_0-z_{n+1}} \; .
    \end{split}
\end{equation}
Adding these we obtain the series expansion of $\bar{N'}^{00}_{00}$:
\begin{equation}\label{DD.2-2}
	\bar{N'}^{00}_{00}=\log\left(-\frac{\alpha_0}{\partial\rho(z_0)}\right)-1+\left[\frac{1}{2}\frac{\partial^2\rho(z_0)}{(\partial\rho(z_0))^2}+\frac{1}{\partial\rho(z_0)}\frac{1}{z_0-z_{n+1}} \right]\alpha_0+\mathcal{O}(\alpha_0^2) \; .
\end{equation}
We can replace $\alpha_0\to -\alpha_0$, and $z_0\leftrightarrow z_{n+1}$ to obtain the expansion of $\bar{N'}^{(n+1) \, (n+1)}_{00}$:
\begin{equation}\label{DD.2-3}
		\bar{N'}^{(n+1) \, (n+1)}_{00}=\log\left(\frac{\alpha_0}{\partial\rho(z_{n+1})}\right)-1-\left[\frac{1}{2}\frac{\partial^2\rho(z_{n+1})}{(\partial\rho(z_{n+1}))^2}+\frac{1}{\partial\rho(z_{n+1})}\frac{1}{z_{n+1}-z_{0}} \right]\alpha_0+\mathcal{O}(\alpha_0^2)\; .
\end{equation}

\paragraph{ $\partial^2\rho (Z'_I, {Z'}_{I^{(0)}}, {Z'}_{I^{(n+1)}} )$ as series in $\alpha_0$: } Now we consider the variation of $\partial^2\rho(Z_I)$. We do it in the following steps. First, we aim to compute $\partial^2\rho'(Z'_I)$. From \eqref{here12}, we obtain
\begin{equation}\label{here13}
	\partial^2 \rho'(z)=\partial^2\rho(z)+\alpha_0\left(-\frac{1}{(z-z_0)^2}+\frac{1}{(z-z_{n+1})^2}\right) \; .
\end{equation}
Putting $z=Z'_I$ here and expanding $\partial^2\rho(Z'_I)=\partial^2\rho(Z_I)+\partial^3\rho(Z_I)\,(Z'_I-Z_I)$ with using \eqref{DD.1-1}, we get 
\begin{equation}\label{DD.4-1}
\begin{split}
	\partial^2\rho'(Z'_I) & =\partial^2\rho(Z_I)+\left(-\frac{1}{(Z_I-z_0)^2}+\frac{1}{(Z_I-z_{n+1})^2}\right)\alpha_0-\frac{\partial^3\rho(Z_I)}{\partial^2\rho(Z_I)} \left(\frac{1}{Z_I-z_0}-\frac{1}{Z_I-z_{n+1}}\right) \alpha_0 \\
    & \quad+ \, \mathcal{O}(\alpha_0^2) \; .
\end{split}    
\end{equation}
Next to compute $\partial^2\rho'({Z'}_{I^{(0)}})$, we fist put $z={Z'}_{I^{(0)}}$ in \eqref{here13} to obtain
 \begin{equation}
 	\begin{split}
 		\partial^2\rho'({Z'}_{I^{(0)}}) = \partial^2\rho({Z'}_{I^{(0)}})-\frac{\alpha_0}{(Z'_{I^{(0)}} -Z_0)^2}+\frac{\alpha_0}{(Z'_{I^{(0)}}-Z_{n+1})^2} \; .
 	\end{split}
 \end{equation}
For the second term on the r.h.s, we use the first relation in \eqref{here14} and keep terms of order $\mathcal{O}(\frac{1}{\alpha_0})$ and $\mathcal{O}(1)$ to get:
\begin{equation}\label{DD.4-2}
	\begin{split}
			\partial^2\rho'({Z'}_{I^{(0)}}) &= \partial^2\rho(z_0) -\frac{(\partial\rho(z_0))^2}{\alpha_0}  \left[ 1+\frac{\partial^2\rho(z_0)}{\partial\rho(z_0)}(Z'_{I^{(0)}}-z_0)-\frac{\alpha_0}{\partial \rho(z_0)} \frac{1}{Z'_{I^{(0)}}-z_{n+1}} \right]^2\\
			&=-\frac{(\partial\rho(z_0))^2}{\alpha_0} \left[1-3\frac{\partial^2\rho(z_0)}{(\partial\rho(z_0))^2}\alpha_0-\frac{2}{\partial\rho(z_0)}\frac{\alpha_0}{z_0-z_{n+1}}\right] +\mathcal{O}\, (\alpha_0)\; .
	\end{split}
\end{equation}
Replacing $\alpha_0\to -\alpha_0$, $z_0\leftrightarrow z_{n+1}$ on the r.h.s above, we obtain the expression of $\partial^2\rho'({Z'}_{I^{(n+1)}})$,
\begin{equation}\label{DD.4-3}
		\partial^2\rho'({Z'}_{I^{(n+1)}})= +\frac{(\partial\rho(z_{n+1}))^2}{\alpha_0} \left[1+3\frac{\partial^2\rho(z_{n+1})}{(\partial\rho(z_{n+1}))^2}\alpha_0+\frac{2}{\partial\rho(z_{n+1})}\frac{\alpha_0}{z_{n+1}-z_{0}}\right]+\mathcal{O}\, (\alpha_0)\; .
\end{equation}

\paragraph{ $\Gamma'[\rho']$ as series in $\alpha_0$:} Having computed the necessary variations of various components, we are now prepared to evaluate the $\alpha_0$–derivative of $\Gamma'[\rho']$. From \eqref{Gamma_for_beta_beta}, we first recall the expression for $\Gamma'[\rho']$ as
\begin{equation}\label{Gamma'_for_beta_beta}
	\begin{split}
		-\frac{1}{q}\Gamma'[\rho'] = & -2\sum_{r=1}^{n}\log|\alpha_r| -4\log |\alpha_0|-4	\log\left|\sum_{s=1}^{n}\alpha_sz_s+\alpha_0z_0-\alpha_0z_{n+1}\right| +2\sum_{r=1}^{n} \text{Re} \, \bar{N'}^{rr}_{00}\\
		& + 2\,\text{Re}\, \bar{N'}^{00}_{00}+2\,\text{Re}\, \bar{N'}^{(n+1)\, (n+1)}_{00}+ \sum_{I=1}^{n-2}\log\left| \partial^2\rho(Z'_I)\right|+ \log\left| \partial^2\rho(Z'_{I^{(0)}})\right|+\log\left| \partial^2\rho(Z'_{I^{(n+1)}})\right|\; .
	\end{split}
\end{equation}
Using the following relation,
\begin{equation}
	\begin{split}
		\log\left|\sum_{s=1}^{n}\alpha_sz_s+\alpha_0z_0-\alpha_0 z_{n+1}\right|=\text{Re}\left[ \log\left(\sum_{s=1}^{n}\alpha_s z_s \right)+\alpha_0\frac{z_0-z_{n+1}}{\sum_{s=1}^{n}\alpha_s z_s}\right] \; ,
	\end{split}
\end{equation}
and variations of \eqref{Gamma'_for_beta_beta} from other components, we find the $\Gamma'[\rho']$ as the following series in $\alpha_0$ :
\begin{equation}
	\begin{split}
		-\frac{1}{q}\Gamma'[\rho']=&-\frac{1}{q}\Gamma[\rho]-2\log\alpha_0-4-4 \text{Re}\left(\alpha_0\frac{z_0-z_{n+1}}{\sum_{s=1}^{n}\alpha_sz_s}\right)\\
		&+ \text{Re} \Big[ 2\alpha_0\sum_{r=1}^{n}\frac{1}{\alpha_r} \log\left|\frac{(Z_I^{(r)}-z_0)(z_r-z_{n+1})}{(z_r-z_0)(Z_I^{(r)}-z_{n+1})}\right| -2\alpha_0\frac{\partial^2\rho(z_0)}{(\partial\rho(z_0))^2}+2\alpha_0\frac{\partial^2\rho(z_{n+1})}{(\partial\rho(z_{n+1}))^2} \\& -\alpha_0\sum_{I=1}^{n-2}\frac{\partial^3\rho(Z_I)}{(\partial^2\rho(Z_I))^2} \left(\frac{1}{Z_I-z_0}-\frac{1}{Z_I-z_{n+1}}\right)-\alpha_0\sum_{I=1}^{n-2}\frac{1}{\partial^2\rho(Z_I)}\left(\frac{1}{(Z_I-z_0)^2}-\frac{1}{(Z_I-z_{n+1})^2}\right) \Big] \; .
	\end{split}
\end{equation}
Next we use the following relation 
\begin{equation}
	-2\frac{\partial^2\rho(z_0)}{(\partial\rho(z_0))^2}+2\frac{\partial^2\rho(z_{n+1})}{(\partial\rho(z_{n+1}))^2} -4\, \frac{z_0-z_{n+1}}{\sum_{s=1}^{n}\alpha_sz_s}=-2\sum_{I=1}^{n-2}\frac{1}{\partial^2\rho(Z_I)}\left(\frac{1}{(Z_I-z_0)^2}-\frac{1}{(Z_I-z_{n+1})^2}\right) \; ,
\end{equation}
to rewrite the CLD action $-\frac{1}{q}\Gamma'[\rho']$ as:
\begin{equation}\label{Gamma-rho'}
		\begin{split}
	-\frac{1}{q}\Gamma'[\rho']=& -\frac{1}{q}\Gamma[\rho]-2\log\alpha_0+ \text{Re} \Big[ 2\alpha_0\sum_{r=1}^{n}\frac{1}{\alpha_r} \log\left|\frac{(Z_I^{(r)}-z_0)(z_r-z_{n+1})}{(z_r-z_0)(Z_I^{(r)}-z_{n+1})}\right| \\& -\alpha_0\sum_{I=1}^{n-2}\frac{\partial^3\rho(Z_I)}{(\partial^2\rho(Z_I))^2} \left(\frac{1}{Z_I-z_0}-\frac{1}{Z_I-z_{n+1}}\right)-3\alpha_0\sum_{I=1}^{n-2}\frac{1}{\partial^2\rho(Z_I)}\left(\frac{1}{(Z_I-z_0)^2}-\frac{1}{(Z_I-z_{n+1})^2}\right) \Big]\; .
	\end{split}
\end{equation}

\subsubsection*{One-point function $\langle \beta(z_0) \rangle_\rho$:}
Taking two consecutive derivatives $\partial_{z_0}\partial_{\alpha_0}$ on \eqref{Gamma-rho'}, we compute the one-point function of $\beta(z_0)$:
\begin{equation}\label{why_this_life_2}
	\begin{split}
	\langle \beta(z_0) \rangle_\rho &=	\partial_{z_0}\partial_{\alpha_0} 	\left( -\Gamma'[\rho']\right) \Big|_{\alpha_0=0} \\
    &=q\,\partial_{z_0}\Big[ \sum_{r=1}^{n}\frac{1}{\alpha_r} \log\left(\frac{Z_I^{(r)}-z_0}{z_r-z_0}\right)  -\frac{1}{2}\sum_{I=1}^{n-2}\frac{\partial^3\rho(Z_I)}{(\partial^2\rho(Z_I))^2} \frac{1}{Z_I-z_0}-\frac{3}{2}\sum_{I=1}^{n-2}\frac{1}{\partial^2\rho(Z_I)}\frac{1}{(Z_I-z_0)^2}\Big] \; .
	\end{split}
\end{equation}
Note that we have an overall factor of $1/2$ to \eqref{Gamma-rho'} since we only need the holomorphic part.

\subsubsection*{Two-point function $\langle \beta(z)\beta(z_0)\rangle_{\rho}$:}
We want to evaluate the two-point function,
\begin{equation}\label{life_goal_2}
	\begin{split}
		\langle \beta(z)\, \beta(z_0)\rangle_{\rho}=\partial_{z_0}\partial_{\alpha_0} \langle \beta(z)\rangle_{\rho'}\Big|_{\alpha_0=0}=-
		\partial_{z_0}\partial_{\alpha_0}\left[ \partial_z\widetilde{\Gamma}[\rho']\right]\Big|_{\alpha_0=0} \; ,
	\end{split}
\end{equation}
where $\partial_z\widetilde{\Gamma}[\rho']$ is obtained from  $\partial_{z_0}\partial_{\alpha_0} 	\Gamma'[\rho']\Big|_{\alpha_0=0} $ in \eqref{why_this_life_2} by substituting in the r.h.s, $z_0\to z,\; \rho(Z_I)\to \rho'(Z'_{I'}),\; I\to I'$, and sum over $r$ going from $0$ to $(n+1)$.  Since we have to act $\partial_{z_0}$ on it, we should just consider the $r=0$ of the first term in \eqref{why_this_life_2}, $I'=I^{(0)}$ of the second and third  terms (recall that $Z'_{I^{(0)}}$ becomes $z_0$ as $\alpha_0\to 0$) in \eqref{why_this_life_2}.  i.e, the version of \eqref{why_this_life_2}, which will be useful for us is the following: 
\begin{equation}\label{why_this_life_3}
	\begin{split}
	\langle \beta(z)\rangle_{\rho'} =& q\,\partial_z\Big[ \frac{1}{\alpha_0} \log \frac{z-Z'_{I^{(0)}}}{z-z_0} -\frac{1}{2}\frac{\partial^3\rho'(Z'_{I^{(0)}})}{(\partial^2\rho'(Z'_{I^{(0)}}))^2} \frac{1}{Z'_{I^{(0)}}-z} -\frac{3}{2}\frac{1}{\partial^2\rho'(Z'_{I^{(0)}})}\frac{1}{(Z'_{I^{(0)}}-z)^2} \Big]+\cdots 
	\end{split} 
\end{equation}
Since we only have the first derivative $\partial_{\alpha_0}$, we need to keep linear term $\mathcal{O}(\alpha_0)$ in $\alpha_0$ in the expansion around $Z'_{I^{(0)}}=z_0$. The first term in \eqref{why_this_life_3} gives
\begin{equation}
	\begin{split}
&\frac{1}{\alpha_0} \log \frac{z-Z'_{I^{(0)}}}{z-z_0}=\frac{1}{\alpha_0} \log\Big( 1+\frac{z_0-Z'_{I^{(0)}}}{z-z_0}\Big) \\
&= \frac{1}{\partial \rho(z_0)(z-z_0)}+\Big[\left(\frac{\partial^2\rho(z_0)}{(\partial\rho(z_0))^3}+\frac{1}{(\partial\rho(z_0))^2(z_0-z_{n+1})}\right)\frac{1}{z-z_0}+\frac{-1/2}{(z-z_0)^2(\partial\rho(z_0))^2}\Big]\,\alpha_0.
\end{split}
\end{equation}
For the second and third terms in \eqref{why_this_life_3}, we evaluate each of its parts separately. For $(z-Z'_{I^{(0)}})^{-1}$, 
using \eqref{here14}, we obtain
\begin{equation}
\frac{1}{z-Z'_{I^{(0)}}}=\frac{1}{z-z_0}\left(1+\frac{z_0-Z'_{I^{(0)}}}{z-z_0}\right)^{-1}=\frac{1}{z-z_0}-\frac{\alpha_0}{\partial\rho(z_0)(z-z_0)^2} +\mathcal{O}(\alpha_0^2) \; .
\end{equation}
To compute $[\partial^2\rho'(Z'_{I^{(0)}})]^{-1}$, from \eqref{DD.4-2}, we obtain
\begin{equation}
	\frac{1}{\partial^2\rho'(Z'_{I^{(0)}})}=-\frac{\alpha_0}{(\partial\rho(z_0))^2}-\left[\frac{3\partial^2\rho(z_0)}{(\partial\rho(z_0))^4}+\frac{2}{(\partial\rho(z_0))^3(z_0-z_{n+1})}\right]\alpha_0^2 +\mathcal{O}(\alpha_0^3) \; .
\end{equation}
Lastly $\partial^3\rho'(Z'_{I^{(0)}})$ is computed as 
\begin{equation}
\partial^3\rho'(Z'_{I^{(0)}})=\partial^3\rho(Z'_{I^{(0)}})+2\alpha_0\left(\frac{1}{(Z'_{I^{(0)}}-z_0)^3}-\frac{1}{(Z'_{I^{(0)}}-z_{n+1})^3}\right) \, .
\end{equation}
Keeping terms till the order $\alpha_0^{0}$, we find
\begin{equation}
	\frac{\partial^3\rho'(Z'_{I^{(0)}})}{\partial^2\rho'(Z'_{I^{(0)}})}=2\frac{\partial\rho(z_0)}{\alpha_0}-\frac{2}{z_0-z_{n+1}}\; .
\end{equation}
We put all these together in \eqref{why_this_life_3}, and collect the terms of order $\alpha_0$ to obtain
\begin{equation}
	\begin{split}
		&\Big[ \frac{1}{\alpha_0} \log \frac{z-Z'_{I^{(0)}}}{z-z_0} -\frac{1}{2}\frac{\partial^3\rho'(Z'_{I^{(0)}})}{(\partial^2\rho'(Z'_{I^{(0)}}))^2} \frac{1}{Z'_{I^{(0)}}-z} -\frac{3}{2}\frac{1}{\partial^2\rho'(Z'_{I^{(0)}})}\frac{1}{(Z'_{I^{(0)}}-z)^2} \Big]_{\text{order}\; \alpha_0} \\
		&=  2\alpha_0\left[ \frac{1}{(\partial\rho(z_0))^2}\frac{1}{(z-z_0)^2}-\frac{\partial^2\rho(z_0)}{(\partial\rho(z_0))^3}\frac{1}{z-z_0}\right].
	\end{split}
\end{equation}
Using this, we finally obtain our two-point function: 
\begin{equation}
\begin{split}
		\langle\beta(z) \beta(z_0)\rangle_{\rho} &= 2 q\,\partial_{z}\partial_{z_0}  \left[ \frac{1}{(\partial\rho(z_0))^2}\frac{1}{(z-z_0)^2}-\frac{\partial^2\rho(z_0)}{(\partial\rho(z_0))^3}\frac{1}{z-z_0}\right] \\
        &= 2q\, \partial_{z}\partial_{z_0}\left[\frac{1}{(z-z_0)^2}\frac{1}{\partial\rho(z)\partial\rho(z_0)}\right] \; .
\end{split}        
\end{equation}
This confirms the $\beta$-$\beta$ OPE given in \eqref{OPE_first}.


\section{Conclusion\label{sec:conclusion}}

In this work, we presented a detailed formulation of the proposed worldsheet dual \cite{Komatsu:2025sqo} to chiral two-dimensional Yang–Mills theory at large $N_c$. Despite the unconventional nature of the worldsheet operator product expansions, which involve inverse fields, we have shown that the stress tensor OPE remains fully consistent with conformal invariance. Furthermore, we demonstrated that the string path integral localizes, allowing explicit computation of its observables. In particular, we evaluated the four-point string amplitude using a KLT-like factorization and found precise agreement with the corresponding amplitudes in the dual Yang–Mills theory, providing a nontrivial consistency check and further support for our duality proposal \cite{Komatsu:2025sqo}.

An important question for the future is to understand the relationship with  other proposed string duals of 2d YM \cite{Cordes:1994sd, Horava:1995ic, Benizri:2025xmz, Aharony:2023tam, Aharony:2025owv}, in particular to topological string dual to chiral 2d YM on a torus  \cite{Vafa:2004qa}. Understanding the connection to topological string may also lead to an improved formulation of the CLD action which is more regular and does not involve a logarithm of fields. This in turn may also clarify our proposed dual for symmetric product orbifolds of arbitrary CFT \cite{Komatsu:2025sqo} and its connection with other string duals to symmetric product orbifolds \cite{Eberhardt:2018ouy, Eberhardt:2019ywk, Hikida:2023jyc, Knighton:2024qxd}.

On the 2d YM side, it would be fascinating to explore whether the known non-perturbative correctons to 2d YM \cite{Witten:1991we, Witten:1992xu, Ochiai:1995zp, Griguolo:1998kq} admits a natural interpretation as D-brane/D-instanton effects within our worldsheet theory, and estabilish our duality at the non-perturbative level.

Constructing an  establishing such connections could lead to a deeper understanding of the underlying geometric structure and potentially yield an improved formulation of the composite linear dilaton action that plays a central role in our framework. It would also be fascinating to explore whether the known non-perturbative corrections to 2d YM observables \cite{Okuyama:2019rqn,Chen:2025yhg,Witten:1991we, Gross:1992tu} admit a natural interpretation in terms of D-brane effects within our worldsheet theory, thereby providing a non-perturbative completion of the duality.

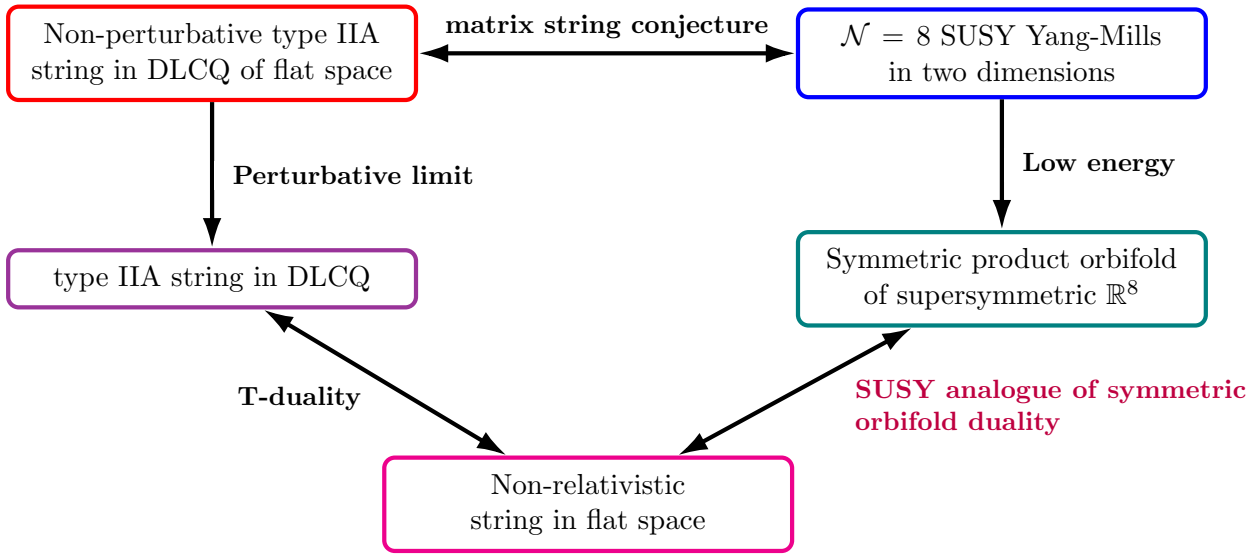
\begin{figure}[!tb]
\vspace{0.7cm}
    \centering
    
\begin{tikzpicture}[
    box/.style={draw, ultra thick, rounded corners, inner sep=6pt, align=center, text width=5cm},
    smalltext/.style={font=\small},
    labelstyle/.style={font=\small, midway, fill=white, inner sep=2pt}]


\vspace{10cm}

\node[box, draw=red] (A) at (0, 6) { Non-perturbative type IIA string in  DLCQ of flat space };

\node[box, draw=blue] (B) at (10.5, 6) {$\mathcal{N}=8$ SUSY Yang-Mills \\ in two dimensions};

\node[box, draw=red!50!blue!80!] (C) at (0, 3) {type IIA string in DLCQ};

\node[box, draw=magenta] (D) at (5, 0) {Non-relativistic string in flat space};

\node[box, draw=teal] (E) at (10.5, 3) {Symmetric product orbifold \\ of supersymmetric $\mathbb{R}^8$};

\draw[ultra thick, ->, >={Latex[length=4mm, width=2.5mm]} ] (A) -- node[labelstyle, right] {\, \textbf{Perturbative  limit} } (C);

\draw[ultra thick, <->, >={Latex[length=4mm, width=2.5mm]} ] (C) -- node[labelstyle, left, xshift=-6pt, yshift=-6pt] { \textbf{T-duality} } (D);


\draw[ultra thick, <->, >={Latex[length=4mm, width=2.5mm]} ] 
  (E) -- node[labelstyle, right, xshift=20pt, yshift=-6pt] 
  {\textcolor{purple}{\textbf{\parbox{5.2cm}{SUSY analogue of symmetric orbifold duality}}}} 
  (D);

\draw[ultra thick, ->, >={Latex[length=4mm, width=2.5mm]} ] (B) -- node[labelstyle, right] {\, \textbf{Low energy} } (E);

\draw[ultra thick, <->, >={Latex[length=4mm, width=2.5mm]} ] (A) -- node[labelstyle, above, yshift=3pt] { \textbf{matrix string conjecture} } (B) ;

\end{tikzpicture}

\caption{Web of dualities and relation to matrix string}
\label{web_of_dualities}
\end{figure}

Finally, let us point out close analogy with the matrix string theory \cite{Motl:1997th, Dijkgraaf:1997vv}. The matrix string conjecture relates the discrete light-cone quantization (DLCQ) of non-perturbative type IIA superstring theory to two-dimensional $\mathcal{N}=8$ super Yang–Mills theory. There, the perturbative limit of IIA string in DLCQ arises as the infrared limit of the gauge theory, where the theory is approximated by a symmetric product orbifold of supersymmetric $\mathbb{R}^8$. On the other hand, it is known  \cite{Harmark:2017rpg, Bergshoeff:2018yvt} that the T-dual of the DLCQ string is a nonrelativistic string theory \cite{Gomis:2000bd, Danielsson:2000gi, Danielsson:2000mu} governed by a $\beta$-$\gamma$ system---remarkably similar to our worldsheet theory. These webs of the duality imply the relation between the non-relativistic string formulated in terms of $\beta$-$\gamma$ system and the symmetric product orbifold of supersymmetric $\mathbb{R}^{8}$, which can be viewed as a supersymmetric counterpart of our proposal \cite{Komatsu:2025sqo} for string dual to symmetric product orbifolds (see Fig.~\ref{web_of_dualities}). All in all, they hint at a broader web of dualities connecting matrix string theory, nonrelativistic string theory, and 2d YM \cite{Billo:1998fb, Billo:1999bv, Blair:2023noj}, which calls for  more thorough examination of these connections.


\subsection*{Acknowledgment}
We thank Ofer Aharony, Alejandra Castro, Lorenz Eberhardt, Matthias Gaberdiel, Rajesh Gopakumar, Victor Gorbenko, David Gross, Bob Knighton, Jorrit Kruthorff, Suman Kundu, Juan Maldacena, Kiarash Naderi, Beat Nairz, Ashoke Sen, Tal Sheaffer, Vit Sriprachyakul, Cumrun Vafa, Spenta Wadia and Ziqi Yan for useful discussions and correspondences. P.M is supported by the Swiss National Science Foundation (SNSF).

\bibliographystyle{JHEP}

\appendix

\section{Two-point amplitudes in critical string}\label{two-point-section}
In this appendix we detail the computation of two-point closed string amplitudes at the critical dimension, in line with the comments in \cite{Erbin:2019uiz}.

We start with the path integral expression of closed string two-point amplitude, which is given by
\begin{equation}\label{two_point_closed}
\mathcal{A}^{\text{closed}}_2=\frac{1}{V_{\text{CKG}}} e^{-2\lambda}\int [\mathfrak{D}X]\, e^{-S[X,\hat{g}]}\, F[\hat{g}]\, \int d^2z_1 d^2z_2\, g_sV_1(z_1,\bar{z}_1)\, g_sV_2(z_2,\bar{z}_2) \, ,
\end{equation}
where $V_{CKG}$ is the volume of conformal killing group (CKG), $e^{-2\lambda}$ comes from Euler number of the closed worldsheet, and $F[\hat{g}]$ is the Fadeev-Popov determinant for fixing the worldsheet metric to $(\hat{g})_{ab}$ in the conformal gauge. To fix the CKG, we insert the following identity in the path integral:
\begin{equation}\label{Delta_closed_string}
1=\Delta_{X^0}\int [\mathfrak{D}\alpha]\,\delta^{(2)}(\alpha\circ z_1-z_1^0)\, \delta^{(2)}(\alpha\circ z_2-z_2^0)\, (\alpha\circ z_2)^{2}\, (\bar{\alpha}\circ \bar{z}_2)^{2}\, \delta\left( \frac{1}{4\pi R_{S^2}^2}\int_{S^2} d^2 z\, \sqrt{|g_{S^2}|}\, X^0\left[\alpha\circ (z,\bar{z})\right]\right) \, .
\end{equation}
Few comments are in order:
\begin{itemize}
\item We have implicitly assumed $(z_2^0,\bar{z}_2^0)\to\infty$, and that explains the extra factor $(\alpha\circ z_2)^{2h}\, (\alpha\circ \bar{z}_2)^{2\bar{h}}$ with $h=\bar{h}=1$, needed to define the corresponding the vertex operator at $\infty$. 

\item Here area of the worldsheet sphere is $\int_{S^2} d^2 z\, \sqrt{|g_{S^2}|}=4\pi R_{S^2}^2$. With this, the last delta function in \eqref{Delta_closed_string} takes the form
\begin{equation}
\delta\left( \frac{1}{4\pi R_{S^2}^2}\int_{S^2} d^2 z\, \sqrt{|g_{S^2}|}\, X^0\left[\alpha\circ (z,\bar{z})\right]\right)=\delta(x^0-\#) \, ,
\end{equation}
where $x^0$ is the zero mode of $X^0[\alpha\circ (z,\bar{z})]$. This then fixes the $\int dx^0$-integral within $X$-path integral: $\int[\mathfrak{D}X]$.

\item  Here $\Delta_{X^0}\equiv\Delta_{X^0}[z_{1}^0,z_2^0,\bar{z}_{1}^0,\bar{z}_2^0]$ should be viewed as an operator, since it depends on the field $X^0$ even though the argument of $X^0(z,\bar{z})$ is integrated on the worldsheet, or in other words, it will be triggered when the identity \eqref{Delta_closed_string} is inserted in the path integral $\int [\mathfrak{D}X]$ of \eqref{two_point_closed}.

\item  Note that, $|g_{S^2}|$ above depends on the vertex operator insertion points $(z_1,\bar{z}_1)$ and $(z_2,\bar{z}_2)$. 

\end{itemize}

After inserting \eqref{Delta_closed_string} in \eqref{two_point_closed}, we relabel $z_1\to \alpha\circ z_1$ and  $z_2\to \alpha\circ z_2$ at the vertex operator insertion points, and then perform $\int d^2z_1 d^2z_2$. Thus we obtain
\begin{equation}
\begin{split}
\mathcal{A}^{\text{closed}}_2=\frac{\left[ \int \mathfrak{D}\alpha\right]}{V_{\text{CKG}}}\, e^{-2\lambda}g_s^2\int [\mathfrak{D}X] \, e^{-S[X,\hat{g}]}\, F[\hat{g}]\, \Delta_{X^0} \, V_1(z_1^0,\bar{z}_1^0)\, V_2(z_2^0,\bar{z}_2^0)\, (z_2^0)^{2}(\bar{z}_2^0)^{2}\,\delta(x^0-\#) \, .
\end{split}
\end{equation}
Within the $\int [\mathcal{D}X]$-path integral, the delta function amounts to the zero mode integral of $X^0$: 
\begin{equation}
i\int dx^0\, \delta(x^0-\#)) \, e^{-i(k_1^0+k_2^0)\,x^0}=i\,e^{i(k_1^0+k_2^0)\, \#} \, ,
\end{equation}
while the zero modes of spatial coordinates give
\begin{equation}
\int \left[\prod_{i=1}^{D-1}dx^i\right]\, e^{i\sum_{i=1}^{25}(k_1^i+k_2^i)\cdot x^i}=(2\pi)^{D-1} \, \delta^{(D-1)}(\vec{k}_1+\vec{k}_2) \, .
\end{equation}
For the two-point amplitudes, we can take $k_{1}^0=\sqrt{m^2+\vec{k}_{1}^2},\quad k_{2}^0=-\sqrt{m^2+\vec{k}_{2}^2}$, with the first excitation incoming and the second outgoing. Momentum conservation, $\delta^{(D-1)}(\vec{k}_1+\vec{k}_2)$ then enforces energy conservation, $k_1^0+k_2^0=0$.
The time integral therefore reduces to a constant, $ie^{i(k_1^0+k_2^0)\, \#}\to i$. In the absence of gauge fixing, the two-point function would produce a divergent energy-conserving delta function, $\delta(k_1^0+k_2^0)$. This divergence is removed by fixing the zero mode of the time coordinate with the gauge condition $\delta\left( \frac{1}{4\pi R_{S^2}^2}\int_{S^2} d^2 z\, \sqrt{|g_{S^2}|}\, X^0\left[\alpha\circ (z,\bar{z})\right]\right)$.

Thus After carrying out the zero-mode integrals and using $\int [\mathfrak{D}\alpha]=V_{CKG}$, the two-point closed string amplitude takes the form
\begin{equation}\label{closed_string_expression}
\mathcal{A}_2^{\text{closed}}=ie^{-2\lambda}g_s^2 \Big\langle \Delta_{X^0}\, V_1(z_1^0,\bar{z}_1^0)\, V_2(z_2^0,\bar{z}_2^0)\Big\rangle'\, (z_2^0)^{2}\,  (\bar{z}_2^0)^{2}\, F[\hat{g}]\, (2\pi)^{D-1}X_0^{-D}\,\delta^{(D-1)}(\vec{k}_1+\vec{k}_2)\, ,
\end{equation}
where the prime indicates that the zero-mode integrals  $\int dx^0 \prod_{k=1}^{D-1}dx^k$ have already been performed. The next step is to evaluate $\Delta_{X^0}$, which we do using a convenient trick:

\paragraph{Decomposing the Mobius transformation:}
To evaluate the $\int [\mathfrak{D}\alpha]$ integral in $\Delta_{X^0}^{-1}$, it is convenient to decompose the Mobius transformation $PSL(2,\mathbb{C})$ as the following (see Fig.~\ref{Mobius_decomposition})
\begin{equation}\label{Mobius_decomposition}
    \alpha \equiv \alpha'\circ \alpha_* \;\Rightarrow 
    \begin{pmatrix}
    a & b\\
    c & d
    \end{pmatrix}\equiv \begin{pmatrix}
        a' & b'\\
        c' & d
    \end{pmatrix}\begin{pmatrix}
        a_* & b_*\\
        c_* & d_*
    \end{pmatrix} \, ,
\end{equation}
with
\begin{equation}
    ( \alpha_*\circ z_1,  \bar{\alpha}_*\circ\bar{z}_1)=(z_1^0,\bar{z}_1^0),\quad \text{and} \quad   (\alpha_*\circ z_2, \bar{\alpha}_*\circ \bar{z}_2)=(z_2^0,\bar{z}_2^0) \, .
\end{equation}
With such decomposition, the measure reduces to $[\mathfrak{D}\alpha] \to [\mathfrak{D}\alpha']$.

In other words, the original integration over all Möbius transformations $\alpha$ constrained by $\delta^{(2)}(\alpha\circ z_{1,2}-z_{1,2}^0)$ is equivalently rewritten as an integration over residual transformations $\alpha'$ around the fixed points $(z^0_{1,2},\bar{z}^0_{1,2})$, with $\alpha_*$ implementing the mapping by group composition.

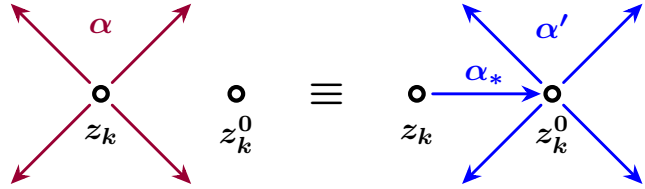
\begin{figure}[!htb]
    \centering
   \begin{tikzpicture}[
  >=Stealth,
  line cap=round,line join=round,
  plum/.style = {draw=purple!80!black, -{Stealth[length=2.6mm,width=2.2mm]}, line width=1.1pt},
  leaf/.style = {draw=blue,  -{Stealth[length=2.6mm,width=2.2mm]}, line width=1.1pt},
  every node/.style={inner sep=0pt}, scale=0.6
]

\draw[black, ultra thick] (0,0) circle (0.15cm);

\draw[black, ultra thick] (3,0) circle (0.15cm);

\draw[black, ultra thick] (7,0) circle (0.15cm);

\draw[black, ultra thick] (10,0) circle (0.15cm);

\draw[plum] (0,0)+(0.25,0.25) -- (2,2)  ;

\draw[plum] (0,0)+(0.25,-0.25) -- (2,-2)  ;

\draw[plum]  (0,0)+(-0.25,0.25) -- (-2,2);

\draw[plum]  (0,0)+(-0.25,-0.25) -- (-2,-2);

\draw[leaf]  (10,0)+(0.25,0.25) -- (12,2) ;

\draw[leaf] (10,0)+(0.25,-0.25) -- (12,-2)  ;

\draw[leaf]  (10,0)+(-0.25,0.25) -- (8,2);

\draw[leaf]  (10,0)+(-0.25,-0.25) -- (8,-2);

\draw[leaf] (7.35,0) -- (9.75,0);


\node at (5,0)  {\scalebox{1.4}{$\boldsymbol{\equiv}$}}; 

\node at (0,-0.9)  {\scalebox{1.1}{$\boldsymbol{z_k}$}}; 

\node at (3,-0.9)  {\scalebox{1.1}{$\boldsymbol{z_k^0}$}};

\node at (7,-0.9)  {\scalebox{1.1}{$\boldsymbol{z_k}$}}; 

\node at (10,-0.9)  {\scalebox{1.1}{$\boldsymbol{z_k^0}$}};

\node at (0,1.5)  {\scalebox{1.1}{$\boldsymbol{\textcolor{purple!80!black}{\alpha}}$}};

\node at (10,1.5)  {\scalebox{1.1}{$\boldsymbol{\textcolor{blue}{\alpha'}}$}}; 

\node at (8.5,0.4)  {\scalebox{1.1}{$\boldsymbol{\textcolor{blue}{\alpha_{*}}}$}};

\end{tikzpicture}
\caption{Decomposition of the Mobius transformation, $\textcolor{purple!80!black}{\alpha}\equiv \textcolor{blue}{\alpha'\circ \alpha_*} \, .$ }
\label{Mobius_decomposition}
\end{figure}

Setting $ (z_1^0,\bar{z}_1^0)=0$ and $(z_2^0,\bar{z}_2^0)=\infty$, we can easily determine the transformation $\alpha_*$:
\begin{equation}
  \alpha_*\circ z_1=\frac{a_* z_1+b_*}{c_* z_1+d_*}=0\Rightarrow b_*=-a_* z_1, \quad 
    \alpha_*\circ z_2=\frac{a_* z_2+b_*}{c_* z_2+d_*}=\infty\Rightarrow d_*=-c_*z_2 \, .   
\end{equation}
Further, the unit determinant condition constraints $a_*$ and $c_*$: $a_*d_*-b_*c_*=1\Rightarrow a_*c_*(z_1-z_2)=1$. Adapting similar decomposition for $PSL(2,\mathbb{R})$, we first compute two-point open string amplitude before analyzing the amplitude for the closed string.

\paragraph{Two-point open string amplitude:} 
The expression similar to \eqref{closed_string_expression} for the two-point open string amplitude reads
\begin{equation}\label{open_string_two_point}
    \mathcal{A}_2^{\text{open}}=ie^{-\lambda}g_o^2 \Big\langle \Delta_{X^0}\, V_1(y_1^0)\, V_2(y_2^0)\Big\rangle'\, (y_2^0)^{2}\, F[\hat{g}]\, (2\pi)^{D-1}\,\delta^{(D-1)}(\vec{k}_1+\vec{k}_2) \, ,
\end{equation}
where \cite{Erbin:2019uiz}
\begin{equation}\label{Delta_open}
\begin{split}
    \Delta_{X^0}^{-1}&=\int [\mathfrak{D}\alpha]\,\delta(\alpha\circ y_1-y_1^0)\, \delta(\alpha\circ y_2-y_2^0)\, (\alpha\circ y_2)^{2}\, \delta\left( X^0(\alpha\circ z_3^0,\alpha\circ \bar{z}_3^0)\right)\\
    &=2\int da\, db\,dc\, dd \, \delta(ad-bc-1)\, \delta\left( \frac{ay_1+b}{cy_1+d}-y_1^0\right)\, \delta\left( \frac{ay_2+b}{cy_2+d}-y_2^0\right)\, \left(\frac{ay_2+b}{cy_2+d} \right)^2\\ &\quad \quad \quad \times \delta\left( X^0\left(\frac{az_3^0+b}{cz_3^0+d},\frac{a\bar{z}_3^0+b}{c\bar{z}_3^0+d}\right)\right) \, .
\end{split}    
\end{equation}
Note that we have used the measure $$ [\mathfrak{D}\alpha] = 4\cdot \frac{1}{2}da\,db\,dc \,dd \,\delta(ad-bc-1) \, ,$$ for $\alpha$ being an element of $PSL(2,\mathbb{R})$: $\alpha \equiv \begin{pmatrix}
    a & b \\
    c & d
\end{pmatrix}$. The extra factor of 4 comes from the standard normnalization in string theory \cite{Liu:1987nz}, as emphasized below eq. (B.10) in \cite{Eberhardt:2021ynh}. 

Setting $y_1^0=0$ and $y_2^0=\infty$, the first two delta functions in \eqref{Delta_open} reduces to
\begin{equation}
\begin{split}
    &\delta\left( \frac{ay_1+b}{cy_1+d}-y_1^0\right)=\delta\left( \frac{a'y^0_1+b'}{c'y^0_1+d'}-y_1^0\right)=\delta\left( \frac{b'}{d'}\right) \, ,\\ & \delta\left( \frac{ay_2+b}{cy_2+d}-y_2^0\right) = \delta\left( \frac{a'y^0_2+b'}{c'y^0_2+d'}-y_2^0\right)=\delta\left( \frac{a'}{c'}-\infty\right) =\left( \frac{c'}{a'}\right)^2 \delta\left( \frac{c'}{a'}\right) \, .
    \end{split}
\end{equation}
Using this, the expression for $\Delta_{X^0}^{-1}$ simplifies to
\begin{equation}
    \begin{split}
        \Delta_{X^0}^{-1}=2\int da'\, db'\, dc'\, dd'\, \delta(a'd'-b'c'-1)\, \delta\left( \frac{b'}{d'}\right)\,  \delta\left( \frac{c'}{a'}\right)\, \delta\left( X^0\left(\frac{a'Z_3^0+b'}{c'Z_3^0+d'},\frac{a'\bar{Z}_3^0+b'}{c'\bar{Z}_3^0+d'}\right)\right) \, ,
    \end{split}
\end{equation}
with $\alpha_*\circ (z_3^0,\bar{z}_3^0)=(Z_3^0,\bar{Z}_3^0)$. Note that, the factor $\left(\frac{ay_2+b}{cy_2+d} \right)^2$ in \eqref{Delta_open} cancels the one coming from $\delta\left( \frac{a'}{c'}-\infty\right)$.

Using $\delta\left( \frac{b'}{d'}\right)=|d'|\,\delta(b')$ and $\delta\left( \frac{c'}{a'}\right)=|a'|\, \delta(c')$, and performing the $b',c'$ integrals, we obtain
\begin{equation}
 \Delta_{X^0}^{-1}=2\int da'dd'\, \delta(a'd'-1)\, |a'||d'| \, \delta\left( X^0\left(\frac{a'}{d'}Z_3^0,\frac{a'}{d'}\bar{Z}_3^0 \right)\right) \, .
\end{equation}
Further we can use $\delta(a'd'-1)=\frac{1}{|a'|}\delta\left(d'-\frac{1}{a'}\right)$ and perform the $d'$-integral to obtain
\begin{equation}
    \begin{split}
         \Delta_{X^0}^{-1} &= 2\int \frac{da'}{|a'|}\,\delta\left[X^0\left((a')^2Z_3^0,(a')^2\bar{Z}_3^0 \right)\right]\\
         &=2\int \frac{da'}{|a'|}\, \frac{\delta(a'-a'_0)}{\Big|2a'_0 Z_3^0\,\partial X^0\left((a'_0)^2Z_3^0,(a'_0)^2\bar{Z}_3^0 \right)+2a'_0 \bar{Z}_3^0\,\bar{\partial} X^0\left((a'_0)^2Z_3^0,(a'_0)^2\bar{Z}_3^0 \right)\Big|}\\
         &=\Big| y_3^0 \, \partial X^0(y_3^0,\bar{y}_3^0)+\bar{y}_3^0 \, \bar{\partial} X^0(y_3^0,\bar{y}_3^0)\Big|^{-1}  \, ,
    \end{split}
\end{equation}
where $X^0\left((a'_0)^2Z_3^0,(a'_0)^2\bar{Z}_3^0 \right)=0$, and we have defined $y_3^0=(a'_0)^2\, Z_3^0$ and $\bar{y}_3^0=(a'_0)^2\, \bar{Z}_3^0$. We thus conclude
\begin{equation}
     \Delta_{X^0}=\left[ y_3^0 \, \partial X^0(y_3^0,\bar{y}_3^0)+\bar{y}_3^0 \, \bar{\partial} X^0(y_3^0,\bar{y}_3^0)\right] \, .
\end{equation}
Inserting this in the worldsheet correlator in \eqref{open_string_two_point}, we get
\begin{equation}\label{two_point_worldsheet_correlator_open}
    \begin{split}
        \Big\langle \Delta_{X^0}\, V_1(y_1^0)\, (y_2^0)^2\,V_2(y_2^0)\Big\rangle' &=\left[ y_3^0 \Big\langle  \partial X^0(y_3^0,\bar{y}_3^0)\, V_1(y_1^0)\,(y_2^0)^2 V_2(y_2^0)\Big\rangle'+\bar{y}_3^0 \Big\langle  \bar{\partial} X^0(y_3^0,\bar{y}_3^0)\, V_1(y_1^0)\,(y_2^0)^2 V_2(y_2^0)\Big\rangle'\right] \, .
    \end{split}
\end{equation}
The three point function on the r.h.s  can be computed as
\begin{equation}\label{3_pt_1}
    \begin{split}
         \Big\langle  \partial X^0(y_3^0,\bar{y}_3^0)\, V_1(y_1^0)\, (y_2^0)^2V_2(y_2^0)\Big\rangle'&=-i\alpha'\left[ \frac{k_1^0}{y_3^0-y_1^0}+\frac{k_2^0}{y_3^0-y_2^0}\right] \Big\langle V_1(y_1^0)\, (y_2^0)^2V_2(y_2^0)\Big\rangle'\\&=-i\alpha'k^0\frac{y_{12}^0}{(y_3^0-y_1^0)(y_3^0-y_2^0)} \Big\langle  V_1(y_1^0)\, (y_2^0)^2V_2(y_2^0)\Big\rangle'\\&=-i\alpha'k^0\,\frac{1}{y_3^0}\Big\langle V_1(0)\, V_2(\infty)\Big\rangle' \, ,
    \end{split}
\end{equation}
where we have used $k_1^0=-k_2^0=k^0$ in the second equality, and $y_1^0=0, \, y_2^0=\infty$ in the last. Similarly,
\begin{equation}\label{3_pt_2}
    \Big\langle  \bar{\partial} X^0(y_3^0,\bar{y}_3^0)\, V_1(y_1^0)\, (y_2^0)^2V_2(y_2^0)\Big\rangle'= -i\alpha'k^0\,\frac{1}{\bar{y}_3^0}\Big\langle V_1(0)\, V_2(\infty)\Big\rangle' \, .
\end{equation}
Using $\Big\langle V_1(z_1^0)\,V_2(z_2^0)\Big\rangle' =C^X_{D_2}$, with $C^X_{D_2}$ being the partition function on the disc $D_2$ from the matter sector, together with the above three-point function computation, the worldsheet correlator in \eqref{two_point_worldsheet_correlator_open} becomes
\begin{equation}
    \Big\langle \Delta_{X^0}\, V_1(y_1^0)\, (y_2^0)^2\,V_2(y_2^0)\Big\rangle' =-2\,i\alpha' k^0 \Big\langle V_1(0)\, V_2(\infty)\Big\rangle'=-2\, i\alpha' k^0\, C^{X}_{D_2} \, .
\end{equation}
The two-point open string amplitude \eqref{open_string_two_point} then reduces to
\begin{equation}
\begin{split}
    \mathcal{A}^{\text{open}}_2 
    &= 2k^0\,(2\pi)^{D-1}\delta^{(D-1)}(\vec{k}_1+\vec{k}_2)\times \left[\alpha' e^{-\lambda} g_0^2\,C^X_{D_2}\,F[\hat{g}]\right]\\
    &= 2k^0\,(2\pi)^{D-1}\delta^{(D-1)}(\vec{k}_1+\vec{k}_2) \, ,
\end{split}    
\end{equation}
where the combination $\alpha'e^{-\lambda} \,g_0^2 \,C^X_{D_2}\,F[\hat{g}]$ has been evaluated in eq. (6.4.14) of \cite{Polchinski:1998rq} (with $F[\hat{g}]$ denoted by $C^{g}_{D_2}$) and equals
\begin{equation}
    \alpha' e^{-\lambda} g_0^2 \,C^X_{D_2}\,F[\hat{g}]=1 \, .
\end{equation}

\paragraph{Two-point closed string amplitude:} 
Having seen how the decomposition \eqref{Mobius_decomposition} works for two-point amplitudes in open string, we now come back to the case for closed string, which is our main concern in this appendix. Using explicit matrix realization of $\alpha$, $\Delta_{X^0}^{-1}$ in \eqref{Delta_closed_string} becomes
\begin{equation}\label{Delta_inverse}
\begin{split}
\Delta_{X^0}^{-1}
&= 4\int d^2a \, d^2b\, d^2c\, d^2d\, \delta^{(2)}(ad-bc-1)\, \delta^{(2)}\left( \frac{a\,z_1+b}{c\, z_1+d}-z_1^0\right)\, \delta^{(2)}\left( \frac{a\,z_2+b}{c\, z_2+d}-z_2^0\right)\, \Big| \frac{az_2+b}{cz_2+d}\Big|^4\\
&\quad \quad\quad\times  \,\delta\left( \frac{1}{4\pi R_{S^2}^2}\int_{S^2} d^2 z\, \sqrt{|g_{S^2}|}\, X^0\left[\alpha\circ (z,\bar{z})\right]\right) \, .
\end{split}
\end{equation}
where we have used the measure $$[\mathfrak{D}\alpha]=8\cdot \frac{1}{2}d^2a \, d^2b\, d^2c\, d^2d \, \delta^{(2)}(ad-bc-1) \, , $$  with the additional factor of $8$, as in the open–string case, arising from the standard string theory normalization \cite{Liu:1987nz}. Setting $(z_1^0,\bar{z}_1^0)=0$, $(z_2^0,\bar{z}_2^0)=\infty$, and using the decomposition of Mobius transformation in \eqref{Mobius_decomposition}, the first delta functions in the above integrand become
\begin{equation}\label{delta_1}
        \delta^{(2)}\left( \frac{a\,z_1+b}{c\, z_1+d}-z_1^0\right) = \delta^{(2)}\left( \frac{a'\,z^0_1+b'}{c'\, z^0_1+d'}-z_1^0\right)= \delta^{(2)}\left( \frac{b'}{d'}\right)
        =\frac{|d'|^2}{|b'|}\, \delta(|b'|)\,\delta(\theta_{b'}-\theta_{d'}) \, .
\end{equation}
Similarly, the second delta function reduces to
\begin{equation}\label{delta_2}
        \delta^{(2)}\left( \frac{a\,z_2+b}{c\, z_2+d}-z_2^0\right)=\delta^{(2)}\left( \frac{a'}{c'}-\infty\right)
        = \delta^{(2)}\left( \frac{c'}{a'}\right)\,\Big| \frac{c'}{a'}\Big|^4= \frac{|a'|^2}{|c'|}\, \delta(|c'|)\,\delta(\theta_{c'}-\theta_{a'}) \,\Big| \frac{c'}{a'}\Big|^4 \, .
\end{equation}
The last delta function becomes
\begin{equation}\label{delta_3}
    \delta\left( \frac{1}{4\pi R_{S^2}^2}\int_{S^2} d^2 z\, \sqrt{|g_{S^2}|}\, X^0\left[\alpha\circ (z,\bar{z})\right]\right)=\delta\left( \frac{1}{4\pi R_{S^2}^2}\int_{S^2} d^2 z\, \sqrt{|g_{S^2}|}\, X^0\left(\frac{a'Z+b'}{c'Z+d'},\frac{\bar{a}'\bar{Z}+\bar{b}'}{\bar{c}'\bar{Z}+\bar{d}'}\right)\right) \, ,
\end{equation}
with
\begin{equation}
    Z=\alpha_{*}\circ z=\frac{a_*}{c_*}\frac{z-z_1}{z-z_2},\quad \bar{Z}=\bar{\alpha}_{*}\circ \bar{z}=\frac{\bar{a}_*}{\bar{c}_*}\frac{\bar{z}-\bar{z}_1}{\bar{z}-\bar{z}_2} \, .
\end{equation}
Using the above delta function evaluations \eqref{delta_1}, \eqref{delta_2}, and \eqref{delta_3} in the expression \eqref{Delta_inverse}, and integrating over $|b'|$ and $|c'|$ in $\int d^2b'\,\int d^2c'=\int |b'|d|b'|d\theta_{b'}\, \int |c'|d|c'|d\theta_{c'}$, we obtain
\begin{equation}
    \begin{split}
        \Delta_{X^0}^{-1}&=4\int d^2a'\, d^2d'\, \delta^{(2)}(a'd'-1)\,|d'|^2|a'|^2\, \delta\left( \frac{1}{4\pi R_{S^2}^2}\int_{S^2} d^2 z\, \sqrt{|g_{S^2}|}\, X^0\left(\frac{a'}{d'}\,Z,\frac{\bar{a}'}{\bar{d}'}\,\bar{Z}\right)\right) \, .
    \end{split}
\end{equation}
Now we evaluate $\delta^{(2)}(a'd'-1)$ as
\begin{equation}
    \begin{split}
        \delta^{(2)}(a'd'-1) &= \delta\left(|a'||d'|\cos(\theta_{a'}+\theta_{d'})-1\right)\, \delta\left(|a'||d'|\sin(\theta_{a'}+\theta_{d'})\right)\\
        &= \delta\left(|a'||d'|\cos(\theta_{a'}+\theta_{d'})-1\right)\, \frac{\delta(\theta_{a'}+\theta_{d'})}{|a'||d'||\cos(\theta_{a'}+\theta_{d'})|}\\
        &=\frac{1}{|a'||d'|} \,\delta(|a'||d'|-1)\, \delta(\theta_{a'}+\theta_{d'})=\frac{1}{|a'|^2|d'|}\,\delta\left(|d'|-\frac{1}{|a'|}\right) \, \delta(\theta_{a'}+\theta_{d'}) \, .
    \end{split}
\end{equation}
We can next perform integrals over $\theta_{d'}$ and $|d'|$ in $\int d^2a'\int d^2d'=\int |a'|d|a'|d\theta_{a'}\int |d'|d|d'|d\theta_{d'}$ to obtain 
\begin{equation}\label{Delta_inverse_2}
    \Delta_{X^0}^{-1}=4\int \frac{d|a'|}{|a'|}\int d\theta_{a'} \, \delta\left( \frac{1}{4\pi R_{S^2}^2}\int_{S^2} d^2 z\, \sqrt{|g_{S^2}|}\, X^0\left(|a'|^2\,e^{2i\theta_{a'}}\,Z,|a'|^2\,e^{-2i\theta_{a'}}\,\bar{Z}\right)\right) \, .
\end{equation}
Directly performing the $\int d\theta_{a'}$ integral is inconvenient, since a change of variables inside the delta function,  
\[
w = |a'|^2 e^{2i\theta_{a'}} Z \, ,
\]  
does not simply translate through the worldsheet measure $d^2z \,\sqrt{|g_{S^2}|}$, which in general depends nontrivially on $(Z,\bar Z)$. To handle this, we choose the worldsheet sphere metric $(g_{S^2})_{ab}$ in $(z,\bar z)$ coordinates so that
\[
d^2z \,\sqrt{|g_{S^2}|} = d^2Z \,\sqrt{|G_{S^2}|} \, ,
\]
where $(G_{S^2})_{AB}$ is the standard stereographic metric on $S^2$, with determinant  
\[
\sqrt{|G_{S^2}|} = \frac{2R_{S^2}^2}{(1+|Z|^2)^2}
= \frac{2R_{S^2}^2}{\left(1+\frac{|a_*|^2}{|c_*|^2}\frac{|z-z_1|^2}{|z-z_2|^2}\right)^2} \, .
\]

Meanwhile, the area elements are related by
\[
d^2Z = 
\det\!\begin{bmatrix}
\frac{\partial Z}{\partial z} & \frac{\partial \bar{Z}}{\partial z}\\[6pt]
\frac{\partial Z}{\partial \bar{z}} & \frac{\partial \bar{Z}}{\partial \bar{z}}
\end{bmatrix} d^2z
= \frac{|a_*|^2}{|c_*|^2}\frac{|z_1-z_2|^2}{|z-z_2|^4}\,d^2z
= \frac{d^2z}{|c_*|^4 |z-z_2|^4} \, ,
\]
where in the last step we used $a_*c_*(z_1-z_2)=1$. Thus, we choose the worldsheet metric such that
\[
\sqrt{|g_{S^2}|} = \frac{2R_{S^2}^2}{\big(|a_*|^2|z-z_1|^2+|c_*|^2|z-z_2|^2\big)^2}\,  .
\]

Now we can make a change of variable $w=e^{2i\theta_{a'}}Z$ in \eqref{Delta_inverse_2}, and the integrand becomes independent of $\theta_{a'}$. Evaluating t $\int d\theta_{a'}=2\pi$, the rest of the integral is
\begin{align}
    \Delta_{X^0}^{-1}&= 8\pi\int \frac{d|a'|}{|a'|}\, \delta\left( \frac{1}{4\pi R_{S^2}^2}\int_{S^2}d^2w\, \sqrt{|g_{S^2}(w)|}\; X^0(|a'|^2w,|a'|^2\bar{w})\right)\\
    &=8 \pi\,\left(\frac{1}{4\pi R_{S^2}^2} \int_{S^2}d^2w\sqrt{|g_{S^2}(w)|}\,\left[2|A|^2w\cdot\partial X^0(|A|^2w,|A|^2\bar{w})+2|A|^2\bar{w}\cdot\bar{\partial} X^0(|A|^2w,|A|^2\bar{w})\right]\right)^{-1} \, ,\nonumber
\end{align}    
where $a^{\prime}=A$ is the support of the delta function. As we will see below, the dependence on $A$ will cancel out in the final answer.

We thus obtain our Fadeev Popov determinant for fixing $PSL(2,\mathbb{C})$ symmetry as:
\begin{equation}
    \Delta_{X^0}=\frac{|A|^2}{4\pi}\, \left(\frac{1}{4\pi R_{S^2}^2} \int_{S^2}d^2w\sqrt{|g_{S^2}(w)|}\,\left[w\cdot\partial X^0(|A|^2w,|A|^2\bar{w})+\bar{w}\cdot\bar{\partial} X^0(|A|^2w,|A|^2\bar{w})\right]\right) \, .
\end{equation}
To evaluate $\mathcal{A}_{2}^{ \text{closed}}$ in \eqref{closed_string_expression}, we compute 
\begin{equation}\label{Correlator_after_decomposition}
    \begin{split}
          \Big\langle \Delta_{X^0}\, V_1(z_1^0,\bar{z}_1^0)\, V_2(z_2^0,\bar{z}_2^0)\Big\rangle'= \frac{|A|^2}{4\pi}\, \int_{S^2} &\frac{d^2w\sqrt{|g_{S^2}(w)|}}{4\pi R_{S^2}^2}\,\Big[ w\,\Big\langle \partial X^0(|A|^2w,|A|^2\bar{w})\, V_1(z_1^0,\bar{z}_1^0)\, V_2(z_2^0,\bar{z}_2^0)\Big\rangle' \\
          &+ \bar{w}\,\Big\langle \bar{\partial} X^0(|A|^2w,|A|^2\bar{w})\, V_1(z_1^0,\bar{z}_1^0)\, V_2(z_2^0,\bar{z}_2^0)\Big\rangle'\Big] \, .
    \end{split}
\end{equation}
As in the open–string three–point functions \eqref{3_pt_1} and \eqref{3_pt_2}, the correlators evaluate to
\begin{equation}
    \begin{split}
        & \Big\langle \partial X^0(|A|^2w,|A|^2\bar{w})\, V_1(z_1^0,\bar{z}_1^0)\, V_2(z_2^0,\bar{z}_2^0)\Big\rangle' = -i\frac{\alpha'}{2}k^0\frac{1}{|A|^2w}\frac{C_{S^2}^X}{|z_2^0|^4} \, ,\\
        &\Big\langle \bar{\partial} X^0(|A|^2w,|A|^2\bar{w})\, V_1(z_1^0,\bar{z}_1^0)\, V_2(z_2^0,\bar{z}_2^0)\Big\rangle' = -i\frac{\alpha'}{2}k^0\frac{1}{|A|^2\bar{w}}\frac{C_{S^2}^X}{|z_2^0|^4} \, ,
    \end{split}
\end{equation}
with $C^X_{S^2}$ being the partition function on worldsheet $S^2$ from the matter sector. Substituting these expressions, one finds
\begin{equation}
\begin{split}
   \Big\langle \Delta_{X^0}\, V_1(z_1^0,\bar{z}_1^0)\, V_2(z_2^0,\bar{z}_2^0)\Big\rangle'(z_2^0)^2(\bar{z}_2^0)^2&=\frac{|A|^2}{4\pi}\,\left( -i\frac{\alpha'}{2}k^0\frac{C_{S^2}^X}{|A|^2}]\times 2\right)\, \int_{S^2} \frac{d^2w\sqrt{|g_{S^2}(w)|}}{4\pi R_{S^2}^2}\\
   &=-\frac{i}{4\pi}\,\alpha' k^0\, C_{S^2}^X \, .
\end{split}   
\end{equation}
The two-point closed string amplitude \eqref{closed_string_expression} then reduces to 
\begin{equation}
    \begin{split}
        \mathcal{A}_2^{\text{closed}}
        &=2k^0\, (2\pi)^{D-1}\delta^{(D-1)}(\vec{k}_1+\vec{k}_2)\times \left[ e^{-2\lambda}g_s^2\,\frac{\alpha'}{8\pi}\, C^{X}_{S^2}\, F[\hat{g}]\right] \\
        &= 2k^0\, (2\pi)^{D-1}\delta^{(D-1)}(\vec{k}_1+\vec{k}_2) \, ,
    \end{split}
\end{equation}
where the combination $\alpha'e^{-2\lambda}g_s^2 \,C^X_{D_2}\,F[\hat{g}]/(8\pi)$ has been computed in eq. (6.6.8) of \cite{Polchinski:1998rq} (with $F[\hat{g}]$ denoted by $C^{g}_{S_2}$) and equals
\begin{equation}
   e^{-2\lambda}g_s^2\,\frac{\alpha'}{8\pi}\, C^{X}_{S^2}\, F[\hat{g}]=1 \, .
\end{equation}

\section{Alternative derivation of Mandelstam formula \label{Alternative_Mandelstam}}
In this appendix, we present an alternative computation of the kinetic term of the CLD action localized at the Mandelstam maps \eqref{Mandelstam_general_maps}. We first rewrite the kinetic term
\begin{equation}
    \left(\Gamma[\varphi] \right)_{\text{kinetic}}=\frac{q}{2} \int_{\mathcal{M}} \frac{d^2z}{2\pi}\,2\partial \varphi \bar{\partial}\varphi \, ,
\end{equation}
where we split the chiral composite linear dilaton field field $\varphi(z,\bar{z})$ into holomorphic and anti-holomorphic components:
\begin{equation}
	\varphi(z,\bar{z})=\log\left[\partial \rho(z)\bar{\partial} \bar{\rho}(\bar{z})\right]=\varphi_h(z)+\varphi_a(\bar{z}) \, ,
\end{equation}
with 
\begin{equation}
	\begin{split}
		\varphi_h(z)=\log\left[\left(\sum_{k=1}^{n}\alpha_k z_k\right)\frac{\prod_{I=1}^{n-2}(z-Z_I)}{\prod_{k=1}^{n}(z-z_k)}\right],\quad \varphi_a(\bar{z})=\log\left[\left(\sum_{k=1}^{n}\bar{\alpha}_k \bar{z}_k\right)\frac{\prod_{I=1}^{n-2}(\bar{z}-\bar{Z}_I)}{\prod_{k=1}^{n}(\bar{z}-\bar{z}_k)}\right] \, .
	\end{split}
\end{equation}
Using \eqref{partial-rho-form}, we obtain the relations
\begin{equation}\label{alpha_k-partial_2-rho}
    \alpha_k= \left( \sum_{i=1}^{n}\alpha_i z_i\right)\frac{ \prod_{I=1}^{n-2}(z_k-Z_I)}{\prod_{i(\neq k)}^{n}(z_k-z_i)} \, , \quad \partial^2\rho(Z_I)=\left( \sum_{k=1}^{n}\alpha_kz_k\right) \frac{\prod_{J(\neq I)}^{n-2}(Z_I-Z_J)}{\prod_{k=1}^{n}(Z_I-z_k)} \, ,
\end{equation}
which will be used in what follows. As before we have set $R=1$ which can be restored later. For convenience, we strip off the factor of $iq/(2\pi)$, and consider the worldsheet integral 
\begin{equation}
	-i\int_{\mathcal{M}}d^2 z\, \partial \varphi \,\bar{\partial}\varphi = \int_{\mathcal{M}} d\varphi_h \wedge d\varphi_a =\frac{1}{2}\int_{\partial \mathcal{M}} \left(\varphi_h(z)  d\varphi_a-\varphi_a(\bar{z}) d\varphi_h\right) \, .
\end{equation}
Although $\varphi (z,\bar{z})$ is smooth and single-valued on $\mathcal{M}$ away from its singularities at $\{ z_k\}_{k=1}^{n}$, $\{ Z_I\}_{I=1}^{n-2}$, and $(z_\infty,\bar{z}_\infty)$, its holomorphic and anti-holomorphic components, $\varphi_h(z)$ and $\varphi_a(\bar{z})$, exhibit branching behavior around these points. 

\begin{figure}[!htb]
    \centering

\resizebox{0.45\linewidth}{!}{%
\begin{tikzpicture}[x=0.75pt,y=0.75pt,yscale=-1,xscale=1]

\draw  [line width=2.25]  (305.01,243.58) -- (334.99,242.42)(319.42,228.01) -- (320.58,257.99) ;
\draw  [line width=2.25]  (135,369) -- (170,369)(152.5,354) -- (152.5,384) ;
\draw  [line width=2.25]  (277.36,422.93) -- (302.64,439.07)(298.07,418.36) -- (281.93,443.64) ;
\draw  [line width=2.25]  (522.47,315.72) -- (551.53,308.28)(533.28,297.47) -- (540.72,326.53) ;
\draw  [line width=2.25]  (113.46,173.69) -- (142.67,166.84)(124.7,155.93) -- (131.42,184.6) ;
\draw  [line width=2.25]  (224.18,69.18) -- (251.82,80.82)(243.82,61.18) -- (232.18,88.82) ;
\draw  [line width=2.25]  (475,117) -- (505,117)(490,102) -- (490,132) ;
\draw[brown!80!black,  ultra thick, snake it]    (240,77.5) -- (318,242.5) ;
\draw[brown!80!black,  ultra thick, snake it]     (128,170.5) -- (319,242.5) ;
\draw[ ultra thick, snake it]     (154,367.5) -- (319,242.5) ;
\draw[ ultra thick, snake it]   (290,431.5) -- (319,242.5) ;
\draw[brown!80!black, ultra thick, snake it]    (491,117.5) -- (319,242.5) ;
\draw[ ultra thick, snake it]    (538,311.5) -- (319,242.5) ;
\draw [color={rgb, 255:red, 74; green, 144; blue, 226 }  ,draw opacity=1 ][line width=1.5]    (164,346.5) -- (294,248.5) ;
\draw [color={rgb, 255:red, 74; green, 144; blue, 226 }  ,draw opacity=1 ][line width=1.5]    (143,188.5) -- (294,248.5) ;
\draw [color={rgb, 255:red, 74; green, 144; blue, 226 }  ,draw opacity=1 ][line width=1.5]    (153,166.5) -- (295,223.5) ;
\draw [color={rgb, 255:red, 74; green, 144; blue, 226 }  ,draw opacity=1 ][line width=1.5]    (237,93.5) -- (295,223.5) ;
\draw [color={rgb, 255:red, 74; green, 144; blue, 226 }  ,draw opacity=1 ][line width=1.5]    (259.27,82.51) -- (320,220.5) ;
\draw [color={rgb, 255:red, 74; green, 144; blue, 226 }  ,draw opacity=1 ][line width=1.5]    (464.3,124.99) -- (335,219.5) ;
\draw [color={rgb, 255:red, 74; green, 144; blue, 226 }  ,draw opacity=1 ][line width=1.5]    (177.01,365.11) -- (307.01,262.61) ;
\draw [color={rgb, 255:red, 74; green, 144; blue, 226 }  ,draw opacity=1 ][line width=1.5]    (284,403.5) -- (307.01,262.61) ;
\draw [color={rgb, 255:red, 74; green, 144; blue, 226 }  ,draw opacity=1 ][line width=1.5]    (478,140.5) -- (345,238.5) ;
\draw [color={rgb, 255:red, 74; green, 144; blue, 226 }  ,draw opacity=1 ][line width=1.5]    (517,295.5) -- (345,238.5) ;
\draw [color={rgb, 255:red, 74; green, 144; blue, 226 }  ,draw opacity=1 ][line width=1.5]    (324,267.5) -- (302,404.5) ;
\draw [color={rgb, 255:red, 74; green, 144; blue, 226 }  ,draw opacity=1 ][line width=1.5]    (510.87,313.5) -- (340,258.5) ;
\draw  [draw opacity=0][line width=1.5]  (143,188.5) .. controls (135.68,194.32) and (125.75,196.64) .. (116.07,193.87) .. controls (100.14,189.32) and (90.92,172.71) .. (95.47,156.78) .. controls (100.02,140.85) and (116.63,131.63) .. (132.56,136.18) .. controls (146.73,140.23) and (155.59,153.81) .. (154.17,167.98) -- (124.31,165.03) -- cycle ; 

\draw  [ green!60!black ,draw opacity=1 ][line width=1.5]  (143,188.5) .. controls (135.68,194.32) and (125.75,196.64) .. (116.07,193.87) .. controls (100.14,189.32) and (90.92,172.71) .. (95.47,156.78) .. controls (100.02,140.85) and (116.63,131.63) .. (132.56,136.18) .. controls (146.73,140.23) and (155.59,153.81) .. (154.17,167.98) ;  
\draw  [draw opacity=0][line width=1.5]  (237.65,94.83) .. controls (227.36,95.02) and (217.37,89.09) .. (212.72,78.98) .. controls (206.49,65.44) and (212.2,49.52) .. (225.46,43.42) .. controls (238.73,37.31) and (254.53,43.35) .. (260.76,56.89) .. controls (264.73,65.51) and (263.85,75.11) .. (259.27,82.51) -- (236.74,67.93) -- cycle ; 

\draw  [ green!60!black ,  ,draw opacity=1 ][line width=1.5]  (237.65,94.83) .. controls (227.36,95.02) and (217.37,89.09) .. (212.72,78.98) .. controls (206.49,65.44) and (212.2,49.52) .. (225.46,43.42) .. controls (238.73,37.31) and (254.53,43.35) .. (260.76,56.89) .. controls (264.73,65.51) and (263.85,75.11) .. (259.27,82.51) ;  
\draw  [draw opacity=0][line width=1.5]  (464.3,124.99) .. controls (459.37,112.28) and (463.68,97.34) .. (475.5,89.24) .. controls (489.28,79.81) and (508.04,83.24) .. (517.4,96.91) .. controls (526.76,110.58) and (523.19,129.31) .. (509.41,138.74) .. controls (499.74,145.37) and (487.61,145.65) .. (478,140.5) -- (492.46,113.99) -- cycle ; 

\draw  [ green!60!black ,  ,draw opacity=1 ][line width=1.5]  (464.3,124.99) .. controls (459.37,112.28) and (463.68,97.34) .. (475.5,89.24) .. controls (489.28,79.81) and (508.04,83.24) .. (517.4,96.91) .. controls (526.76,110.58) and (523.19,129.31) .. (509.41,138.74) .. controls (499.74,145.37) and (487.61,145.65) .. (478,140.5) ;  
\draw  [draw opacity=0][line width=1.5]  (517,295.5) .. controls (523.52,286.96) and (534.43,282.21) .. (545.77,284.09) .. controls (562.11,286.79) and (573.17,302.24) .. (570.47,318.58) .. controls (567.76,334.93) and (552.32,345.99) .. (535.97,343.28) .. controls (521.21,340.84) and (510.76,328) .. (510.87,313.5) -- (540.87,313.68) -- cycle ;

\draw  [ green!60!black ,draw opacity=1 ][line width=1.5]  (517,295.5) .. controls (523.52,286.96) and (534.43,282.21) .. (545.77,284.09) .. controls (562.11,286.79) and (573.17,302.24) .. (570.47,318.58) .. controls (567.76,334.93) and (552.32,345.99) .. (535.97,343.28) .. controls (521.21,340.84) and (510.76,328) .. (510.87,313.5) ;  
\draw  [draw opacity=0][line width=1.5]  (177.01,365.11) .. controls (179.99,377.55) and (174.97,390.91) .. (163.65,397.85) .. controls (149.52,406.5) and (130.92,401.86) .. (122.12,387.48) .. controls (113.32,373.1) and (117.64,354.43) .. (131.77,345.78) .. controls (141.97,339.53) and (154.51,340.22) .. (164,346.5) -- (147.71,371.81) -- cycle ; 

\draw[ green!60!black ,  ,draw opacity=1 ][line width=1.5]  (177.01,365.11) .. controls (179.99,377.55) and (174.97,390.91) .. (163.65,397.85) .. controls (149.52,406.5) and (130.92,401.86) .. (122.12,387.48) .. controls (113.32,373.1) and (117.64,354.43) .. (131.77,345.78) .. controls (141.97,339.53) and (154.51,340.22) .. (164,346.5) ;  
\draw  [draw opacity=0][line width=1.5]  (302.19,404.08) .. controls (315.21,409.85) and (322.64,424.28) .. (319.15,438.6) .. controls (315.23,454.7) and (299,464.57) .. (282.9,460.65) .. controls (266.8,456.73) and (256.93,440.5) .. (260.85,424.4) .. controls (263.77,412.43) and (273.49,403.9) .. (284.9,401.93) -- (290,431.5) -- cycle ; 

\draw  [ green!60!black ,  ,draw opacity=1 ][line width=1.5]  (302.19,404.08) .. controls (315.21,409.85) and (322.64,424.28) .. (319.15,438.6) .. controls (315.23,454.7) and (299,464.57) .. (282.9,460.65) .. controls (266.8,456.73) and (256.93,440.5) .. (260.85,424.4) .. controls (263.77,412.43) and (273.49,403.9) .. (284.9,401.93) ;  
\draw  [color={rgb, 255:red, 208; green, 2; blue, 27 }  ,draw opacity=1 ][line width=1.5]  (222,295.5) -- (233.29,293.74) -- (228.96,304.32) ;
\draw  [color={rgb, 255:red, 208; green, 2; blue, 27 }  ,draw opacity=1 ][line width=1.5]  (252,313.5) -- (239.9,315.39) -- (242.89,303.52) ;
\draw  [color={rgb, 255:red, 208; green, 2; blue, 27 }  ,draw opacity=1 ][line width=1.5]  (124.15,399.57) -- (121.8,387.14) -- (134,390.5) ;
\draw  [color={rgb, 255:red, 208; green, 2; blue, 27 }  ,draw opacity=1 ][line width=1.5]  (219.52,227.35) -- (213.39,216.3) -- (225.82,214.01) ;
\draw  [color={rgb, 255:red, 208; green, 2; blue, 27 }  ,draw opacity=1 ][line width=1.5]  (224.54,186.99) -- (232.9,197.2) -- (219.87,199.26) ;
\draw  [color={rgb, 255:red, 208; green, 2; blue, 27 }  ,draw opacity=1 ][line width=1.5]  (87.13,161.25) -- (97.13,151.3) -- (99.63,165.19) ;
\draw  [color={rgb, 255:red, 208; green, 2; blue, 27 }  ,draw opacity=1 ][line width=1.5]  (261.17,156.93) -- (261.92,145.77) -- (270,153.5) ;
\draw  [color={rgb, 255:red, 208; green, 2; blue, 27 }  ,draw opacity=1 ][line width=1.5]  (223.66,34.89) -- (237.31,40.82) -- (226,50.5) ;
\draw  [color={rgb, 255:red, 208; green, 2; blue, 27 }  ,draw opacity=1 ][line width=1.5]  (291,136.5) -- (289.14,149.62) -- (278.27,142.04) ;
\draw  [color={rgb, 255:red, 208; green, 2; blue, 27 }  ,draw opacity=1 ][line width=1.5]  (389.87,172.01) -- (401.8,169.69) -- (396.38,180.56) ;
\draw  [color={rgb, 255:red, 208; green, 2; blue, 27 }  ,draw opacity=1 ][line width=1.5]  (508.98,81.76) -- (515.64,94.1) -- (501.71,92.45) ;
\draw  [color={rgb, 255:red, 208; green, 2; blue, 27 }  ,draw opacity=1 ][line width=1.5]  (416,192.5) -- (405.81,194.01) -- (409.1,184.25) ;
\draw  [color={rgb, 255:red, 208; green, 2; blue, 27 }  ,draw opacity=1 ][line width=1.5]  (435.08,261.03) -- (440.22,270.53) -- (429.47,271.54) ;
\draw  [color={rgb, 255:red, 208; green, 2; blue, 27 }  ,draw opacity=1 ][line width=1.5]  (579,310.5) -- (570.55,322.76) -- (562.93,309.97) ;
\draw  [color={rgb, 255:red, 208; green, 2; blue, 27 }  ,draw opacity=1 ][line width=1.5]  (431.73,292.98) -- (425.74,285.97) -- (434.49,283.05) ;
\draw  [color={rgb, 255:red, 208; green, 2; blue, 27 }  ,draw opacity=1 ][line width=1.5]  (319,335.5) -- (312.78,341.65) -- (308.92,333.81) ;
\draw  [color={rgb, 255:red, 208; green, 2; blue, 27 }  ,draw opacity=1 ][line width=1.5]  (294.13,467.88) -- (283.95,461.08) -- (294.49,454.83) ;
\draw  [color={rgb, 255:red, 208; green, 2; blue, 27 }  ,draw opacity=1 ][line width=1.5]  (289.92,340.35) -- (296.07,333.82) -- (298.53,342.45) ;


\node at  (222,295.5-17)  {\scalebox{1.5}{$\boldsymbol{a_1}$}}; 

\node at  (252,313.5+20)  {\scalebox{1.5}{$\boldsymbol{a_1^{-1}}$}}; 

\node at  (261.17-25,156.93)  {\scalebox{1.5}{$\boldsymbol{A_2^{-1}}$}}; 

\node at  (291+20,136.5)   {\scalebox{1.5}{$\boldsymbol{A_2}$}}; 

\node at  (396.38-22,180.56-30)   {\scalebox{1.5}{$\boldsymbol{A_{n-2}^{-1}}$}}; 

\node at  (416+30,192.5+10)  {\scalebox{1.5}{$\boldsymbol{A_{n-2}}$}};

\node at  (440.22+10,270.53-25)  {\scalebox{1.5}{$\boldsymbol{a_n^{-1}}$}};

\node at (425.74,285.97+17)  {\scalebox{1.5}{$\boldsymbol{a_n}$}};

\node at (312.78+30,341.65)  {\scalebox{1.5}{$\boldsymbol{a_2^{-1}}$}};

\node at (298.53-22,342.45)   {\scalebox{1.5}{$\boldsymbol{a_2}$}};

\node at (232.9+13,197.2-12)  {\scalebox{1.5}{$\boldsymbol{A_1}$}};

\node at (213.39,216.3+25)  {\scalebox{1.5}{$\boldsymbol{A_1^{-1}}$}};

\node at  (121.8-15,387.14+15) {\scalebox{1.5}{$\boldsymbol{z_1}$}}; 

\node at  (121.8+150,387.14+95) {\scalebox{1.5}{$\boldsymbol{z_2}$}}; 

\node at (237.31+25+320,40.82+300) {\scalebox{1.5}{$\boldsymbol{z_n}$}};

\node at  (237.31-25,40.82-20)    {\scalebox{1.5}{$\boldsymbol{Z_2}$}}; 

\node at  (237.31-25-120,40.82+80)  {\scalebox{1.5}{$\boldsymbol{Z_1}$}};

\node at  (237.31+25+240,40.82+20)    {\scalebox{1.5}{$\boldsymbol{Z_{n-2}}$}};





\node at  (320,220.5) {\scalebox{1.5}{$\boldsymbol{.}$}}; 

\node at  (325,220.167) {\scalebox{1.5}{$\boldsymbol{.}$}}; 

\node at  (330,219.833) {\scalebox{1.5}{$\boldsymbol{.}$}}; 

\node at  (335,219.5) {\scalebox{1.5}{$\boldsymbol{.}$}};

\node at  (324,267.5) {\scalebox{1.5}{$\boldsymbol{.}$}}; 

\node at  (329,264.688) {\scalebox{1.5}{$\boldsymbol{.}$}}; 

\node at  (334,261.875) {\scalebox{1.5}{$\boldsymbol{.}$}}; 

\node at  (340,258.5) {\scalebox{1.5}{$\boldsymbol{.}$}};

\end{tikzpicture}}

\caption{ Octopus diagram for computing the kinetic term of the CLD action at the Mandelstam maps, expressed as a boundary integral along the branch cuts and around the branch points of $\varphi_h$ and $\varphi_a$. }
    \label{fig:Octopus_Mandelstam_formula}
\end{figure}
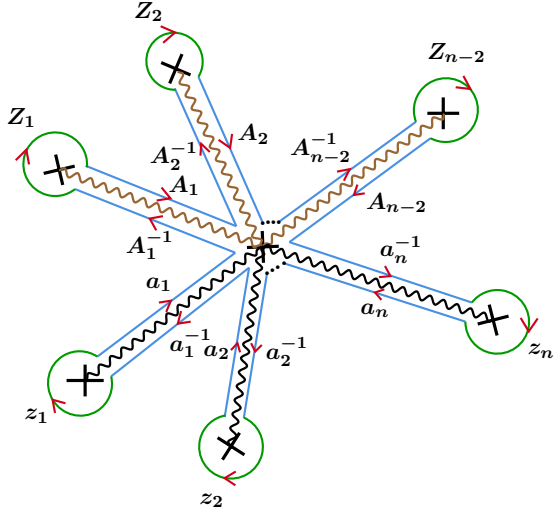

We introduce $n$ branch cuts $\{C_j\}_{j=1}^{n}$ joining $(z_\infty,\bar{z}_\infty)$ to each of the insertion points $(z_k,\bar{z}_k),\; k=1,\cdots,n$, and $(n-2)$ cuts $\{\widetilde{C}_J\}_{J=1}^{n-2}$ joining $(z_\infty,\bar{z}_\infty)$ to each of the interaction points $(Z_I,\bar{Z}_I),\; I=1,\cdots,(n-2)$. This way we make the integrand of the above area integral single-valued on $\mathcal{M}\backslash C$ with $C=\cup_j C_j\cup_J \widetilde{C}_J$. Then $\partial \mathcal{M}$ consists of both sides of the cuts, and boundaries of small circles around $\{(z_k,\bar{z}_k)\}_{k=1}^{n}$, $\{(Z_I,\bar{Z}_I)\}_{I=1}^{n-2}$ and $(z_\infty,\bar{z}_\infty)$ --- all traversed with positive orientation (see Fig.~\ref{fig:Octopus_Mandelstam_formula}). In the following, we compute the boundary integral from each component of $\partial \mathcal{M}$. As in the other computation of the CLD action in section \ref{Mandelstam_formula_and_its_derivation}, we introduce regulators given by the radii $\epsilon_{z_j}$ and $\epsilon_{Z_I}$ of small circles around $(z_j,\bar{z}_j)$ and $(Z_I,\bar{Z}_I)$ respectively.

\paragraph{Along $C_j$ :} Along  $C_j\equiv a_j+a_j^{-1}$ in fig. \ref{fig:Octopus_Mandelstam_formula}, we have
\begin{equation}
	\int_{a_j+a_j^{-1}}\varphi_h\, d\varphi_a=\int_{a_j}\left(\oint_{C_{z_j}}d\varphi_h\right) d\varphi_a = \oint_{C_{z_j}}d\varphi_h\cdot\int_{\bar{z}_j}^{\bar{z}_{\infty}}d\varphi_a=2\pi i \log \left[\frac{1}{\bar{z}_{\infty}^2} \frac{\epsilon_{z_j}e^{-i\theta_j}\prod_{k(\neq j))}^{n}(\bar{z}_j-\bar{z}_k)}{\prod_{I=1}^{n-2}(\bar{z}_j-\bar{Z}_I)}\right] \, ,
\end{equation}
where the first equality above follows from 
\begin{equation}
	\begin{split}
		&\int_{a_1} \varphi_h(a_1)\,d\varphi_a+\int_{a_1^{-1}} \varphi_h(a_1^{-1})\, d\varphi_a =\int_{a_1} \varphi_h(a_1)\,d\varphi_a-\int_{a_1} \varphi_h(a_1^{-1})\, d\varphi_a = \int_{a_1} \left[\varphi_h(a_1)-\varphi_h(a_1^{-1})\right]\, d\varphi_a\\
		&\text{and}\quad \delta \varphi_h=\left[ \varphi_h(a_1)-\varphi_h(a_1^{-1})\right] = \oint_{C_{z_1}}d\varphi_h.=2\pi i \, .
	\end{split}
\end{equation}
Likewise, for the other term of the boundary integral, we get
\begin{equation}
	\int_{a_j+a_j^{-1}}\varphi_a\, d\varphi_h=\int_{a_j}\left(\oint_{C_{\bar{z}_j}}d\varphi_a\right) d\varphi_h =-2\pi i \log \left[\frac{1}{z_{\infty}^2} \frac{\epsilon_{z_j}e^{i\theta_j}\prod_{k(\neq j))}^{n}(z_j-z_k)}{\prod_{I=1}^{n-2}(z_j-Z_I)}\right] \, .
\end{equation}
Hence, the net contribution from all $C_j$ with $j=1,\cdots,n$ reads
\begin{equation}\label{first}
		\frac{1}{2}\sum_{k=1}^{n}\oint_{a_k+a_k^{-1}}(\varphi_h \, d\varphi_a-\varphi_a\,d\varphi_h)=\pi i \,\log\left[\frac{1}{|z_{\infty}|^{4n}}\frac{\prod_{j\neq k}^{n}|z_j-z_k|^2}{\prod_{I=1}^{n-2}\prod_{k=1}^{n}|z_k-Z_I|^2}\,\prod_{k=1}^{n}\epsilon_{z_k}^2\right] \, .
	\end{equation}
\paragraph{Along $\widetilde{C}_J$:} Along $\widetilde{C}_J\equiv A_J+A_J^{-1}$, we have similar expressions,
\begin{equation}
	\int_{A_J+A_J^{-1}}\varphi_h\, d\varphi_a=\int_{A_J}\left(\oint_{C_{Z_J}}d\varphi_h\right) d\varphi_a = \oint_{C_{Z_J}}d\varphi_h \cdot\int_{\bar{Z}_J}^{\bar{z}_{\infty}}d\varphi_a=-2\pi i \log \left[\frac{1}{\bar{z}_{\infty}^2} \frac{\prod_{k=1}^{n}(\bar{Z}_J-\bar{z}_k)}{\epsilon_{Z_J}e^{-i\Theta_J}\prod_{I(\neq J)}^{n-2}(\bar{Z}_J-\bar{Z}_I)}\right] \, ,
\end{equation}
and similarly, the other term of the boundary integral takes the form
\begin{equation}
	\int_{A_J+A_J^{-1}}\varphi_a\, d\varphi_h=2\pi i \log \left[\frac{1}{z_{\infty}^2} \frac{\prod_{k=1}^{n}(Z_J-z_k)}{\epsilon_{Z_J}e^{i\Theta_J}\prod_{I(\neq J)}^{n-2}(Z_J-Z_I)}\right] \, .
\end{equation}
Net contribution from all $\widetilde{C}_J$ with $J=1,\cdots,(n-2)$ reads
\begin{equation}\label{second}
		\frac{1}{2}\sum_{I=1}^{n-2}\oint_{A_I+A_I^{-1}}(\varphi_h \, d\varphi_a-\varphi_a\,d\varphi_h)=-\pi i \, \log\left[\frac{1}{|z_{\infty}|^{4(n-2)}}\left|\sum_{k=1}^{n}\alpha_k z_k\right|^{n-2}\frac{\prod_{k=1}^{n}\prod_{I=1}^{n-2}|z_k-Z_I|}{\prod_{I\neq J}^{n-2}|Z_I-Z_J|}\frac{1}{\prod_{J=1}^{n-2}(2r_J)}\right] \, ,
	\end{equation}
where we have translated regulator $\epsilon_{Z_I}$ around $(Z_I,\bar{Z}_I)$ from the $\angle z$-plane to the $r_I$ on the $\angle\rho$-plane using \eqref{regulator_map} and \eqref{alpha_k-partial_2-rho}:
\begin{equation}\label{regulator_map_2}
\begin{split}
	 \log \epsilon_{Z_I}&= \frac{1}{2}\left(\log(2r_I)-\log|\partial^2 \rho(Z_I)|\right) =-\frac{1}{2}\log\left[\left|\sum_{k=1}^{n}\alpha_k z_k\right|\frac{\prod_{J(\neq I)}|Z_I-Z_J|}{\prod_{k=1}^{n}|Z_I-z_k|}\right]+\frac{1}{2}\log(2r_I) \, .
\end{split}     
\end{equation}

\paragraph{Around $(z_i,\bar{z}_i)$: } Next we compute the contour integral around $(z_i,\bar{z}_i)$:
\begin{equation}
	\oint_{C_{z_i}} \varphi_h \, d\varphi_a =\log\left[\left(\sum_{k=1}^{n}\alpha_k z_k\right)\frac{\prod_{I=1}^{n-2}(z_i-Z_I)}{\prod_{k(\neq 1)}^{n}(z_i-z_k)}\right]\oint_{C_{z_i}}d\varphi_a-\oint_{C_{z_1}}\log(z-z_1)\,d\varphi_a \, .
\end{equation}
The first contour integral gives $\oint_{C_{z_i}}d\varphi_a = -2\pi i$, while the second one yields
\begin{equation}
	\oint_{C_{z_i}}\log(z-z_1)\,d\varphi_a =	\oint_{C_{z_i}}\log(z-z_i)\, \left(\sum_{I=1}^{n-2}\frac{d\bar{z}}{\bar{z}-\bar{Z}_I}-\sum_{k=1}^{n}\frac{d\bar{z}}{\bar{z}-\bar{z}_k}\right) \,  ,
\end{equation}
 where all terms other than the one with $(\bar{z}-\bar{z}_i)^{-1}$, vanishes: 
\begin{equation}
	\oint_{C_{z_i}} d\bar{z}\, \frac{\log(z-z_i)}{\bar{z}-U}=\int_{0}^{-2\pi} (-i\epsilon_{z_i} e^{-i\theta_i}) d\theta_i \frac{\log(\epsilon_{z_i}e^{i\theta_i})}{\bar{z}_i+\epsilon_{z_i}e^{-i\theta_i}-U} \to 0,\quad \text{as}\quad \epsilon_{z_i}\to 0 \, . 
\end{equation}
The only term that contributes is the following:
\begin{equation}
	\oint_{C_{z_i}} \log(z-z_i)\frac{-d\bar{z}}{\bar{z}-\bar{z}_i} = +i\int_{0}^{-2\pi} d\theta_i \left[\log( \epsilon_{z_i})+i\theta_i\right]=-2\pi i \log (\epsilon_{z_i}) -2\pi^2  \, .
\end{equation}
Collecting these contributions, we obtain
\begin{equation}
	\oint_{C_{z_i}}\varphi_h \, d\varphi_a=-2\pi i\log\left[\left(\sum_{k=1}^{n}\alpha_kz_k\right)\frac{\prod_{I=1}^{n-2}(z_i-Z_I)}{\prod_{k\left(\neq i\right)}^{n}(z_i-z_k)}\right]+2\pi i \log \epsilon_{z_i}+2\pi^2 \, .
\end{equation}
Similarly, for the other term of the boundary integral, we get
\begin{equation}
	\oint_{C_{z_i}}\varphi_a \, d\varphi_h=2\pi i\log\left[\left(\sum_{k=1}^{n}\bar{\alpha}_k\bar{z}_k\right)\frac{\prod_{I=1}^{n-2}(\bar{z}_i-\bar{Z}_I)}{\prod_{k\left(\neq i\right)}^{n}(\bar{z}_i-\bar{z}_k)}\right]-2\pi i \log \epsilon_{z_i}+2\pi^2 \, .
\end{equation}
Thus net contribution from the contour circles around vertex operator insertion points reads
\begin{equation}\label{third}
		\frac{1}{2}\sum_{k=1}^{n}\oint_{C_{z_k}}(\varphi_h \, d\varphi_a-\varphi_a\,d\varphi_h)=-2\pi i\,\log\left[\left|\sum_{k=1}^{n}\alpha_kz_k\right|^{n}\frac{\prod_{k=1}^{n}\prod_{I=1}^{n-2}|z_k-Z_I|}{\prod_{k\neq j}^{n}|z_j-z_k|\prod_{k} \epsilon_{z_k}}\right] \, .
	\end{equation}

\paragraph{Around $(Z_J,\bar{Z}_J)$:}  Contribution from contour integral around the interaction points is similar:  
\begin{equation}
	\oint_{C_{Z_J}} \varphi_h \, d\varphi_a =\log\left[\left(\sum_{k=1}^{n}\alpha_k z_k\right)\frac{\prod_{I(\neq J)}^{n-2}(Z_J-Z_I)}{\prod_{k=1}^{n}(Z_J-z_k)}\right]\oint_{C_{Z_1}}d\varphi_a+\oint_{C_{Z_J}}\log(z-Z_J)\,d\varphi_a \, .
\end{equation}
With $\oint_{C_{Z_1}}d\varphi_a=2\pi i$, and so the last term on the rhs of the above expression reads
\begin{equation}
	\oint_{C_{Z_J}}\log(z-Z_J)\,d\varphi_a=\oint_{C_{Z_1}} \log(z-Z_J)\frac{d\bar{z}}{\bar{z}-\bar{Z}_J} = -i\int_{0}^{-2\pi} d\Theta_J \left[\log( \epsilon_{Z_J})+i\Theta_J\right]=2\pi i \log \epsilon_{Z_J} +2\pi^2 \, .
\end{equation}
Thus, we obtain
\begin{equation}
	\oint_{C_{Z_J}}\varphi_h \, d\varphi_a=2\pi i\log\left[\left(\sum_{k=1}^{n}\alpha_k z_k\right)\frac{\prod_{I(\neq J)}^{n-2}(Z_J-Z_I)}{\prod_{k=1}^{n}(Z_J-z_k)}\right]+2\pi i \log \epsilon_{Z_J}+2\pi^2 \, .
\end{equation}
Similarly, for other term of the boundary integral, we obtain
\begin{equation}
	\oint_{C_{Z_J}}\varphi_a \, d\varphi_h=-2\pi i\log\left[\left(\sum_{k=1}^{n}\bar{\alpha}_k \bar{z}_k\right)\frac{\prod_{I(\neq J)}^{n-2}(\bar{Z}_J-\bar{Z}_I)}{\prod_{k=1}^{n}(\bar{Z}_J-\bar{z}_k)}\right]-2\pi i \log \epsilon_{Z_J}+2\pi^2 \, .
\end{equation}
Hence net contribution from the contours around the interaction point reads
\begin{equation}\label{fourth}
		\frac{1}{2}\sum_{I=1}^{n-2}\oint_{C_{Z_I}}(\varphi_h \, d\varphi_a-\varphi_a\,d\varphi_h)=\pi i\, \log\left[\left|\sum_{k=1}^{n}\alpha_k \,z_k\right|^{n-2}\frac{\prod_{I\neq J}^{n-2}|Z_I-Z_J|}{\prod_{I=1}^{n-2}\prod_{k=1}^{n}|z_k-Z_I}\,\prod_{I=1}^{n-2}(2r_I)\right] \, ,
	\end{equation}
 where we have used \eqref{regulator_map_2}.

\paragraph{Around $(z_\infty,\bar{z}_\infty)$:}  To compute $\oint_{C_{z_\infty}}\varphi_h \, d\varphi_a$, 
we first note that near $(z_\infty,\bar{z}_\infty)$, $\varphi_h$ and $d\varphi_a$ behaves as 
\begin{equation}
	\begin{split}
		\varphi_h \sim \log\left(\sum_{k=1}^{n}\alpha_k z_k\right)-2\log(z)+\mathcal{O}\left(\frac{1}{z}\right),\quad \text{and}\quad d\varphi_a \sim -\frac{2}{\bar{z}}\,d\bar{z}+\mathcal{O}\left(\frac{1}{\bar{z}^2}\right) d\bar{z} \, .
	\end{split}
\end{equation}
Using this, the contour integral becomes
\begin{equation}
	\begin{split}
		\oint_{C_{z_\infty}}\varphi_h\, d\varphi_a&=-2\log\left(\sum_{k=1}^{n}\alpha_k z_k\right)\underbrace{\oint_{C_{z_\infty}}\frac{d\bar{z}}{\bar{z}}}_{-2\pi i}+4\underbrace{\oint_{C_{z_\infty}} \frac{d\bar{z}}{\bar{z}}\, \log(z)}_{-2\pi i \log|z_{\infty}|-2\pi^2}\\
        &=2\pi i \log\left(\sum_{k=1}^{n}\alpha_k z_k\right)^2-2\pi i \log|z_{\infty}|^4-8\pi^2 \, .
	\end{split}
\end{equation}
Similarly, for the other term around $(z_\infty,\bar{z}_\infty)$, we have
\begin{equation}
	\oint_{C_{z_\infty}}\varphi_a\, d\varphi_h=-2\pi i \log\left(\sum_{k=1}^{n}\bar{\alpha}_k \bar{z}_k\right)^2+2\pi i \log|z_{\infty}|^4-8\pi^2 \, .
\end{equation}
Net contribution from the contour around $(z_\infty,\bar{z}_\infty)$ is thus
\begin{equation}\label{fifth}
		\frac{1}{2}\oint_{C_{z_\infty}}(\varphi_h \, d\varphi_a-\varphi_a\,d\varphi_h)=2\pi i \, \log \left|\sum_{k=1}^{n}\alpha_k z_k\right|^2-2\pi i\, \log \left|z_{\infty}\right|^4 \, .
	\end{equation}
Adding all the contributions \eqref{first}, \eqref{second}, \eqref{third}, \eqref{fourth}, and \eqref{fifth}, we obtain the kinetic term of CLD action localized at the Mandelstam maps:
\begin{equation}
		\begin{split}
			-i\int_{\mathcal{M}}d^2z\, \partial\varphi\, \bar{\partial}\varphi &=\frac{1}{2}\left[\sum_{k=1}^{n} \int_{a_k+a_k^{-1}+C_{z_k}} +\sum_{k=1}^{n}\oint_{C_{z_k}}+\sum_{I=1}^{n-2}\int_{A_I+A_I^{-1}} +\sum_{I=1}^{n-2}\oint_{C_{Z_I}}+\oint_{C_{z_\infty}}\right](\varphi_h \, d\varphi_a-\varphi_a\, d\varphi_h)\\
			&=-2\pi i\log\left[\left|\sum_{k=1}^{n}\alpha_k z_k\right|^{n-2}\frac{\prod_{k=1}^{n}\prod_{I=1}^{n-2}|z_k-Z_I|^3}{\prod_{j\neq k}^{n}|z_j-z_k|^2 \prod_{I\neq J}^{n-2}|Z_I-Z_J|}\,\frac{|z_{\infty}|^8}{\prod_{I=1}^{n-2}(2r_I)\prod_{k=1}^{n}\epsilon_{z_k}^2}\right] \, .
	\end{split}
\end{equation}
Using \eqref{alpha_k-partial_2-rho}, it matches with what has been computed in the alternative way in \eqref{canonical_1}.

 \small
 \baselineskip=.75\baselineskip
 \bibliography{ref}

\end{document}